**Strategic Responses to Technological Change: Evidence from an Online Labor Market**

Shun Yiu (Wharton), Rob Seamans (NYU Stern), Manav Raj (Wharton), and Teng Liu (UC Santa Cruz)[1]

May 2025

**Abstract**

In this project, we examine how freelancers changed their strategic positioning on an online work platform following the launch of ChatGPT in November 2022 – a major advance in AI technologies. We document that post-ChatGPT, freelancers bid on fewer jobs and reposition themselves by differentiating their distribution of bids (i.e., job applications) relative to their prior behavior. We disentangle heterogeneity in strategic responses by exploring how exposure to changes in demand or supply shape incumbent repositioning. We find that the launch of ChatGPT was associated with a short-term decrease in labor demand and an increase in labor supply, though these changes vary across work domains. In response to decreases in labor demand, workers changed their horizontal positioning and withdrew from the platform. In response to increases in labor supply, workers were less likely to decrease bidding or reposition horizontally but shifted their vertical position by targeting lower-value jobs. We further show that repositioning is less likely for high-skill freelancers who face greater adjustment costs. This research contributes to our understanding of how and why workers respond to technological change in the context of recent advances in AI technologies.

Keywords: AI, generative AI, skills, labor, positioning, freelance, online work platforms

[1] Authors are listed in reverse alphabetical order and contributed equally. The authors thank Peter Capelli, Ozge Demirci, Anil Gupta, John Horton, Dmitry Sorokin, Prasanna Tambe, Cedric Xu, Valery Yakubovich, Xinrong Zhu, and seminar participants at the HBS TOM unit, the Strategy & Entrepreneurship Area at the Washington University in St. Louis and the Smith Entrepreneurship Research Conference for thoughtful feedback. Seamans thanks the Smith-Richardson Foundation, Raj and Yiu thank the Mack Center, and Liu thanks the Upwork Research Institute for generous funding and support. Corresponding author email: manavraj@wharton.upenn.edu. Any errors are our own.

## Introduction

In this project, we examine whether and how workers respond to advances in artificial intelligence (AI) technologies. Recent advances in AI have generated both excitement about its ability to increase worker productivity and concerns regarding its potential to cause job loss. Given such excitement and concern, scholars have been eager to document how advances in AI technologies change how work is done and affect labor markets (Capelli, Tambe, and Yakubovich, 2024). Much of the early research in this area has focused on understanding how AI may reshape labor demand by either augmenting (e.g., Brynjolfsson, Li, and Raymond, 2023; Choi and Schwarcz, 2023; Dell'Acqua *et al.*, 2023; Noy and Zhang, 2023; Peng *et al.*, 2023; Teutloff *et al.*, 2025) or automating labor (Demirci, Hannane, and Zhu, 2023; Hui, Reshef, and Zhou, 2023; Liu *et al.*, 2023). However, understanding how AI affects labor requires attention to supply-side responses as well (e.g., Acemoglu and Restrepo, 2019; Autor, 2015; Hampole *et al.*, 2025), as workers are agentic and will adapt to new technologies by altering their behaviors strategically. Despite these considerations, extant research says little about how workers may strategically reposition themselves following technological change, whether horizontally across domains or vertically within them. Through this study, we generate insight into whether and how technological change elicits worker repositioning and the mechanisms underlying such responses.

We study this question by examining how freelance workers respond to a sudden advancement in generative AI technologies – the launch of ChatGPT in December 2022 – using proprietary data from a leading online labor market platform. Online labor markets offer a uniquely apt setting for this question: they are large, diverse, and fast-moving, allowing us to observe granular changes in worker behavior in ways that may be obscured in traditional labor market data (Agrawal *et al.*, 2015; Horton and Tambe, 2015). We begin by asking a simple but foundational question: do freelance workers respond to the launch of ChatGPT, and if so, how? From there, we consider why these responses may differ across workers, focusing on two sources of heterogeneity: a) differences in exposure to technology-induced changes in the demand for and supply of labor, which may both alter competitive conditions (albeit in different ways); and b) variation in worker skill, which may shape adjustment costs associated with repositioning.

We document significant changes in activity on the platform and in freelancer behavior following the launch of ChatGPT. At the platform-level, we document a decrease in the number of jobs posted (i.e., a decrease in labor demand), but an increase in the number of active freelancers on the platform (i.e., an increase in labor supply). We find evidence that incumbent freelancers' strategic positioning, as captured by the jobs they applied for, shifted in response to technological change and the resulting changes to on-platform competition.[1] On average, incumbent freelancers bid on fewer jobs and altered their horizontal positioning, as captured by the distribution of bids across work categories. We do not find evidence that, on average, incumbent freelancers altered their vertical positioning, as captured by the distribution of bids to high-value jobs.

In outlining the changes in strategic positioning post-ChatGPT, we document that such changes vary across freelancers that specialize in different work domains, and further, that work domains differ in the extent to which the demand for labor decreased and the supply of labor increased. Accordingly, to shed light on differential responses to technological change, we consider how heterogeneous changes in the demand for and supply of labor may drive repositioning. To do so, we construct a measure of ex-ante exposure to post-ChatGPT demand and supply change based on pre-period bidding behavior and explore how such exposure moderates the post-ChatGPT effect. We find evidence that exposure to technology-induced demand contraction or supply expansion led to different strategic responses. Workers that were more exposed to demand contraction were more likely to respond by differentiating horizontally relative to prior behavior, while workers more exposed to supply expansion were less likely to differentiate horizontally but altered their vertical positioning, bidding for lower value jobs perhaps in response to a more competitive market.

Finally, we explore heterogeneity in strategic repositioning based on freelancer skill level—an individual-level characteristic often discussed in considering the effects of AI on worker outcomes (e.g., Brynjolfsson *et al.*, 2023; Dell'Acqua *et al.*, 2023; Noy and Zhang, 2023) that may also determine workers' re-positioning-related adjustment costs. We find that higher-skilled freelancers were less likely to decrease their bidding on the platform and reposition horizontally but were more likely to bid for lower value jobs. We view these responses as reflecting a higher cost of repositioning across domains for such workers. High-skill freelancers that have invested more to develop expertise in a given domain may be reticent to switch work areas, and instead may alter their vertical positioning to navigate the more competitive, post-ChatGPT market.

[1] We use "incumbent" to refer to freelancers that were already on the platform before the launch of ChatGPT. The sample definition is provided later in the manuscript.

With this research, we contribute to a recent body of work that seeks to understand how AI affects labor and firm strategy. Building on work that studies how AI technologies may augment (e.g., Choi and Schwarcz, 2023; Dell'Acqua *et al.*, 2023; Noy and Zhang, 2023) or automate labor (e.g., Demirci *et al.*, 2023; Hui *et al.*, 2023; Liu *et al.*, 2023), we examine the supply-side responses from freelance workers following an advance in generative AI technology. To our knowledge, we are the first to consider how generative AI technologies drive workers to strategically reposition themselves, adding to a body of work that considers how AI affects strategic decision making (e.g., Csaszar, Ketkar, and Kim, 2024; Krakowski, Luger, and Raisch, 2023). In doing so, we answer a call for research that considers how AI may affect the future of work (e.g., Anthony, Bechky, and Fayard, 2023; Bailey *et al.*, 2022).

Importantly, beyond providing descriptive evidence of repositioning, we integrate and extend theory regarding how technology-enabled changes in the demand for or supply of labor may drive repositioning in different ways. We build on literature that suggests that workers may respond to changes in the demand for labor by repositioning horizontally (e.g., Autor, Levy, and Murnane, 2003) and literature that documents how firms may respond to increases in competition by repositioning vertically (e.g., Dixit and Stiglitz, 1977) by showing how technological change may also elicit vertical repositioning among workers within a domain. Finally, we add to a body of literature that considers whether the benefits of AI technologies are more likely to be captured by high- or low-skilled workers (e.g., Dell'Acqua *et al.*, 2023; Noy and Zhang, 2023; Peng *et al.*, 2023) and literature that considers how skill or tenure may shape adjustment costs (Kambourov and Manovskii, 2009; Neal, 1995) by documenting how high vs. low skill workers are likely to alter their strategic behaviors in response to advances in AI.

## Related Literature

*Supply-Side Responses to Technological Change*

Technological change has the potential to affect the allocation of tasks or skills between capital and labor (Acemoglu and Restrepo, 2019). When novel technologies make it possible for capital to replace labor for certain tasks, the resulting "displacement effect" can reduce demand for certain skills (Acemoglu and Restrepo, 2019). For example, the adoption of mechanical call-switching technologies in the twentieth century reduced the returns to telephone operating skills and thus resulted in a decrease in employment and wages among incumbent telephone operators (Feigenbaum and Gross, 2024a, 2024b). As technology substitutes for labor in certain tasks, labor may be reallocated to other tasks where it has a comparative advantage (Autor *et al.*, 2003). Such changes will have implications for how workers position themselves, in terms of which jobs they apply for and what skills they develop.

Existing research on workers' adaptation to technological change tends to focus on the results of adaptation rather than the adaptive behavior itself. For example, Feigenbaum and Gross (2024b) show that after the automation of telephone operations, many women telephone operators transitioned into jobs that required similar or less skill, and Furman and Teodoridis (2020) suggest that automating technology cause scientists to innovate in more distant and diverse domains. Autor and Dorn (2013) also document how the rise of automating technologies resulted in the reallocation of low-skill labor into service occupations. The displacement effects of technological change are often conceptualized and measured as a combination of demand-side shifts and supply-side responses. However, as Autor (2015) noted, how labor supply responds to technological change is a significant contributor to the net effect of technological change on employment. Accordingly, we believe there is value in documenting supply-side worker re-positioning in response to technological change and documenting the drivers of such actions.

In understanding how workers may strategically re-position themselves in response to technological change, this study is related to research that documents how knowledge workers within organizations strategically adjust their behaviors to best take advantage of novel technologies like AI (Beane and Anthony, 2024; Lebovitz, Lifshitz-Assaf, and Levina, 2022) and when and how worker positioning can shape career outcomes (e.g., Ferguson and Hasan, 2013; Narayanan, Balasubramanian, and Swaminathan, 2009; Zuckerman *et al.*, 2003). A portion of this literature considers how technologies like AI may differentially affect more- and less-experienced or skilled workers and finds evidence suggests that less-skilled and less-experienced workers are more likely to benefit from advances in AI technologies (e.g., Dell'Acqua *et al.*, 2023; Noy and Zhang, 2023; Peng *et al.*, 2023). Further, particularly as we study a freelance labor market platform, and each freelancer may be considered an entrepreneur operating their own firm, we draw from a literature that studies how incumbent firms change strategies in the face of technological change (e.g., Eggers and Park, 2018; Klepper, 1996; Tushman and Anderson, 1986). Such research has found that firms may attempt to differentiate themselves in new ways following technological change (Miller and Wang, 2024); that technological change can spur entry of new kinds of products and producers (Benner and Waldfogel, 2023); and that firms' responses to technological change may depend on their prior strategic positioning (Benner and Waldfogel, 2016).

*Our Approach*

In the extant research, we build on literature studying workers' responses to technological change (e.g., Beane and Anthony, 2024; Feigenbaum and Gross, 2024b; Furman and Teodoridis, 2020) and literature on incumbent adaptation to technological change (e.g., Benner and Waldfogel, 2023; Eggers and Park, 2018; Miller and Wang, 2024) by investigating how freelance workers reposition themselves following advances in generative AI technologies. In recent years, scholars and practitioners have begun to consider how advances in AI—particularly generative AI, which can produce novel content (Berg, Raj, and Seamans, 2023)—may affect labor. Early evidence highlights both the transformative promise of these tools and their potential to displace workers. Generative AI tools appear to increase productivity across a broad range of settings and may provide lower-skill or less-experienced workers with a way to make up for skill deficits (e.g., Dell'Acqua *et al.*, 2023; Noy and Zhang, 2023; Peng *et al.*, 2023). At the same time, they may reduce demand for more exposed or affected occupations and thus reduce earnings and employment for some workers (Demirci *et al.*, 2023; Hui *et al.*, 2023; Liu *et al.*, 2023). Given these potential changes to the demand for and supply of labor, we expect that workers will respond to advances in generative AI technologies. However, because the technologies are so new and labor market data is scarce, little research has considered how such strategic responses might unfold. Research on this important question will help us anticipate how the broader labor market will evolve as generative AI technologies become prevalent and further shed light on how workers may reposition themselves in response to future technological advancements.

We examine supply-side responses to the emergence of generative AI technologies in the context of a freelance labor market platform. Freelance workers account for a significant portion of the U.S. labor force, estimated to be between 6.9% and 15% (Abraham *et al.*, 2023). The rise in importance of online labor markets in recent years makes this a useful setting for studying labor market behavior (Agrawal *et al.*, 2015; Horton and Tambe, 2015). Further, relative to studying workers' responses in organizational settings, given the more flexible nature of freelance work, we may be able to more quickly capture changes around the discontinuity in this setting. We study how freelance workers responded to the launch of ChatGPT, a recent advancement in generative AI technologies using granular internal data from a leading international online labor market platform. The data allow us to observe a large sample of workers with different occupations and their labor market behavior (e.g., job applications) over time.

Ex ante, it is theoretically ambiguous how workers may respond to the sudden advance in generative AI technologies, particularly as technological change may impose a number of counteractive effects on workers. The sudden advance in generative AI technologies may decrease the demand for certain skills (e.g., Demirci *et al.*, 2023; Hui *et al.*, 2023), while also decreasing the cost of carrying out certain tasks, thus inducing new entry and increasing competition (e.g., Autor *et al.*, 2003; Klepper, 1996). Put differently, technology change may alter workers' strategic repositioning by changing the demand for skills, as well as the supply of skills. It is plausible that workers could respond to these changes in different ways. For example, freelancers could have reduced their activity on the platform because the demand for their skills declined or increased their activity on the platform because they could leverage the new technology to increase their productivity. Further, freelancers might have taken advantage of new AI tools to exploit domains that they were already familiar with or instead leveraged AI as a tool to lower barriers to entry to enter new domains. Of course, such responses may differ based on the freelancers' ex-ante specialization as well as individual-level characteristics.

Because making predictions is challenging in this setting, we instead adopt a question-based and data-driven approach. We first document changes in freelancer behavior following the launch of ChatGPT to characterize the ways in which freelancers reposition themselves following technological change. Then to better understand heterogeneity in freelancers' response to technological change and to shed light on the drivers of technology-induced strategic repositioning, we investigate the mechanisms underlying them. Specifically, we consider how worker repositioning may be shaped by exposure to changes in the demand for skills, as well as changes in the supply of labor (caused by new entry and peer re-positioning). Through this investigation, we not only describe how workers reposition themselves following technology change but shed light on why they may do so and the drivers of their choices. We then investigate how individual-level characteristics may also shape responses, holding work domain constant. Following our empirical analyses, we integrate our results to generate theoretical contributions to literature studying strategic responses to technological change.

We provide more detail on our empirical setting and study design in the following section.

**Empirical Setting and Study Design**

*Online Labor Market Platform*

Our research setting is one of the largest online labor market platforms that facilitates matching between freelancers and hiring clients. The hiring process starts with clients writing a job posting on the platform. Freelancers can then submit applications to those job postings as a bid. After vetting the applications, clients decide whom to hire. An important characteristic of this hiring process is that job bids are costly. Every month, each freelancer is allowed to make a fixed number of bids for free. They pay a small fee for any additional job bids. The extent of bidding and kinds of job posts freelancers bid on are thus important labor market behaviors that reveal freelancers' strategic positioning. Put another way, freelancers' bidding behavior should capture how freelancers are making strategic decisions regarding what is best for them on the platform, given the cost of applications and the expected benefits. We draw from the labor market platform's internal database, which records details about each freelancer's bids on the platform. We discuss how we use the data to measure strategic behavior later in the text.

*ChatGPT as a Discontinuous Shock*

We use the public launch of ChatGPT on 30 November 2022 as a sudden and discontinuous advance in the availability of generative AI technologies. ChatGPT is an AI-based conversational agent developed by the company OpenAI. The original release of ChatGPT was built on the GPT (Generative Pre-Trained Transformer) architecture, specifically GPT-3.5, that OpenAI had been developing internally for years. The GPT-3.5 architecture was made available by OpenAI via API access in March 2022 and was originally described as an update to the previous GPT-3 architecture (OpenAI, 2022). While API access allowed those with more technical knowledge an early ability to access this tool, ultimately, OpenAI hoped to create a user-friendly interface that would also allow the general public to use the underlying GPT technology (Roose, 2023). OpenAI originally planned to release a new model, GPT-4, with a chatbot interface that would allow users to more easily access the tool in early 2023. However, fearing that rival companies might beat them to the market, OpenAI executives decided in November 2022 to update and release a chatbot using GPT-3.5 (Roose, 2023). Accordingly, the public release of ChatGPT was unexpected, even to those working within OpenAI, until mere weeks before the launch.

Partially owing to the unexpected nature of the launch, neither the public nor developers of ChatGPT expected it to garner such widespread attention and adoption, let alone have an immediate and widespread impact on labor markets. The release of ChatGPT in November 2022 was intended to be a "research preview" that would provide feedback and generate interest from the public (Heaven, 2023). John Schulman, one of the leading developers of ChatGPT, said "it was definitely a surprise for all of us how much people began using it. We work on these models so much, we forget how surprising they can be for the outside world sometimes" (Heaven, 2023). Within two months of its debut, ChatGPT had more than 30 million users and more than 5 million visits a day, making it one of the fastest-growing software products in recorded history (Roose, 2023).

The sudden launch of ChatGPT and its unexpected speed of adoption represented an unanticipated technological change that introduced discontinuous advancement to generative AI capabilities. The release of ChatGPT also spurred a boom in interest and coverage in generative AI by media outlets, which also contributed to its adoption and diffusion (Baldassarre *et al.*, 2023). For these reasons, previous studies have used the launch of ChatGPT as a discontinuous shock to AI capabilities (e.g., Demirci *et al.*, 2023; Hui *et al.*, 2023; Saggu and Ante, 2023; Yuan and Chen, 2023). Given the unexpected nature of its launch and rapid adoption, we believe that the release of ChatGPT can be viewed as quasi-exogenous with respect to activity on the online labor market platform. In later analyses, we probe the plausibility of this assumption by examining pre-trends in platform activity and freelancer behavior before the release of ChatGPT.

*Platform-Level Descriptive Data*

We begin with a set of non-parametric analyses to document how activity across the platform changed following the launch of ChatGPT before considering changes in incumbents' strategic positioning. We construct a panel dataset at the month level measuring total demand- and supply-side activity on the platform across a number of dimensions. Figure 1 presents binned scatterplots showing how these measures changed over the course of the sample period.

We capture aggregate demand on the platform by measuring the log total job posts that received at least one bid on the platform in a month (Panel A). There was an immediate drop in job postings on the platform in December 2022 right after the launch of ChatGPT, followed by a rebound and then a decrease again at a steeper rate toward the end of the sample period. This suggests that the launch of ChatGPT was accompanied by a contraction in demand on the platform. We then probed how demand across vertical market segments changed by plotting the ratio of high over low value job posted on the platform (Panel B). High value jobs are defined as jobs with earnings of over $1,000, jobs with earnings below that level are considered low value jobs. In the one year before the launch of ChatGPT, we see that this ratio hover around 0.5. There was a sharp drop in high value job posts immediately after the launch of ChatGPT, followed by a rebound, then a decrease again. These patterns suggest that the decrease in job postings on

the platform is particularly strong among high value jobs. We then examined how the breadth of demand shifted (Panel C). We calculated the Herfindahl-Hirschman Index of job postings across the 12 work domains on the platform for each month. We see that the level of concentration in job postings had a downward trend prior to the launch of ChatGPT. When ChatGPT launched, there was a sharp drop in concentration, but it quickly reversed and formed an upward trend. This suggests that while the overall number of job postings decreased, such a drop in demand did not impact the entire platform equally but differentially affected specific work domains. We explore this heterogeneity in a later section.

We then probe changes in the supply of labor on the platform. First, we document an immediate increase in unique bidding freelancers (Panel D). We see an immediate one-month drop in December 2022 that quickly rebounded and accompanied with a sustained elevation in the number of bidding freelancers on the platform for around 10 months after the launch of ChatGPT. This increase returned to levels close to periods before the launch of ChatGPT after 10 months. This seems to suggest that the launch of ChatGPT was accompanied by a wave of supply expansion. We then document the role of entrants, defined as freelancers who made their first bid on the platform in the last six months, in shaping supply dynamics on the platform. Specifically, we look at the ratio of entrants over incumbents over time (Panel E). We document that pre-ChatGPT, there was a downward trend in the entrant-to-incumbent ratio stabilized to a value around 1, i.e., for every incumbent, there was one entrant entering the platform. After the launch of ChatGPT, there was a sharp increase in the number of entrants. This upward trend sustained for many months after the launch and stabilized at an elevated level. We further characterize entrant-related supply dynamics by examining the entrants' bid to high over low value job ratio (Panel F). Entrants who entered the platform immediately after the launch of ChatGPT targeted low value jobs more so than high value jobs. However, within a few months, this trend switched, and entrants targeted high value jobs at a rate that exceeded that in the pre-period. Finally, we document that, post-ChatGPT, the ratio of international (i.e., non-US) freelancers increases on the platform, suggesting that ChatGPT may have increased participation particularly among international workers (Panel G).

In the aggregate, ChatGPT seems to have engendered both demand-side changes (i.e., contraction) and supply-side changes (i.e., expansion) on the online labor market platform. In this paper, we remain agnostic as to why the launch of ChatGPT engender these changes, as recent research is just beginning to unpack the kinds of jobs, skills, and tasks that generative AI technologies automate and augment (Acemoglu, 2025; Acemoglu *et al.*, 2022). Our approach is to treat these changes as exogenous to a given freelancer and focus on understanding how incumbent freelancers responded to these and repositioned themselves on the platform.

## Sample, Data, and Methodology

### *Sample and Measures*

Our main empirical analyses focus on identifying trends in incumbent freelancer strategic behavior following the launch of ChatGPT and then disentangling the mechanisms that underly changes in freelancer behavior. Our goal is to shed light on how technological change shapes strategic re-positioning among workers. Our main sample period runs from June 2022 through March 2024, providing us with a six-month pre-launch window and sixteen months post-ChatGPT, thereby allowing us to examine changes in freelancer behavior in the short and medium term. We restrict our analysis to incumbent freelancers who are active on the platform, as we would otherwise lack the data required to analyze labor market behavior before and after the event of interest. In order to be included in our sample, a freelancer must have: 1) completed at least one contract on the platform before the launch of ChatGPT, 2) submitted at least one job bid in the three months before the launch of ChatGPT, and 3) submitted on average at least one job bid per month from January through November 2022. These inclusion criteria yield a sample of 312,143 unique freelancers.[2] Aggregated at the monthly level, our data constitutes a freelancer-month panel of approximately 6.6 million freelancer-month observations.

[2] There are in total approximately 816K unique freelancers who have completed at least one job on the platform during our sample period.

We utilize the data to measure freelancer strategic positioning before and after the launch of ChatGPT, as measured by the volume and characteristics of bids (i.e., job applications). We construct a number of variables to measure freelancer behavior and strategic positioning on the platform.

***Bids***. A primary outcome of interest relates to changes in freelancers' bidding behavior following the launch of ChatGPT. We count the total number of bids a freelancer made each month. Freelancer bidding behavior can often be lumpy and uneven, resulting in freelancer-months with zero bids that we could not use to construct strategic positioning variables. To account for this lumpiness and to better allow us to view freelancers' strategies over medium-term windows, we apply a three-month rolling window when summing the number of bids. Since the nature of the data-generation process is count-based and the data are heavily skewed, we log-transform the resulting count variable.

***Bid Similarity.*** We measure bid similarity to examine differentiation in freelancers' positioning. We measure the similarity of the distribution of bids across specializations relative to the focal freelancer's pre-period bidding behavior. We first transform a freelancer's three-month rolling bids into vector representations, reflecting freelancers' distribution of bids across specializations, and normalize this vector by the total number of bids submitted by the freelancer. We then construct a referent vector that measures distribution of freelancers' bids across specializations over the previous three-month rolling period. We compute a cosine similarity comparing a freelancer's bids during time period *t* against each of these referent vectors to construct measures that capture similarity/differentiation relative to the focal freelancer's prior bidding behavior. Bid similarity relative to prior behavior is missing if a freelancer did not apply for any jobs in either the current or prior three-month rolling window. This captures the extent to which the distribution of bids across specializations is similar to or different from the freelancer's distribution of bids in the prior three months.[3]

***Percent High Value Bid.*** Contracts on the online labor market platform are associated with different levels of earnings. Contracts that pay well are attractive to freelancers for several reasons. First, freelancers earn more by virtue of taking on a better-paying job. Second, since the platform charges a flat fee for contract initiation, securing a large contract instead of multiple smaller ones means freelancers pay less of their earnings to the platform. Third, as the initiation of a contract often involves upfront communication and negotiation with clients, securing large contracts means freelancers can spend less time on administrative tasks. We define high-value contracts as those that pay more than US$1,000; we measure what percentage of freelancers' bids were for high-value contracts. This variable is missing if a freelancer did not apply for any posted jobs in the current three-month rolling window.

[3] For concision, the remaining references to "differentiation" in the text refer to differentiation in terms of the distribution of bids across specialties as described here.

While there are many ways to characterize strategic positioning by freelancers on the platform, for ease of interpretation and to provide a cohesive framework, we focus on these three measures to capture changes in the volume of activity (number of bids), in horizontal positioning (bid similarity), and in vertical positioning (percent of high-value bids). Table 1 shows summary statistics and other variables of interest.[4] In supplementary analyses, we examine a broader range of strategic positioning variables.

*Empirical Strategy*

To document how incumbent freelancers change their behavior, we first look at changes in behavior across the sample of incumbent freelancers. We construct a simple relative time model with freelancer-level fixed effects and a categorical month-by-year variable:

$$Y_{it} = \sum_{k=T_0}^{k=-1} Pre_k + \sum_{k=1}^{k=T_1} Post_k + \alpha_i + \varepsilon_{it} \qquad (1)$$

where $Pre_k$ and $Post_k$ are indicator variables that are equal to 1 for observations that are $k$ months before and after the launch of ChatGPT, and $\alpha_i$ denotes freelancer fixed effects. By plotting the coefficients of the time categorical variable, we can graphically analyze changes in freelancers' strategic behavior over time while accounting for individual-level freelancer variation. This allows us to probe for any trends in freelancer activity leading up to the launch of ChatGPT, as well as how any changes following the launch of ChatGPT manifest temporally. For this analysis, the omitted group ($t = 0$) is November 2022, representing the month prior to the launch of ChatGPT on November 30, 2022. We control for the log-transformed total count of bids in all models except when *Bids* is the dependent variable, as the measures of strategic positioning are mechanically related to the count of bids. We also estimate the average change in strategic positioning using a simple pre- vs. post-comparison:

$$Y_{it} = \beta Post_t + \alpha_i + \varepsilon_{it} \qquad (2)$$

where $Post_t$ is an indicator variable equal to 1 for observations after the launch of ChatGPT, and $\alpha_i$ denotes freelancer fixed effects. As with the relative time model, we also control for the log-transformed total count of bids in all models except when *Bids* is the dependent variable.

[4] See *Appendix Table A1* for pairwise correlations.

For both the relative time models and the simple pre- vs. post-comparison, we note that we are capturing simple time trends. In an ideal natural experiment, we would have a counterfactual set of freelancers that were not exposed to the launch of ChatGPT that we could use as a control group. Such a control group would allow us to draw causal inferences about how the launch of ChatGPT affected freelancers' strategic positioning. However, creating a counterfactual is not feasible given that the launch of ChatGPT happened internationally and we do not have a sufficient or appropriate control sample of freelancers on the online labor market platform that were not exposed to this technological change.

Instead, we adopt a panel differences (PD) approach that utilizes data from the previous year to construct a pseudo-control group (e.g., Eichenbaum *et al.*, 2020; Goldberg, Johnson, and Shriver, 2024; Han *et al.*, 2022). Akin to a difference-in-differences model, this approach relies on the assumption that pre-treatment trends—i.e., the trends in freelancer strategic behavior for both 2021 and 2022 from June to November—are parallel. If this condition is met, it suggests that the introduction of ChatGPT was unlikely to have been anticipated by freelancers on platform and that the documented patterns are unlikely to be driven by any cyclical time trends. For this analysis, we consider the "treated" period to be June 2022 through May 2023, while the "control" period is June 2021 through May 2022. We use this sample criteria so that: a) we have an equal number of periods in both the treated and control periods, allowing for a comparison; and b) there are no freelancer-month observations that are included in both the treated and control periods. We utilize specifications similar to those outlined above, except that the categorical time or post-period indicator is interacted with an indicator variable for whether the observation is part of the treatment period. Standard errors are clustered at the freelancer- level.

After exploring incumbent freelancers' strategic responses following the launch of ChatGPT, we examine heterogeneity in these responses based on exposure to changes in demand for labor and exposure to changes in supply of labor following the launch of ChatGPT. We employ a continuous difference-in-difference design (Angrist and Pischke, 2009; D'Haultfœuille, Hoderlein, and Sasaki, 2023) and treat the public release of the large language modeling chatbot ChatGPT by OpenAI on November 30, 2022, as an exogenous shock that represents a sharp and discontinuous advancement in generative AI technologies. We estimate regression models in the following form:

$$Y_{it} = \beta_1 DemandContractionExposure_i \times Post_t + \beta_2 SupplyExpansionExposure_i \times Post_t + \alpha_i + \tau_t + \varepsilon_{it} . \qquad (3)$$

$DemandContractionExposure_i$ and $SupplyExpansionExposure_i$ refer to a freelancer's exposure to contraction in the demand for labor and exposure to expansion in the supply of labor post-ChatGPT based on pre-period bidding activity. To construct these measures, we leverage heterogeneity in the change in demand and supply across work domains. While all twelve of the platform's defined work categories experience a decline in demand (as captured by the average number of jobs each month) and eleven of twelve experience an increase in supply (as captured by the average number of unique bidding freelancers each month) from the pre- vs. post-ChatGPT, there is significant heterogeneity in the size of these changes.[5] We graphically display this heterogeneity by displaying the change in measures of supply and demand in the year post-ChatGPT relative to the year pre-ChatGPT in Figure 2. We construct a freelancer-specific measure by weighing the work domain-level percent change by each freelancer's distribution of bids across work categories during the pre-period. The goal of these measures is to identify the extent to which a freelancers' bidding behavior in the pre-period would have exposed them to changes in the demand for labor or supply of labor following the launch of ChatGPT. As we are interested in capturing how repositioning may be shaped by a contraction in demand, we invert the percentage change in demand (i.e., multiply by negative one), such that effect estimates inform us as to how a demand contraction may shape repositioning. Because supply and demand are related and determined simultaneously, our models consider the moderating effect of demand contraction exposure and supply expansion exposure simultaneously in an attempt to disentangle these effects.

[5] We acknowledge that these are just one possible way to capture supply or demand on the platform and are noisy measures but believe that they should broadly capture changes in supply and demand at the domain-level.

As with the specifications above, we include freelancer fixed effects, denoted by $\alpha_i$. The inclusion of freelancer fixed effect accounts for time-invariant unobserved characteristics such as ability and demographics. For this analysis, as we are focused on capturing the moderating role of demand contraction and supply expansion exposure rather than baseline time trends, we include month-by-year fixed effects, denoted by $\tau_t$.

We then attempt to examine how, holding work domain constant, freelancer skill may shape strategic repositioning post-ChatGPT. To do so, we estimate equations of the following form:

$$Y_{it} = \beta Skill_i \times Post_t + \alpha_i + \Gamma_i \times \tau_t + \varepsilon_{it} \,. \qquad (4)$$

$Skill_i$ refers to the freelancer's skill level, as captured using the average skill level for all jobs completed from January 2021 through November 2022. The skill-level field is assigned to each job by the client. Because clients have the option to not specify a skill level, this field is missing for some jobs. For freelancers who did not complete a job with a skill field, average skill level is missing, and they are excluded from the analysis. As with the specifications above, we include freelancer fixed effects, denoted by $\alpha_i$. For this analysis, we also account for work domain-specific changes in strategic responses driven by idiosyncratic changes in supply and demand using freelancer pre-ChatGPT modal work domain-by-month fixed effect, denoted by $\Gamma_i \times \tau_t$. The inclusion of freelancer pre-ChatGPT modal work domain-by-month fixed effect accounts for any domain-specific time trends that might skew demand for or supply of skills, as well as any general time-specific unobserved factors. $Y_{it}$ denotes outcomes of interest as defined above. All count outcome variables are log-transformed. $Post_t$ is an indicator variable that is equal to 0 in the months before the launch of ChatGPT (November 2022 and earlier) and 1 after (December 2022 onward). In models where measures of specialization or differentiation are the outcome variables, we also include *log bids* as a control variable. Standard errors are clustered at the freelancer level.

For these analyses, we acknowledge that our empirical strategy is unable to perfectly identify a causal effect, and in discussing our results we are careful to avoid overinterpretation and to document their limitations. Nevertheless, given the sharpness of the responses and the timeliness and importance of the phenomenon, we believe that the results we present are interesting and informative.

**Incumbent Repositioning Post-ChatGPT**

*Changes in Strategic Positioning Pre- vs. Post-ChatGPT*

We first examine changes in freelancer behavior using relative time models that track changes in behavior pre- and post-ChatGPT. The results of this analysis are presented in Figure 3, Panel A. In Panel A1, we show patterns in the count of bids for freelancers in the months before and after the launch of ChatGPT. We note that this chart has clear pre-trends, as freelancer bidding activity increased from the start of the sample until the launch of ChatGPT, likely because of our sample restrictions, which dictate that freelancers must be active in the three months leading up to the launch of ChatGPT. However, following the launch of ChatGPT, we see an immediate, large, and persistent decrease in the number of bids a freelancer submits on the platform. Furthermore, we document a decrease in the level of bid similarity relative to their prior bidding behavior (Panel A2), suggesting that freelancers differentiated their distribution of bids across specialties post-ChatGPT. Further, though there is an initial dip in the percent of high-value bids immediately following the launch of ChatGPT, this reverses and we document an increase in the percent of high-value bids in the post-period (Panel A3).

The patterns we document are simple time trends that may also reflect seasonal patterns in bidding behavior on the online labor market platform. To further probe the nature of these effects and to shed more light on whether these changes were plausibly driven by the launch of ChatGPT, we utilize the panel differences (PD) approach discussed above, using a treated sample from June 2022 through May 2023 and a control sample from the same freelancers from June 2021 through May 2022. In Figure 3 Panel B, we present evidence of a discontinuity in the treated sample relative to the previous year's trend using binned scatterplots that include freelancer-specific fixed effects and additionally control for the count of bids in Panel B2 and B3. We document sharp changes in the treated sample relative to the control sample following the launch of ChatGPT for both number of bids and similarity relative to their prior bidding behavior that mirrors that patterns we see in the simple time trend figures. Considering the percent of high-value bids, we see similar positive trends across both samples, and it appears that the growth in high-value bids is actually faster in the control sample. This likely reflects the growth of the platform in earlier years that led to a proliferation of high-value jobs. This figure suggests that there is no discontinuous change in the percent of high-value bids across freelancers post-ChatGPT.

In Table 2, we present regression models that estimate the size of the post-effect using both the baseline sample (Model 1-3) and the PD approach (Model 4-6). The baseline estimates suggests that, following the launch of ChatGPT, freelancers submitted 61.8% fewer bids, while the PD estimates in Panel B suggest a 51.2% reduction in the number of bids.[6] The magnitude of this effect may be particularly large because our sample selection criteria identify freelancers that were highly active just prior to the shock, who may have been unlikely to sustain that level of activity. Both the baseline and PD estimates suggest that freelancers differentiate in terms of the distribution of bids across specializations relative to their own prior behavior. The baseline model suggests that freelancers' bidding similarity relative to their prior behavior fell by 2.2% post-ChatGPT, while the PD model suggests a 2.0% decrease in similarity. Estimates on the post-indicator considering the percent high value bids seem to be driven more by differential pre-trends than by the effect of ChatGPT; while the baseline model estimates a 1.6% increase in the post-period, we document a 10.5% decrease in the post-period in the PD model. We do not interpret this estimate as a post-ChatGPT effect given the pre-trends documented in Figure 2.

[6] For non-logged dependent variables, effect sizes are calculated by dividing the coefficient estimate by the unconditional sample mean. For logged dependent variables, effect sizes are calculated by subtracting one from the exponentiated coefficient estimate. In further analyses, we document that this decrease in the number of bids for both hourly and fixed contract jobs, and there is no clear change in the composition of bids to fixed vs. hourly jobs (*Appendix Figure A1*).

These results present clear evidence of changes in strategic behavior among freelancers on the platform following the launch of ChatGPT. Incumbent freelancers appear to have decreased their bidding activity on the platform following the shock. This withdrawal from the platform could be due to a decrease in the demand for jobs on the platform but can also result from increased competition associated with the influx of new entrants. We examine these possibilities in the next set of analyses. Further, following the launch of ChatGPT, incumbent freelancers were on average more likely to alter their horizontal positioning by changing the distribution of their bids across specialties relative to their own prior behavior, though we find no evidence of sharp changes in vertical positioning as captured by the percentage of high-value bids post-ChatGPT.

*Repositioning Across Domains*

While the analyses above document aggregated changes in freelancer positioning across our sample, it does not reveal differences in responses across domains or shed light on mechanisms that drive such repositioning. As the advances in AI technologies will likely affect some domains more than others (e.g., Felten, Raj, and Seamans, 2021), we expect there to be heterogeneity in repositioning responses across skill domains. As a first step to explore such heterogeneity descriptively, we construct matrices that document freelancers' propensity to move across work domains from the pre- to post-period. To construct these matrices, we identify each freelancer's modal work domain based on their bidding behavior in the pre- and post-period. We then aggregate across work domains to document the propensity of freelancers to shift across (or exit from or enter) work domains. We finally normalize the post-ChatGPT transitions relative to prior trends by measuring how the values differ from those obtained conducting the same analyses in a placebo period the year before to construct a normalized transition matrix. We construct such a matrix for two separate samples – the first of which includes any freelancer that submits a bid on the platform during the relative time period, which allows us to investigate changes in entry by taking advantage of freelancers not on the platform in the pre-period, and the second of which focuses on our main sample of incumbents. Both normalized matrices are presented in Table 3.

In the matrix constructed using all freelancers that submit a bid on the platform during the relevant sample period (Panel A), we see that post-ChatGPT, freelancers appear less likely to remain in the same domain (i.e., negative change on the diagonal values of the matrix) and more likely to become inactive on the platform (i.e., positive changes across almost all rows in column 13 of the matrix). Further, we can track patterns among new entrants by examining transitions from those who were inactive in the pre-period (i.e., row 13). Among new entrants, we see greater entry into Web, Mobile, & Software Development and Data Science & Analytics, and less entry into Administrative Support, Design & Creative, and Translation. In the matrix focusing on the main sample of incumbent freelancers, we similarly see more switching across work domains and an increase in the likelihood of exit across all work categories. Transitions to a new work domain are particularly common for freelancers that focused on Translation, Customer Service, and Writing. We document that incumbents may be slightly more likely to transition into Administrative Support or Data Science & Analytics in the post-period, and perhaps slightly less likely to transition into Web, Mobile, & Software Development.

The matrices begin to document how freelancers in different domains may respond differently to the launch of ChatGPT. We present further evidence of differences in responses in Figure 3, which uses binned scatter plots to document changes in strategic behavior post-ChatGPT across work domains by assigning freelancers to one of five broad groups based on their modal work domain from January 2022 through November 2022: 1) Technical & Engineering (consisting of the work domains Data Science & Analytics, Engineering & Architecture, IT & Networking, and Web, Mobile, & Software Development), 2) Creative & Design (work domain Creative & Design), 3) Business & Administrative Services (work domains Accounting & Consulting, Administrative Support, Customer Service, and Legal), 4) Sales & Marketing (work domains Sales & Marketing), and 5) Writing & Translation (work categories Writing and Translation). Across domains, we see similar patterns as in the population of sample freelancers considering the number of bids (Panel A); however, interesting differences emerge considering bid similarity (Panel B) and the percent of high-value bids (Panel C). For example, freelancers specialized in Writing & Translation appear to horizontally differentiate to the largest extent relative to their prior positioning. On the other hand, freelancers who specialized in Design & Creative domains did not adjust their position by much. At the same time, Panel C suggests that freelancers specialized in certain clusters of work domains (i.e., Writing & Translation, Sales & Marketing, Business & Administrative Services) shifted their bidding to high-value jobs, while freelancers in the other work domains did not.

The heterogeneity across domains naturally raises questions regarding what factors may cause differential responses. One possibility is that these differences may be driven by heterogeneous changes in the demand for given tasks as well as heterogeneous changes in the supply of labor that alters competitive conditions across work domains. To explore this possibility, we visualize the average levels of percentage change in both demand (i.e., number of job postings) and supply (i.e., number of bidding freelancers) associated with each work domain in Figure 4. Overall, these patterns conform to the demand contraction and supply expansion dynamics we see at the platform level. However, there is considerable heterogeneity in how those demand and supply changes manifest for different work domains on the online labor market platform. In particular, we see that supply expansion is the most pronounced in the Data Science & Analytics and Web, Mobile, & Software Development domains. On the other hand, areas that saw the steepest contraction in demand were Writing and Translation.

These post-ChatGPT changes to demand and supply may have shaped freelancers' strategic responses by altering competitive conditions. Put differently, freelancers may strategically reposition themselves in different ways when technology change causes a contraction in demand vs. an expansion in supply for skills that they previously specialized in. We explore these dynamics in the following section.

**How Do Changes in Demand and Supply Shape Repositioning?**

We examine how changes in the demand for and supply of labor shape incumbent strategic repositioning. Specifically, we consider how incumbent freelancers' post-period response is shaped by ex-ante exposure to demand contraction and supply expansion associated with the launch of ChatGPT. To do so, we construct freelancer-specific measures that combine freelancers' bidding behavior in the pre-period and the changes in the demand for labor and the supply of labor across domains. As noted above, we measure the percentage change in the average monthly demand (as captured by job posts) and the change in the average monthly supply of labor (as captured by the number of unique freelancers that submit a bid to a job) in the pre- vs. post-period at the work domain level (see Figure 4). With demand contracting and supply expanding, we construct freelancer-level measures of *demand contraction exposure* and *supply expansion exposure* by weighing the work category level changes by the freelancer's distribution of bids across work domains in the pre-period.[7] Across the sample, the mean demand contraction exposure is 12.1% and the mean supply expansion exposure is 14.5%.

We examine how demand contraction and supply change exposure shape strategic re-positioning by considering models that simultaneously interact the post-ChatGPT indicator with both demand contraction exposure and supply expansion exposure. While the two measures are negatively correlated ($r = -0.219$, $p < 0.01$)[8], by including both interaction terms simultaneously, we hope to disentangle how these dimensions may differentially shape repositioning. We present baseline relative time models documenting how strategic responses post-ChatGPT are shaped by demand contraction and supply expansion exposure in Figure 5, Panel A. Corresponding regression estimates documenting how these dimensions shape the post-ChatGPT effect are presented in Table 4, Panel A.

We find evidence that exposure to demand contraction and supply expansion simultaneously shape freelancers' responses to the launch of ChatGPT, and interestingly, it appears that exposure to post-ChatGPT demand contraction and supply expansion shape freelancers' strategic responses in different ways. Relative to other freelancers on the platform, freelancers who are more exposed to demand contraction decrease their bidding activity more, are more likely to differentiate relative to prior behavior, and bid for a higher percent of high-value jobs. A one standard deviation increase in demand contraction is associated with a 5.2% decrease in the number of bids, a 1.8% decrease in similarity relative to prior behavior, and a 6.7% increase in the percent of high-value bids. On the other hand, freelancers exposed to a post-ChatGPT supply expansion increase their bidding activity, differentiate less relative to prior behavior, and bid for a lower percent of high-value jobs. A one standard deviation increase in supply expansion exposure is associated with a 2.7% increase in the number of bids, a 1.1% increase in similarity relative to prior behavior, and a 4.5% decrease in the percent of high-value bids.

[7] For *demand contraction exposure*, we scale by -1 such that the measure represents exposure to a demand contraction.
[8] *Appendix Figure A2* visually documents the relationship between supply and demand change exposure.

The relative time estimates in the baseline model display some extent of pre-trends. In an effort to mitigate concerns regarding the presence of these pre-trends, we utilize a coarsened exact matching (CEM) approach that matches freelancers with a high (top quartile) vs. low (bottom quartile) demand contraction exposure and supply expansion exposure based on pre-period bidding characteristics. Specifically, we match on: 1) the number of bids they submitted one, three, and six months prior to ChatGPT, and 2) the number of jobs completed in the pre-period. While ideally, we would also match freelancers that are high vs. low in terms of demand (supply) change exposure based on their supply (demand) change exposure, the correlation between these measures means that requiring matches to differ on demand (supply) change exposure but be similar in supply (demand) change exposure results in an insufficient sample. Accordingly, for this analysis, we identify the post-period effect for freelancers who are high vs. low on demand (supply) change exposure, while still controlling for the interaction of the supply (demand) change exposure and time categorical variables. The results of this analysis, presented in Figure 5 Panel B, suggest that the matching approach is effective in mitigating pre-trends. Comparing figures from the baseline sample versus the matched sample, our results are qualitatively similar. Further, the matched sample post-period regression estimates continue to document that demand contraction exposure and supply expansion exposure strongly moderate the post-period effect (Table 4, Panel B).

*Takeaways from this Analysis*

These results provide evidence regarding how and why the launch of ChatGPT may elicit changes in worker strategic positioning. Relative to other freelancers on the platform, workers that are exposed to a post-ChatGPT contraction in labor demand are more likely to decrease activity on the platform and horizontally reposition themselves. Concurrent with patterns seen in Table 3, this likely reflects a tendency to reposition into areas that experience a lower degree of demand contraction. In contrast, workers that are exposed to a post-ChatGPT expansion in the labor supply are less likely to disengage from the platform or reposition horizontally, but more likely to reposition vertically by targeting lower-value jobs, which may be a response to navigate an increasingly competitive and crowded domain. In documenting these patterns, we begin to generate a theoretical framework that helps explain how workers may respond to technological change in generative AI depending on how such change alters the demand for and supply of labor across domains. We expand on this in the Discussion section.

**How Does Skill Shape Repositioning?**

We next consider how worker-specific characteristics may also inform the ways that workers reposition post-technological change. Specifically, we examine whether there is heterogeneity in the post-ChatGPT effect based on freelancer skill level. The logic underlying this investigation is that workers that have a higher skill-level within a given domain may be less likely to reposition from that domain given ex-ante investments to develop expertise. In other words, skill may reflect a proxy for an individual freelancer's adjustment costs. We present relative time models documenting how freelancers' strategic responses following the launch of ChatGPT differ by freelancer skill level in Figure 6, Panel A. For this analysis, to account for the heterogeneous responses at the domain-level outlined above, we include pre-period modal work category-by-month fixed effects to account for time-varying effects at the work category level. We note these estimates use other freelancers on the platform as a reference group, so should be contextualized amidst the full sample results documented above. Corresponding regression estimates of the estimated post-ChatGPT effect are presented in Table 5.

We indeed find evidence that freelancer repositioning post-ChatGPT is importantly shaped by freelancer skill level. While there is some evidence of pre-trends considering the log bids as a dependent variable, the parallel trends assumption appears to be met considering the other dependent variables. A one-unit increase in average skill pre-treatment is associated with a 12.7% increase in bids, a 1.6% increase in similarity to their own prior bidding behavior, and a 7.2% decrease in the percentage of high-value bids relative to other freelancers on the platform. To mitigate concerns regarding pre-trends, we conduct a version of this test utilizing coarsened exact matching to match high-skill freelancers (top quartile) with low-skill freelancers (bottom quartile) based on the same matching procedure described in the previous section. We then examine how the post-period effect differs for high- vs. low-skill freelancers utilizing this matched sample. The results of this analyses, presented graphically in Figure 6, Panel B and in columns 4-6 of Table 5, appear consistent with the baseline results.

*Takeaways from this Analysis*

Synthesizing these results, we document that, holding work domain constant, freelancer skill level shapes strategic responses to advances in AI technologies. Higher-skill freelancers are less likely to decrease activity on the platform and make less stark changes to the distribution of bids across work domains than their peers. Further, we find evidence that the percentage of bids on high value jobs fell more among higher-skilled freelancers than among their peers. As our non-parametric evidence in Figure 1 suggests that new entrants may have been more likely to bid on high-value jobs post-ChatGPT, one plausible explanation is that high-skill freelancers face greater competition for high-value jobs and thus shift their vertical positioning in response.

**Additional Analyses and Robustness**

We conduct additional analyses to probe other forms of strategic positioning with an additional set of dependent variables. Specifically, we examine 1) *bid for job* which is a binary variable for whether a freelancer bid for at least one job; 2) *new modal work domain* which is a binary variable for whether a freelancer has a new modal work domain in the current 3-month rolling period as compared to the last; 3) *specialization HHI* which measures the concentration of the freelancer's bid across the two hundred work specializations on the platform; and 4) *AI profile* which is a binary variable for whether a freelancer's public profile contains a set of AI-related keywords. We replicate all our main analyses for this set of outcome measures. These results are presented visually in *Appendix Figures A3-A5.*

The effects for bid for job and new modal work domain largely mirror the results for log bids and similarity relative to prior behavior: on average, incumbents are more likely to withdraw from the platform and switch work domains, these effects are heightened when demand contraction exposure is greater and mitigated when supply contraction exposure is greater and are smaller for higher-skill freelancers. Results are not entirely clear considering specialization HHI – there is some evidence that that demand contraction exposure is associated with a decrease in concentration and supply expansion exposure is associated with an increase in concentration post-ChatGPT. However, the average effect across sample incumbents is not clear, and baseline and CEM estimates considering the moderating effect of freelancer skill are not consistent. This could be because specialization HHI is very much tied to the number of bids, and even controlling for that, changes in bidding activity may underlie patterns in concentration. Finally, perhaps unsurprisingly, we find that on average, incumbents increase the use of AI keywords in their profile post-ChatGPT. This response is particularly large among freelancers exposed to supply expansion and among high-skill freelancers. This could be because, as supply expansion increases competition, incumbent freelancers try to stick out by embracing AI in their public profiles. These effects may be strongest among high-skill freelancers who are more reticent to switch domains due to the higher adjustment costs that they face.

We further replicate our analyses considering the moderating role of freelancer skill using freelancer experience, as captured by the number of jobs a freelancer completes in the pre-period to capture experience (*Appendix Figure A6*).[9] Experience appears to play a similar role as skill: greater experience appears to result in a smaller decrease in the number of bids in the post-period, decreases differentiation relative to prior behavior, and results in a lower percent of high-value bids. We document that our results are robust to limiting our sample only to freelancers with an above median number of bids in a given three-month window, limiting concerns that patterns are driven by freelancers with sparse activity (*Appendix Figures A7-A9*). Further, we document that our main results are robust to clustering standard errors at the month-by-year rather than freelancer-level (*Appendix Tables A2-A4*), and that the supply and demand change exposure results are robust to constructing the measures of freelancer-specific supply and demand change exposure using more granular work specializations rather than broader work categories used in our main analyses (*Appendix Figure A10*), though results are noisier, likely because the more granular work categories are more subject to noise in the construction of the measure.

[9] Note that as pre-period experience is a matching criterion for the CEM analysis, we do not conduct our matched sample analysis and instead just present baseline estimates. There are significant pre-trends considering the count of bids as a dependent variable, but little evidence of pre-trends for the other dependent variables.

**Discussion**

*New Theoretical Perspectives*

In this project, we explore how freelance workers responded to advances in generative AI. Leveraging internal data from a large online labor market platform, we first examine how freelancers changed their behavior in response to the launch of ChatGPT and find evidence that they submitted fewer bids and increase their differentiation (as captured by the similarity in the distribution of bids across specializations) relative to their own prior behavior. However, we do not find evidence across the sample that freelancers alter their vertical positioning and target more or less high value jobs. We then explore the mechanisms driving such repositioning, and document how freelancer-specific exposure to demand contraction and supply expansion post-ChatGPT and freelancer skill drive repositioning in specific ways.

We believe that our results have a number of generalizable and theoretically important implications. First, we extend literature that considers how economic actors change strategies in the face of technological change (e.g., Benner and Waldfogel, 2023; Eggers and Park, 2018; Miller and Wang, 2024) by showing how freelance workers altered their competitive strategies following the launch of ChatGPT. We find evidence that the advances in the accessibility and capabilities of AI technology facilitated by the public launch of ChatGPT increased entry among new freelancers and led to differentiation by incumbent freelancers. We believe these findings are likely to generalize beyond this setting. Prior literature has documented that one impact of digital technologies on competitive dynamics is reducing production costs and thus lowering entry barriers to existing markets (e.g. Waldfogel, 2017). The patterns we observe are broadly consistent with such literature, as incumbents become more selective in the competitive position they adopt when facing demand contraction (Benner and Waldfogel, 2016), and disruptive technologies enable new entrants to capture demand that was once served exclusively by incumbents in the high-end market (Benner and Waldfogel, 2023). We believe a theoretical implication stemming from our results is the applicability of models of competition and technological disruption in understanding the effects of AI, which complements the automation-based theoretical perspectives often employed in the literature.

Second, our results suggest that the effects of advances in AI technologies manifest differently across work domains. These differences emerge even in domains that ex-ante could have been expected to face a similar level of exposure to AI. For example, while both writing and software development are exposed to language modeling (Felten, Raj, and Seamans, 2023), our empirical results document disparate changes in behavior by freelancers who specialize in those domains. A straightforward but important takeaway from this work is that exposure to AI comes in different forms. The literature discusses how exposure may lead to substitution or complementarities (e.g., Autor *et al.*, 2003; Frank *et al.*, 2019), but it is important to note that workers will respond to those different forms of exposure in different ways.

Relatedly, our platform-level results highlight the dual nature of generative AI as both automating and augmenting technology. We observe that the launch of ChatGPT was accompanied by both demand contraction and supply expansion, suggesting that advances in generative is simultaneously displacing labor and lowering entry barrier in some skill domains. Prior studies tend to highlight either automation (e.g. Demirci, Hannane, and Zhu, 2023; Hui, Reshef, and Zhou, 2023; Liu *et al.*, 2023) or augmentation (e.g. Brynjolfsson, Li, and Raymond, 2023; Choi and Schwarcz, 2023; Dell'Acqua *et al.*, 2023; Noy and Zhang, 2023; Peng *et al.*, 2023) as the primary technological capability associated with generative AI. Our study integrates these perspectives and suggests that both forms of capabilities can manifest simultaneously. As noted in Restrepo (2024), "One fundamental question that remains unanswered is whether it is appropriate to model narrow AI systems as automation technologies" (p. 22). Our study suggests that for some skill domains, it may indeed be appropriate to conceptualize AI as an automating technology that displaces labor, while in others, it may be more appropriate to model AI as a technology that augment worker abilities and widen labor supply.

Third, perhaps most importantly, in exploring the heterogeneity in responses across work domains, we provide more insight into *why* and *how* technology change elicits repositioning (e.g., Beane and Anthony, 2024; Benner and Waldfogel, 2023) by exploring how exposure to post-ChatGPT changes in the demand for and supply of labor shapes freelancers' post-period responses. We document that technology can simultaneously induce both demand and supply changes, and those changes may elicit different responses for workers. In doing so, we generate generalizable insights that may inform how workers respond to future generations of technological change. When technological change leads to a contraction in demand for certain tasks and skills, we may expect workers to horizontally differentiate and seek out new work domains. This seems a natural response, as workers adapt by moving away from a less promising work domain. If instead, labor demand remains constant but technological change lowers barriers to entry, decreases the cost of supplying labor, and facilitates new entry, we may expect workers to alter their vertical positioning, as greater competition forces incumbents to move to different parts of the market, but may be unlikely to change work domains. Our findings appear consistent with extant research in labor economics which suggests both that workers reallocate tasks based on changes in demand (Autor *et al.*, 2003), as well as literature on competition which suggests that firms may alter their vertical positioning in response to changes in market competition (Dixit and Stiglitz, 1977). The joint consideration of technology induced demand and supply allow us to integrate these disparate streams of the literature, and the juxtaposition of horizontal and vertical positioning help generate a conceptual framework of how technological change affect different forms of repositioning.

Finally, our heterogeneous results by freelancer skill level add to a body of literature that discusses the distributional effects of advances in AI technologies, including work that suggests that lower-skill workers may be best positioned to take advantage of AI (e.g., Brynjolfsson *et al.*, 2023; Dell'Acqua *et al.*, 2023; Noy and Zhang, 2023; Peng *et al.*, 2023). Our findings suggest that skill level shape responses to technological change. In the online labor market setting, high-skill individuals changed the areas in which they work less than their less-skilled counterparts, perhaps due to higher adjustment costs (Argyres, Mahoney, and Nickerson, 2019; Kambourov and Manovskii, 2009; Neal, 1995). At the same time, the decrease in the percentage of high-value bids among more-skilled freelancers suggests that, those freelancers may find that the value of the jobs that they are competing for falls as AI tools enable new entry and increase competition. We also find that high-skilled freelancers are quicker to add AI keywords to their public-facing profiles, echoing prior literature that suggests that higher skill workers may be quicker to embrace such tools (Humlum and Vestergaard, 2024). To the extent that the adoption of AI technologies is beneficial for productivity (and/or makes workers more attractive to potential employers), this result may also have implications for whether and to what extent advances in AI tools can mitigate or exacerbate inequality across workers.

*Limitations, Future Research, and Conclusion*

Our project has limitations that have implications for the interpretation of our results. We study a relatively short time period following the launch of ChatGPT (16 months), and it is unclear whether the patterns we document will hold or evolve over the longer term. For concision, we focus on a small number of measures to capture positioning, and there are many other dimensions of positioning that could be interesting to study. Further, while conducting this study among freelance workers on an online labor market platform has advantages — namely that we can leverage detailed and granular internal data and are able to capture immediate changes in behavior around the discontinuity given the more flexible nature of freelance work, it is unclear how these results might translate to organizational settings. We may expect worker repositioning to take longer to manifest. While our results around strategic positioning suggest implications for entrepreneurial firms, it is possible that collective decision-making by individual firms could result in different versions of strategic positioning following technological change. Similarly, it is unclear how our results may generalize to workers within firms, whose decisions around adoption and task content will be shaped by firm-specific characteristics rather than exclusively by their own decisions. Finally, while we provide evidence that freelance workers altered their strategic positioning post-ChatGPT, we do not consider the downstream implications of their responses for their performance on the online labor market platform. Limitations to the scope of our data access prevent us from exploring this in detail. Future studies that identify what kinds of strategic responses are associated with better performance following technological change would be a critical extension of this research.

Headlines often highlight the potential for generative AI to replace humans en masse (e.g., Mitchell, 2023; Scheiber, 2025). While generative AI may lead to automation in some settings, our study suggests that workers are also quick to adapt and respond strategically to technological change. Such responsiveness among workers, while intuitive, is under discussed in the literature. By documenting that workers respond quickly to technological change (and that technological change can lead to increased labor supply and competition), our work highlights that future theoretical and empirical efforts to understand the effects of AI should incorporate supply-side considerations. To the extent that AI does lead to automation in certain domains, understanding the net effects of employment and labor markets will require a knowledge of whether and how workers will adjust in response. We hope that this research contributes to understanding how advances in AI technologies can affect the strategic positioning of workers and provides a foundation that future work can build on.

**References**

Abraham KG, Hershbein B, Houseman SN, Truesdale B. 2023, March. The Independent Contractor Workforce: New Evidence On Its Size and Composition and Ways to Improve Its Measurement in Household Surveys. Working Paper Series. Working Paper, National Bureau of Economic Research. Available at: https://www.nber.org/papers/w30997.

Acemoglu D. 2025. The simple macroeconomics of AI*. *Economic Policy* **40**(121): 13–58.

Acemoglu D, Autor D, Hazell J, Restrepo P. 2022. Artificial Intelligence and Jobs: Evidence from Online Vacancies. *Journal of Labor Economics*. The University of Chicago Press **40**(S1): S293–S340.

Acemoglu D, Restrepo P. 2019. Automation and New Tasks: How Technology Displaces and Reinstates Labor. *Journal of Economic Perspectives* **33**(2): 3–30.

Agrawal A, Horton J, Lacetera N, Lyons E. 2015. Digitization and the Contract Labor Market: A Research Agenda. In *Economic Analysis of the Digital Economy*. University of Chicago Press: 219–250. Available at: https://www.nber.org/books-and-chapters/economic-analysis-digital-economy/digitization-and-contract-labor-market-research-agenda.

Angrist JD, Pischke J-S. 2009. *Mostly Harmless Econometrics: an Empiricist's Companion*. Princeton University Press: Princeton.

Anthony C, Bechky BA, Fayard A-L. 2023. "Collaborating" with AI: Taking a System View to Explore the Future of Work. *Organization Science*. INFORMS **34**(5): 1672–1694.

Argyres N, Mahoney JT, Nickerson J. 2019. Strategic responses to shocks: Comparative adjustment costs, transaction costs, and opportunity costs. *Strategic Management Journal* **40**(3): 357–376.

Autor DH. 2015. Why Are There Still So Many Jobs? The History and Future of Workplace Automation. *Journal of Economic Perspectives* **29**(3): 3–30.

Autor DH, Dorn D. 2013. The Growth of Low-Skill Service Jobs and the Polarization of the US Labor Market. *American Economic Review* **103**(5): 1553–1597.

Autor DH, Levy F, Murnane RJ. 2003. The Skill Content of Recent Technological Change: An Empirical Exploration. *The Quarterly Journal of Economics* **118**(4): 1279–1333.

Bailey DE, Faraj S, Hinds PJ, Leonardi PM, von Krogh G. 2022. We Are All Theorists of Technology Now: A Relational Perspective on Emerging Technology and Organizing. *Organization Science*. INFORMS **33**(1): 1–18.

Baldassarre MT, Caivano D, Fernandez Nieto B, Gigante D, Ragone A. 2023. The Social Impact of Generative AI: An Analysis on ChatGPT. In *Proceedings of the 2023 ACM Conference on*

*Information Technology for Social Good*, GoodIT ’23. Association for Computing Machinery: New York, NY, USA: 363–373. Available at: https://dl.acm.org/doi/10.1145/3582515.3609555.
Beane M, Anthony C. 2024. Inverted Apprenticeship: How Senior Occupational Members Develop Practical Expertise and Preserve Their Position When New Technologies Arrive. *Organization Science*. INFORMS **35**(2): 405–431.
Benner MJ, Waldfogel J. 2016. The Song Remains the Same? Technological Change and Positioning in the Recorded Music Industry. *Strategy Science*. INFORMS **1**(3): 129–147.
Benner MJ, Waldfogel J. 2023. Changing the channel: Digitization and the rise of “middle tail” strategies. *Strategic Management Journal* **44**(1): 264–287.
Berg JM, Raj M, Seamans R. 2023. Capturing Value from Artificial Intelligence. *Academy of Management Discoveries*. Academy of Management **9**(4): 424–428.
Brynjolfsson E, Li D, Raymond LR. 2023, April. Generative AI at Work. Working Paper Series. Working Paper, National Bureau of Economic Research. Available at: https://www-nber-org.proxy.library.upenn.edu/papers/w31161.
Capelli P, Tambe P, Yakubovich V. 2024, March 4. Will Large Language Models Really Change How Work Is Done? *MIT Sloan Management Review*. Available at: https://sloanreview.mit.edu/article/will-large-language-models-really-change-how-work-is-done/ [24 April 2024].
Choi JH, Schwarcz D. 2023, August 13. AI Assistance in Legal Analysis: An Empirical Study. SSRN Scholarly Paper, Rochester, NY. Available at: https://papers.ssrn.com/abstract=4539836.
Csaszar FA, Ketkar H, Kim H. 2024, August 16. Artificial Intelligence and Strategic Decision-Making: Evidence from Entrepreneurs and Investors. arXiv. Available at: http://arxiv.org/abs/2408.08811.
Dell’Acqua F *et al.* 2023, September 15. Navigating the Jagged Technological Frontier: Field Experimental Evidence of the Effects of AI on Knowledge Worker Productivity and Quality. SSRN Scholarly Paper, Rochester, NY. Available at: https://papers.ssrn.com/abstract=4573321.
Demirci O, Hannane J, Zhu X. 2023, October 15. Who is AI Replacing? The Impact of Generative AI on Online Freelancing Platforms. SSRN Scholarly Paper, Rochester, NY. Available at: https://papers.ssrn.com/abstract=4602944.
D’Haultfœuille X, Hoderlein S, Sasaki Y. 2023. Nonparametric difference-in-differences in repeated cross-sections with continuous treatments. *Journal of Econometrics* **234**(2): 664–690.
Dixit AK, Stiglitz JE. 1977. Monopolistic Competition and Optimum Product Diversity. *The American Economic Review*. American Economic Association **67**(3): 297–308.
Eggers JP, Park KF. 2018. Incumbent Adaptation to Technological Change: The Past, Present, and Future of Research on Heterogeneous Incumbent Response. *Academy of Management Annals*. Academy of Management **12**(1): 357–389.
Eichenbaum M, Matos MG de, Lima F, Trabandt M, Rebelo S. 2020. How do people respond to small probability events with large, negative consequences? *National Bureau of Economic Research (NBER)* **27988**. Available at: https://repositorio.ucp.pt/handle/10400.14/32950.
Feigenbaum J, Gross DP. 2024a. Organizational and Economic Obstacles to Automation: A Cautionary Tale from AT&T in the Twentieth Century. *Management Science*. INFORMS. Available at: https://pubsonline.informs.org/doi/abs/10.1287/mnsc.2022.01760.
Feigenbaum J, Gross DP. 2024b. Answering the Call of Automation: How the Labor Market Adjusted to Mechanizing Telephone Operation*. *The Quarterly Journal of Economics* **139**(3): 1879–1939.
Felten E, Raj M, Seamans R. 2021. Occupational, industry, and geographic exposure to artificial intelligence: A novel dataset and its potential uses. *Strategic Management Journal* **42**(12): 2195–2217.
Felten EW, Raj M, Seamans R. 2023, April 10. Occupational Heterogeneity in Exposure to Generative AI. SSRN Scholarly Paper, Rochester, NY. Available at: https://papers.ssrn.com/abstract=4414065.
Ferguson J-P, Hasan S. 2013. Specialization and Career Dynamics: Evidence from the Indian Administrative Service. *Administrative Science Quarterly*. SAGE Publications Inc **58**(2): 233–256.

Frank MR *et al.* 2019. Toward understanding the impact of artificial intelligence on labor. *Proceedings of the National Academy of Sciences of the United States of America* **116**(14): 6531–6539.
Furman JL, Teodoridis F. 2020. Automation, Research Technology, and Researchers' Trajectories: Evidence from Computer Science and Electrical Engineering. *Organization Science*. INFORMS **31**(2): 330–354.
Goldberg SG, Johnson GA, Shriver SK. 2024. Regulating Privacy Online: An Economic Evaluation of the GDPR. *American Economic Journal: Economic Policy* **16**(1): 325–358.
Hampole M, Papanikolaou D, Schmidt LDW, Seegmiller B. 2025, February. Artificial Intelligence and the Labor Market. Working Paper Series. Working Paper, National Bureau of Economic Research. Available at: https://www.nber.org/papers/w33509.
Han BR, Sun T, Chu LY, Wu L. 2022. COVID-19 and E-Commerce Operations: Evidence from Alibaba. *Manufacturing & Service Operations Management*. INFORMS **24**(3): 1388–1405.
Heaven WD. 2023, March 3. The inside story of how ChatGPT was built from the people who made it | MIT Technology Review. Available at: https://www.technologyreview.com/2023/03/03/1069311/inside-story-oral-history-how-chatgpt-built-openai/ [18 March 2024].
Horton JJ, Tambe P. 2015. Labor Economists Get Their Microscope: Big Data and Labor Market Analysis. *Big Data*. Mary Ann Liebert, Inc., publishers **3**(3): 130–137.
Hui X, Reshef O, Zhou L. 2023, July 31. The Short-Term Effects of Generative Artificial Intelligence on Employment: Evidence from an Online Labor Market. SSRN Scholarly Paper, Rochester, NY. Available at: https://papers.ssrn.com/abstract=4527336.
Humlum A, Vestergaard E. 2024, May 15. The Adoption of ChatGPT. SSRN Scholarly Paper, Rochester, NY. Available at: https://papers.ssrn.com/abstract=4827166.
Kambourov G, Manovskii I. 2009. Occupational Specificity of Human Capital. *International Economic Review* **50**(1): 63–115.
Klepper S. 1996. Entry, Exit, Growth, and Innovation over the Product Life Cycle. *American Economic Review* **86**(3): 562–583.
Krakowski S, Luger J, Raisch S. 2023. Artificial intelligence and the changing sources of competitive advantage. *Strategic Management Journal* **44**(6): 1425–1452.
Lebovitz S, Lifshitz-Assaf H, Levina N. 2022. To Engage or Not to Engage with AI for Critical Judgments: How Professionals Deal with Opacity When Using AI for Medical Diagnosis. *Organization Science*. INFORMS **33**(1): 126–148.
Liu J, Xu X, Li Y, Tan Y. 2023, August 9. 'Generate' the Future of Work through AI: Empirical Evidence from Online Labor Markets. arXiv. Available at: http://arxiv.org/abs/2308.05201.
Miller CD, Wang RD. 2024. Product digitization and differentiation strategy change: Evidence from the book publishing industry. *Strategic Management Journal* **n/a**(n/a). Available at: https://onlinelibrary.wiley.com/doi/abs/10.1002/smj.3586.
Mitchell A. 2023, January 25. ChatGPT could make these jobs obsolete. *New York Post*. Available at: https://nypost.com/2023/01/25/chat-gpt-could-make-these-jobs-obsolete/.
Narayanan S, Balasubramanian S, Swaminathan JM. 2009. A Matter of Balance: Specialization, Task Variety, and Individual Learning in a Software Maintenance Environment. *Management Science*. INFORMS **55**(11): 1861–1876.
Neal D. 1995. Industry-Specific Human Capital: Evidence from Displaced Workers. *Journal of Labor Economics*. [University of Chicago Press, Society of Labor Economists, NORC at the University of Chicago] **13**(4): 653–677.
Noy S, Zhang W. 2023, March 1. Experimental Evidence on the Productivity Effects of Generative Artificial Intelligence. SSRN Scholarly Paper, Rochester, NY. Available at: https://papers.ssrn.com/abstract=4375283.
OpenAI. 2022, March. New GPT-3 capabilities: Edit & insert. *OpenAI*. Available at: https://openai.com/blog/gpt-3-edit-insert [18 March 2024].

Peng S, Kalliamvakou E, Cihon P, Demirer M. 2023, February 13. The Impact of AI on Developer Productivity: Evidence from GitHub Copilot. arXiv. Available at: http://arxiv.org/abs/2302.06590.
Restrepo P. 2024. Automation: Theory, Evidence, and Outlook. *Annual Review of Economics*. Annual Reviews **16**(Volume 16, 2024): 1–25.
Roose K. 2023, February 3. How ChatGPT Kicked Off an A.I. Arms Race. *The New York Times*. Available at: https://www.nytimes.com/2023/02/03/technology/chatgpt-openai-artificial-intelligence.html.
Saggu A, Ante L. 2023. The influence of ChatGPT on artificial intelligence related crypto assets: Evidence from a synthetic control analysis. *Finance Research Letters* **55**: 103993.
Scheiber N. 2025, March 25. Has the Decline of Knowledge Work Begun? *The New York Times*. Available at: https://www.nytimes.com/2025/03/25/business/economy/white-collar-layoffs.html.
Teutloff O *et al.* 2025. Winners and losers of generative AI: Early Evidence of Shifts in Freelancer Demand. *Journal of Economic Behavior & Organization* : 106845.
Tushman ML, Anderson P. 1986. Technological Discontinuities and Organizational Environments. *Administrative Science Quarterly* **31**(3): 439.
Waldfogel J. 2017. How Digitization Has Created a Golden Age of Music, Movies, Books, and Television. *Journal of Economic Perspectives* **31**(3): 195–214.
Yuan Z, Chen H. 2023. The Impact of ChatGPT on the Demand for Human Content Generating and Editing Services: Evidence from an Online Labor Market. *ICIS 2023 Proceedings*. Available at: https://aisel.aisnet.org/icis2023/techandfow/techandfow/1.
Zuckerman EW, Kim T, Ukanwa K, von Rittmann J. 2003. Robust Identities or Nonentities? Typecasting in the Feature-Film Labor Market. *American Journal of Sociology*. The University of Chicago Press **108**(5): 1018–1073.

**Figure 1. Binned scatterplots of platform-level activity on the online labor market platform pre- and post-ChatGPT.**

The following figures are binned scatterplots showing trends in platform-level demand (i.e., log total job posts, high over low value job ratio, and job post concentration across domains) and supply (i.e., log total bidding freelancers, entrants over incumbent ratio, and entrant bid to high over low value job ratio) dynamics from January 2021 through March 2024.

**Panel A. Log total job posts.**

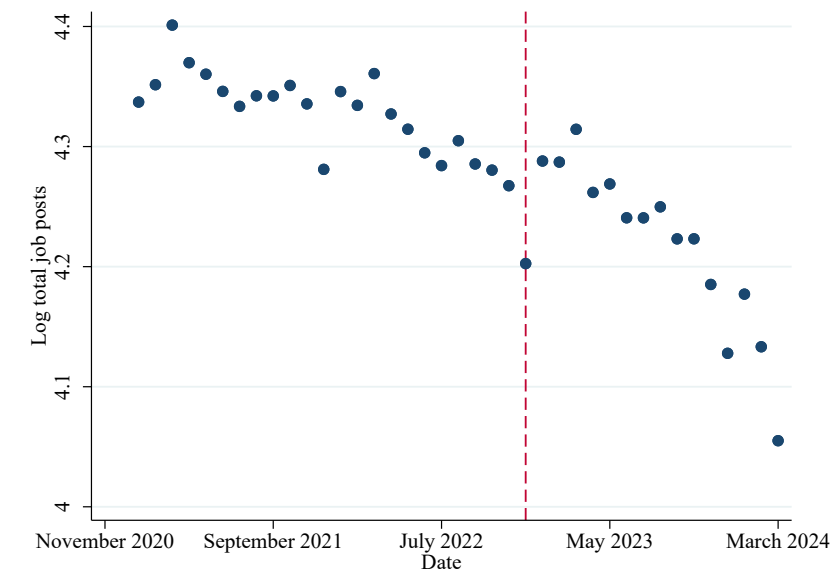

**Panel B. High over low value job posts ratio.**

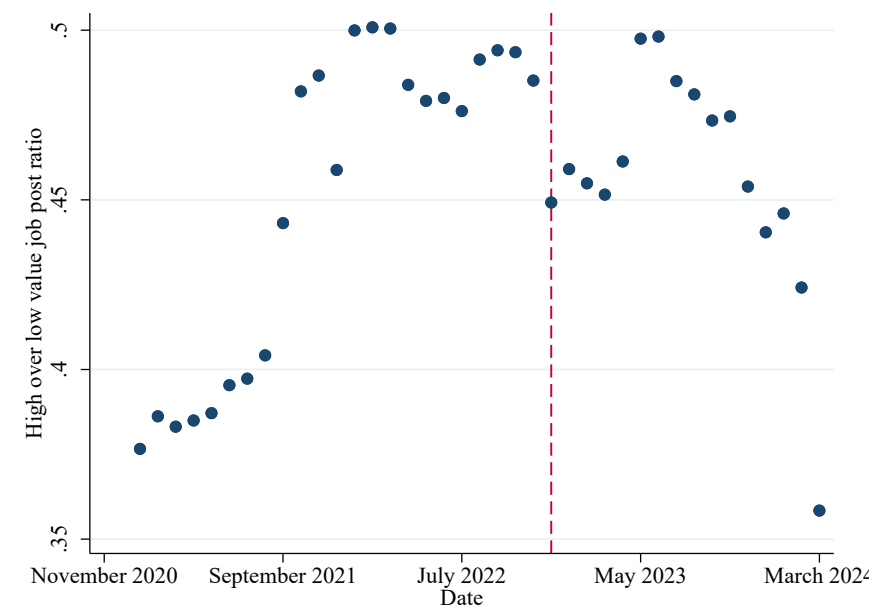

**Panel C. Job post concentration across domains.**

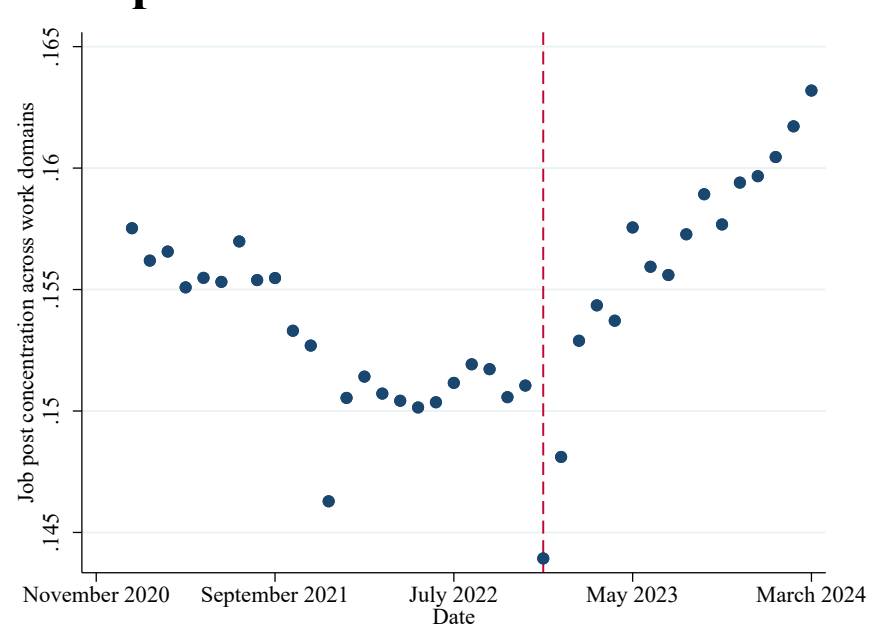

**Panel D. Log total unique bidding freelancers.**

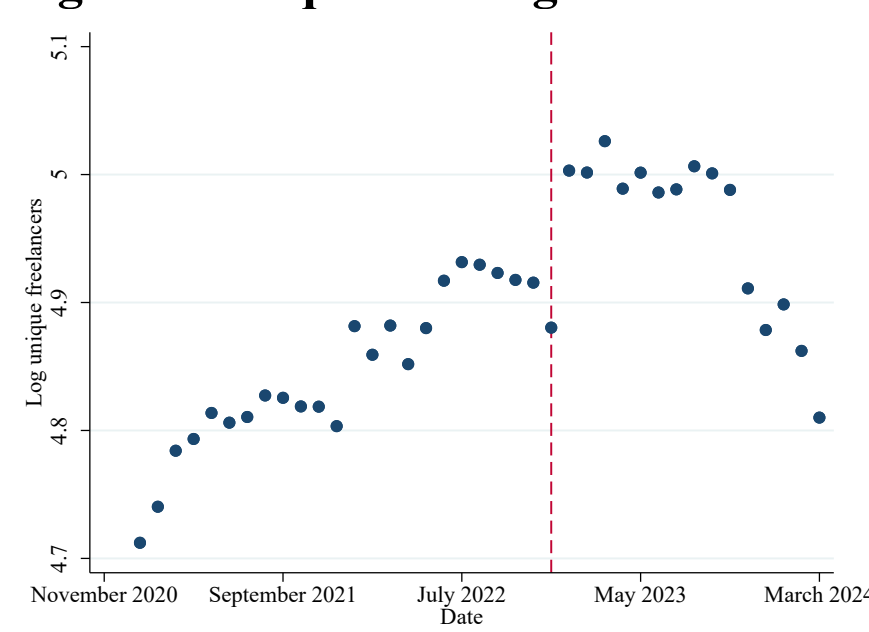

**Panel E. Entrants over incumbents ratio.**

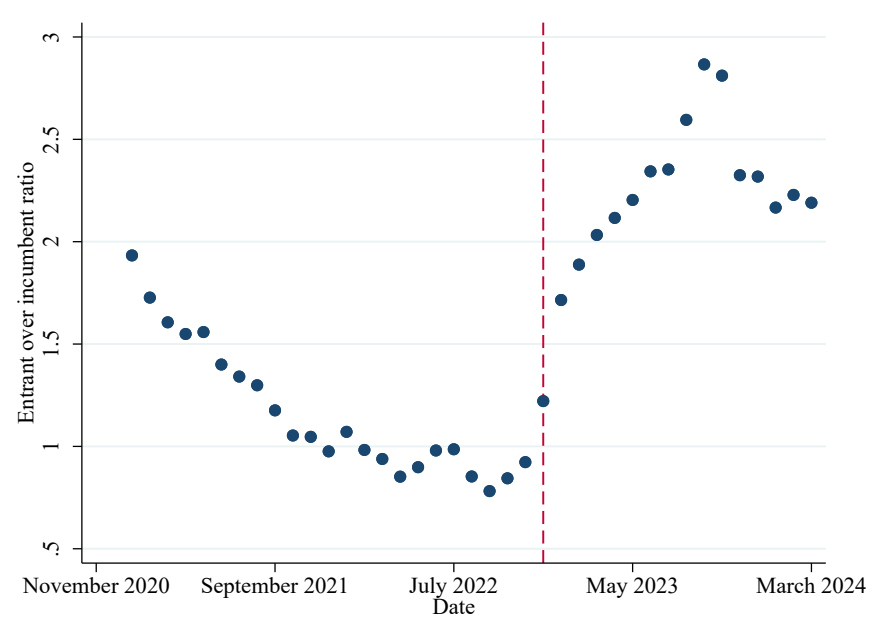

**Panel F. Entrant bid to high over low value job ratio.**

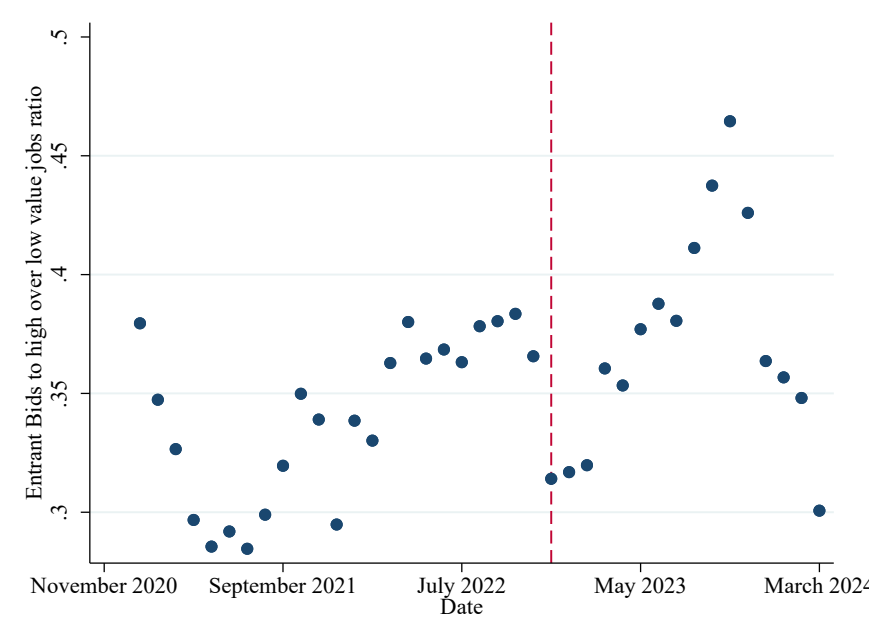

**Panel G. International to US Freelancer Ratio.**

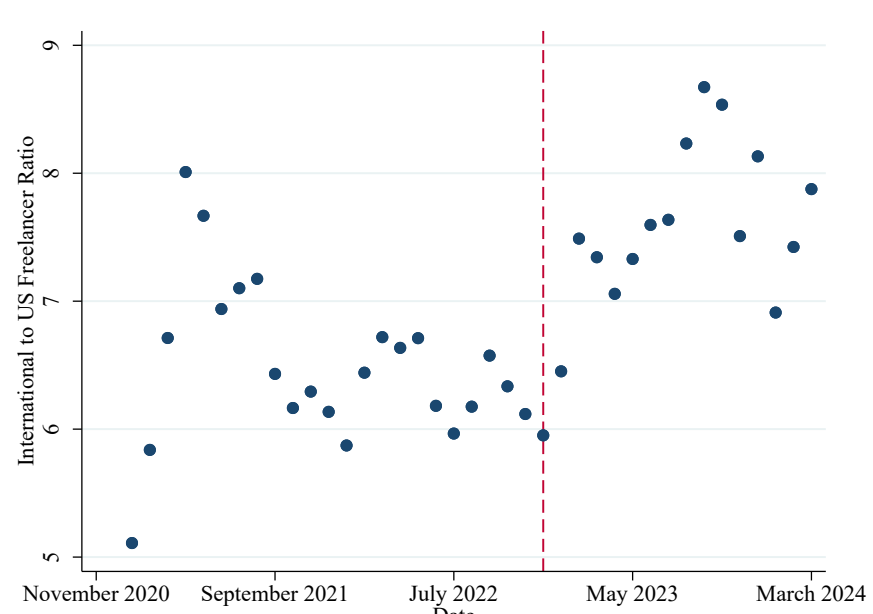

**Figure 2. Demand and supply changes in the short term by work domain.**

This figure shows bar charts of demand (i.e., average number of job postings per month) and supply (i.e., average number of bidding freelancers per month) change associated with each of the twelve domains on the online labor market platform in the year pre-ChatGPT (December 2021 through November 2022) vs. post-ChatGPT (December 2022 through November 2023).

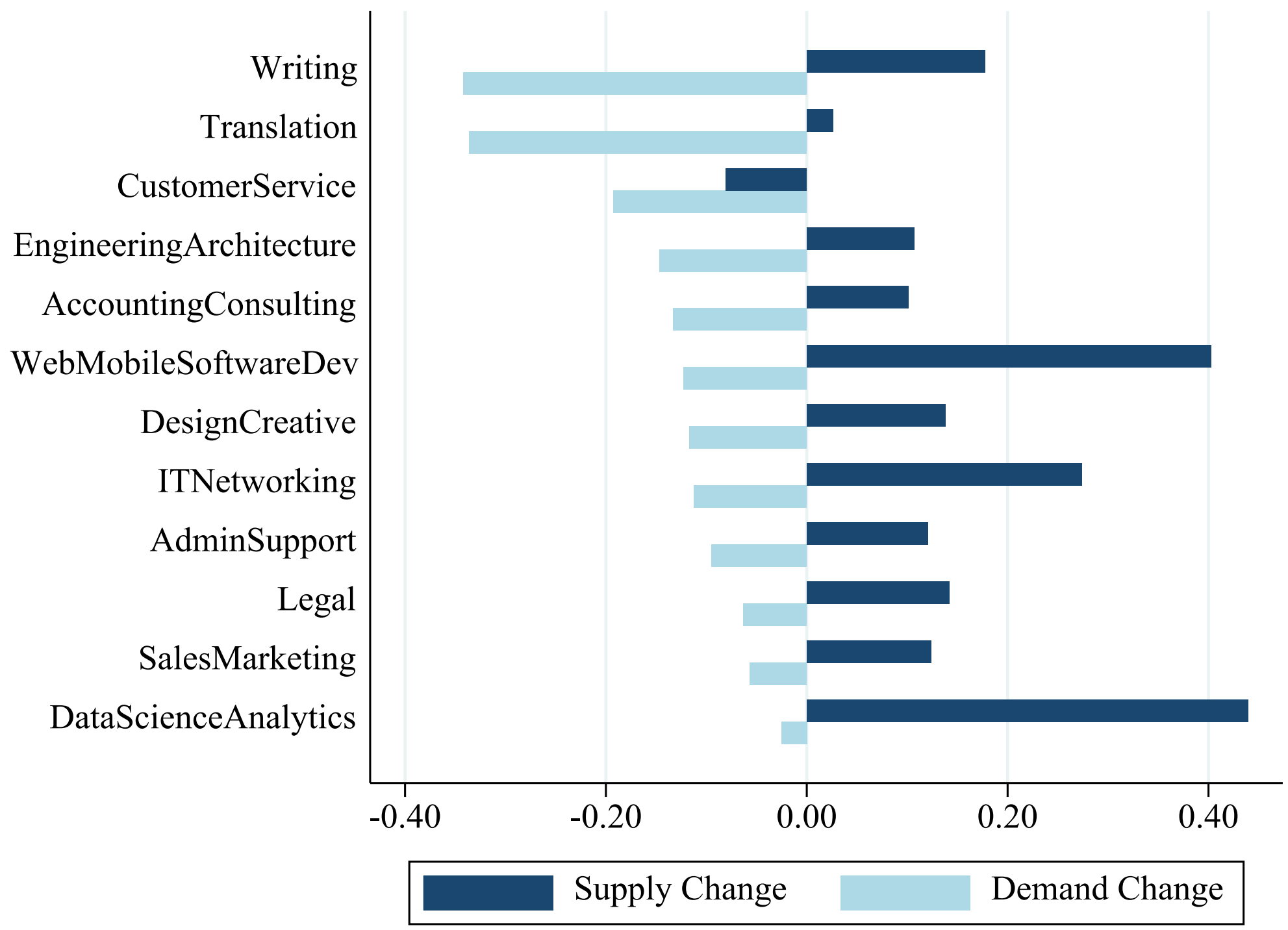

**Figure 3. Changes in freelancer strategic positioning pre- and post-ChatGPT.**

The following figures present the results of relative time models documenting changes in freelancer positioning from June 2022 through March 2024, following equation (1). The x-axis represents months relative to the launch of ChatGPT. For a description of the PD estimation approach, please see the draft. Panel A contains simple time trends with freelancer fixed effects for sample freelancers. Corresponding regression models are found in *Appendix Table A3*. Panel B documents binned scatterplots with freelancer fixed effects using the PD approach discussed in the draft. In both panels, we control for *log bids* when *cosine similarity to prior behavior* or *percent high value bids* are dependent variables.

**Panel A. Sample time trends.**

A1. Log bids.

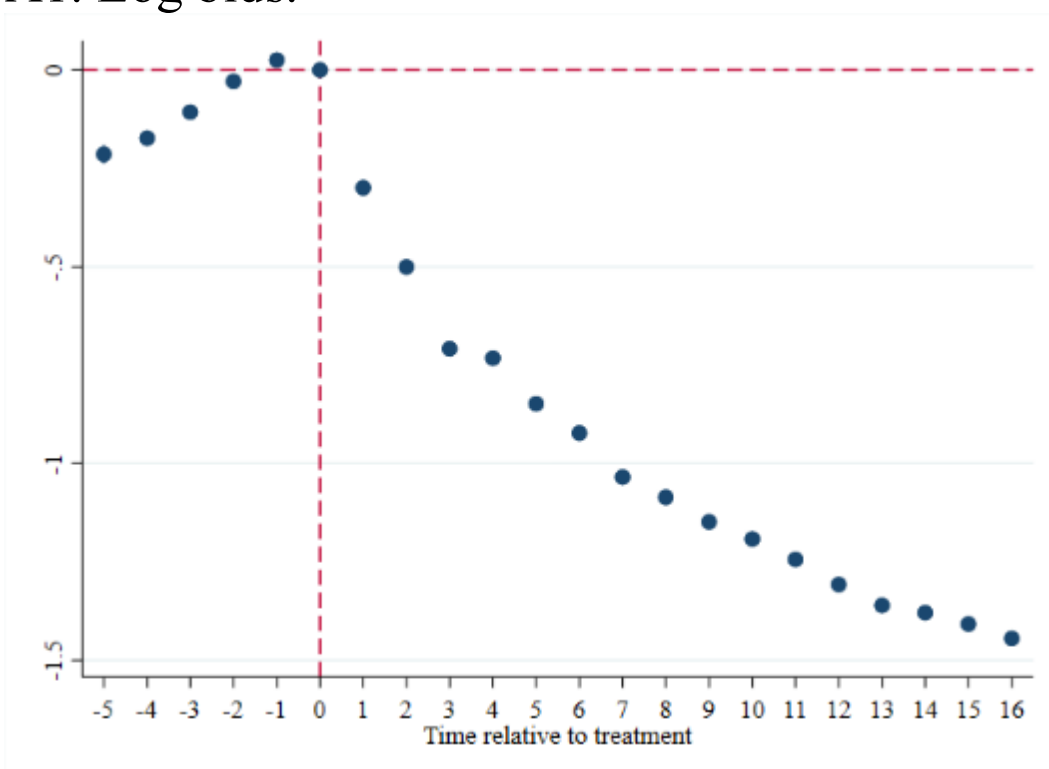

A2. Similarity to prior behavior.

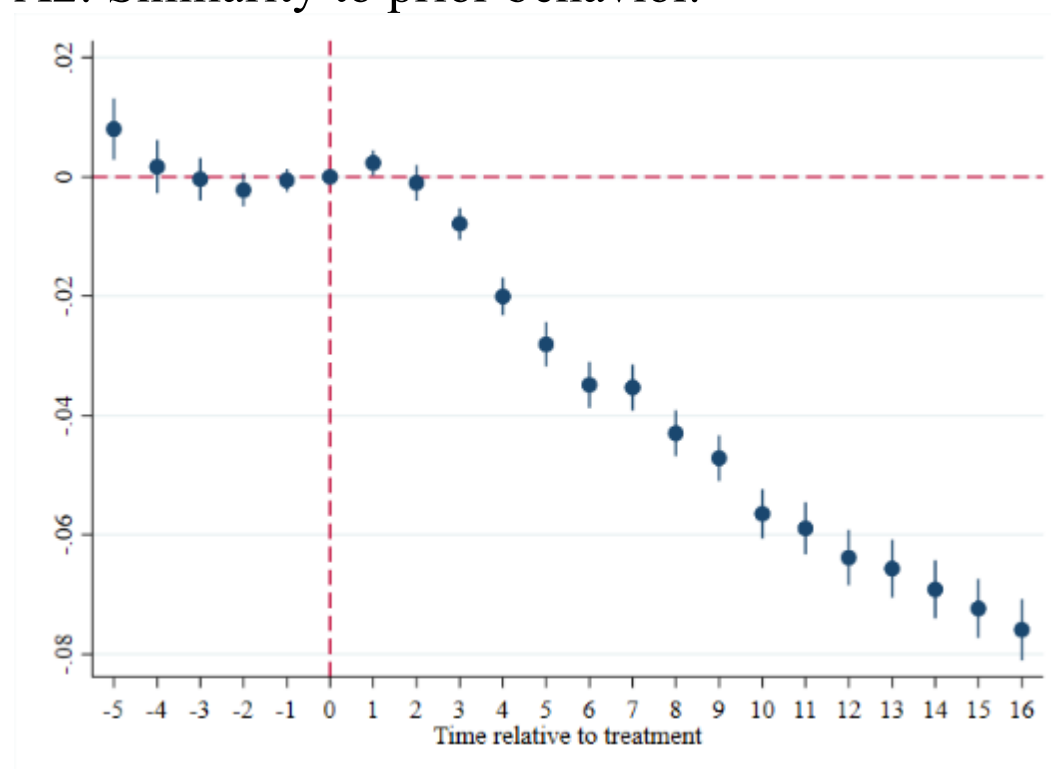

A3. Percent high value bids.

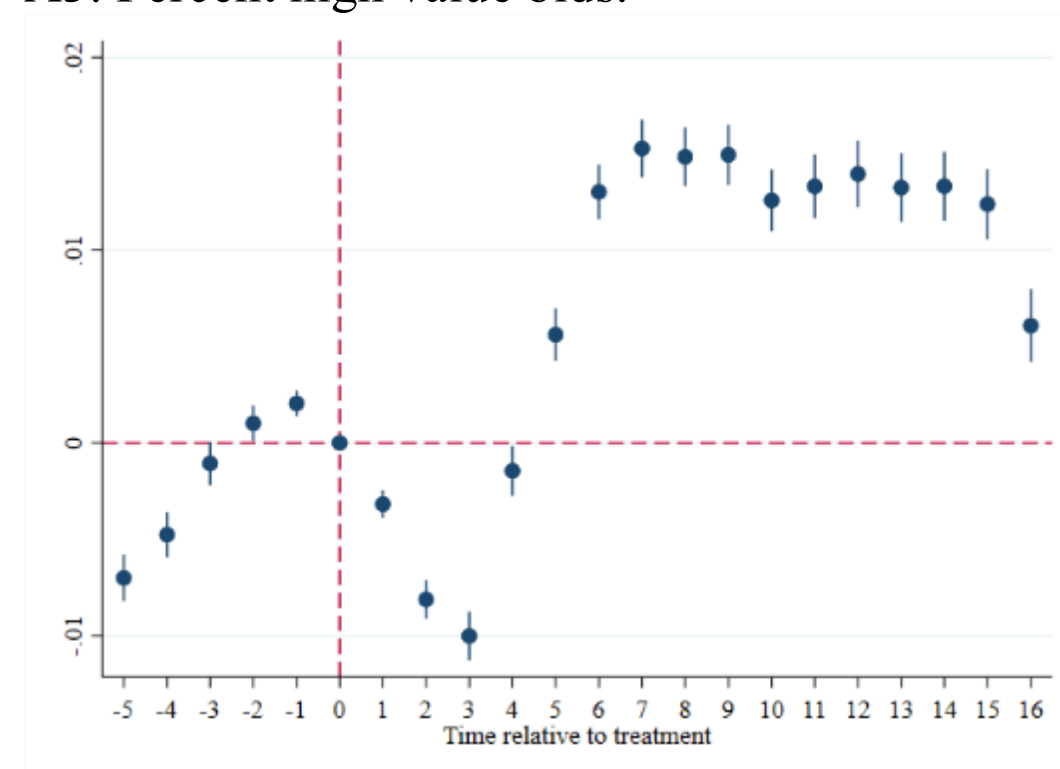

**Panel B. Panel differences binned scatterplots.**

B1. Log bids.

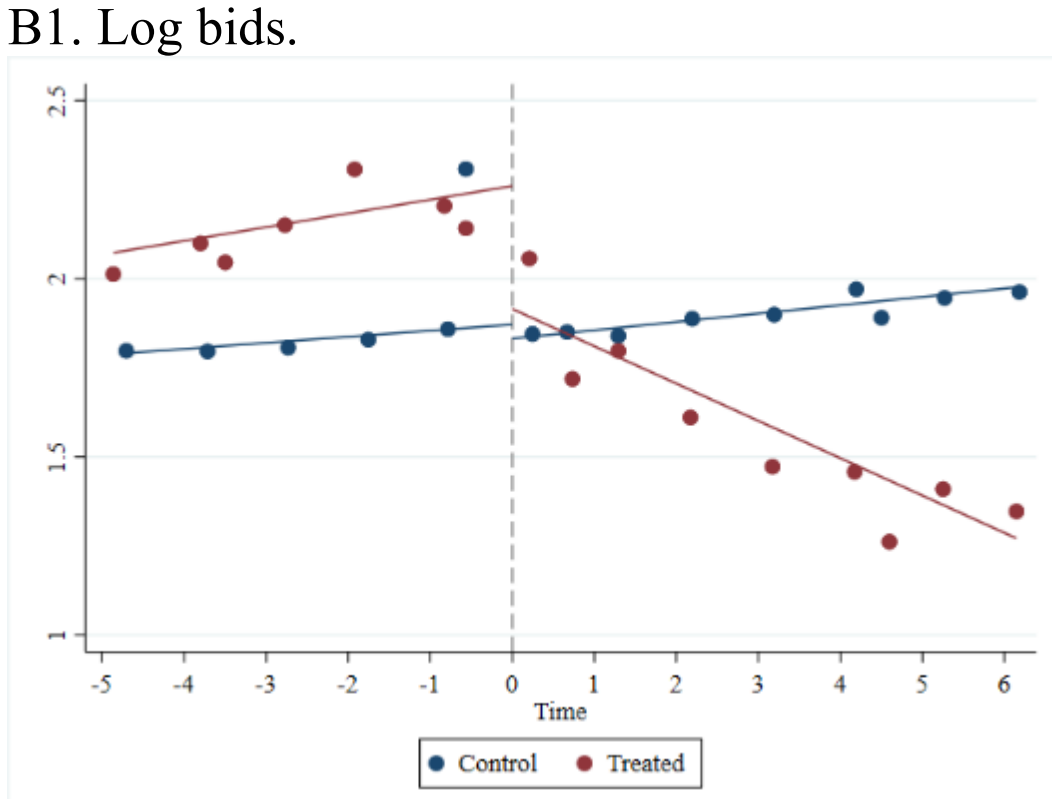

B2. Similarity to prior behavior.

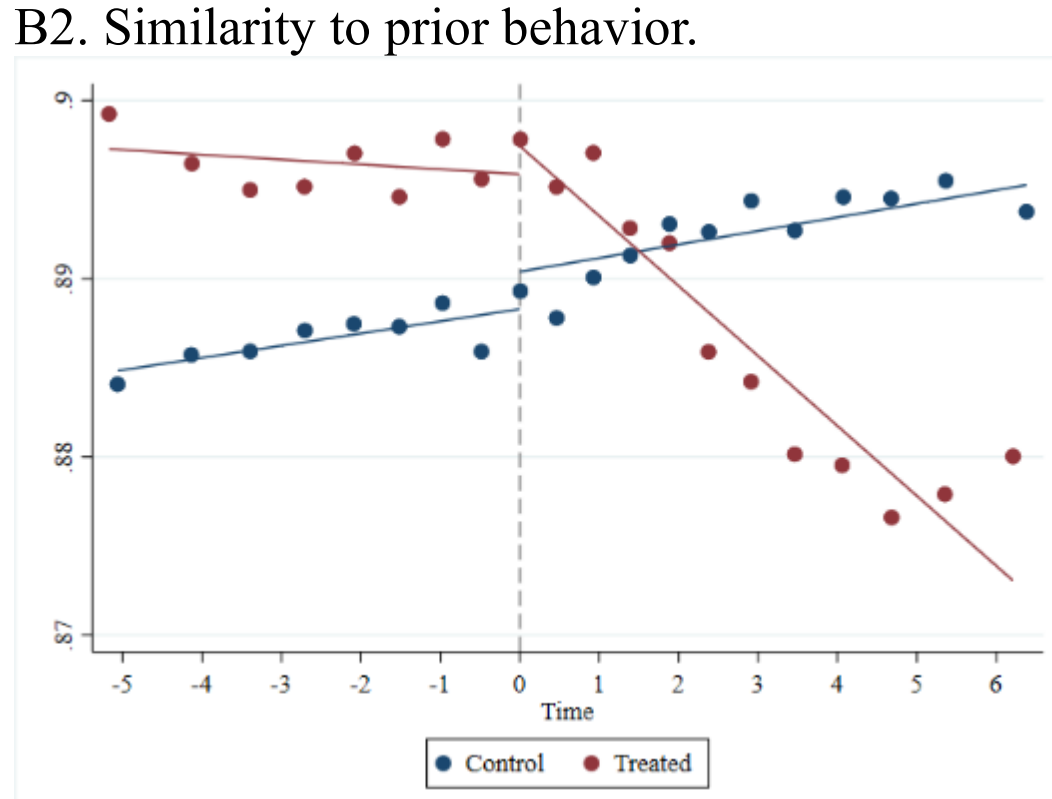

B3. Percent high value bids.

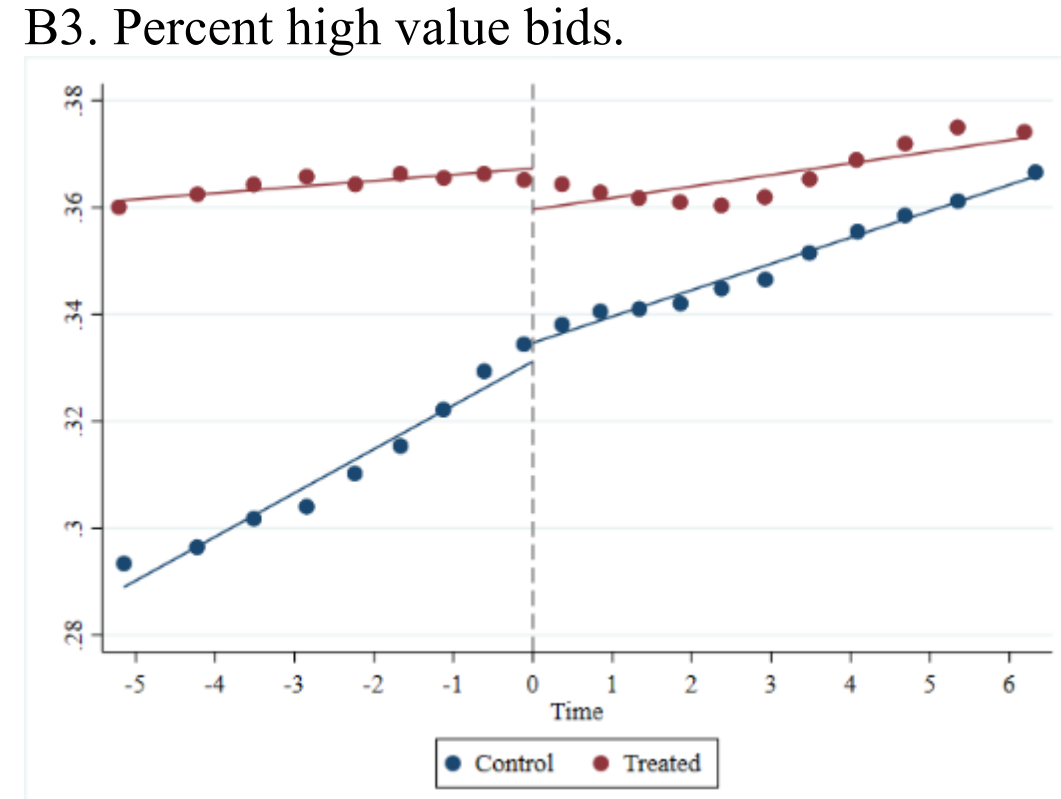

**Figure 4. Changes to freelancer activity pre- and post-ChatGPT by domain.**

This figure contains binned scatterplots with freelancer fixed effects documenting changes in freelancer activity by pre-period modal work category. Panels B and C include a control for the count of bids. The grouping of the twelve platform work categories into five broader domains is defined in the text.

**Panel A. Log bids.**

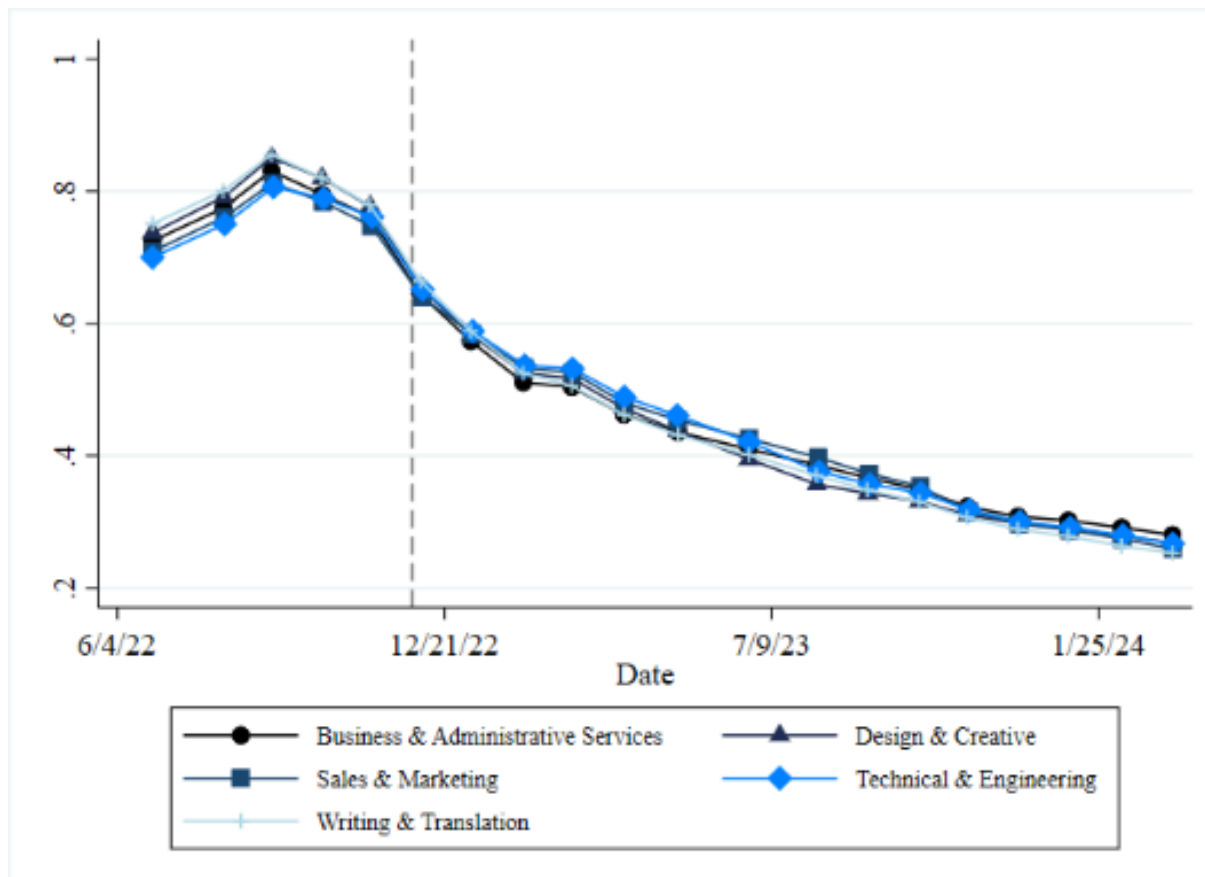

**Panel B. Similarity to prior behavior.**

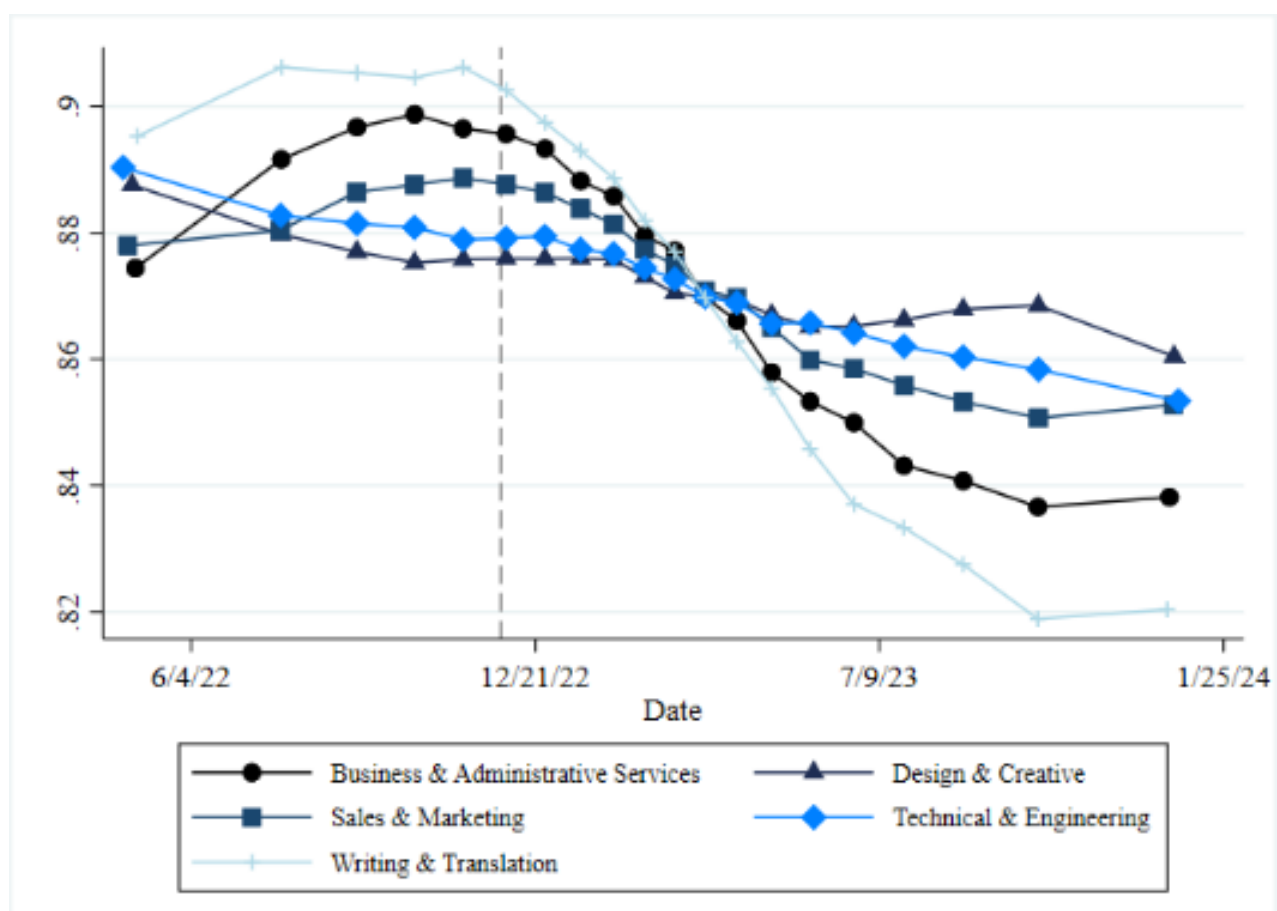

**Panel C. Percent high value bids.**

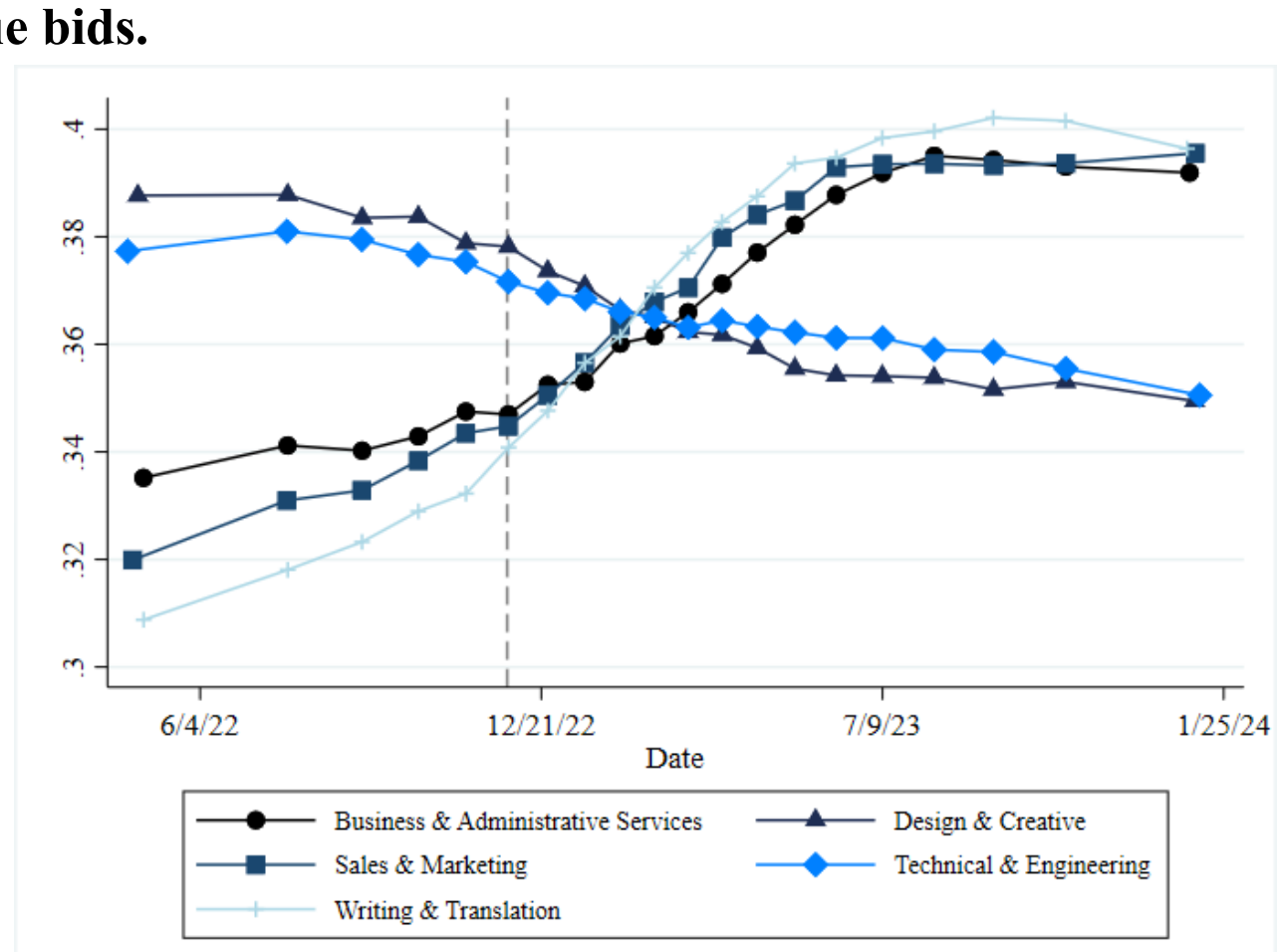

**Figure 5. Changes in freelancer strategic positioning pre- and post-ChatGPT by demand contraction and supply expansion exposure.**

The following figures present the results of relative time models documenting changes in freelancer positioning from June 2022 through March 2024 by demand contraction exposure and supply expansion exposure. The CEM procedure is described in the manuscript. The x-axis represents months relative to the launch of ChatGPT. Corresponding regression models are found in *Appendix Tables A6-A8*.

**Panel A. Full sample estimates.**
**Panel A1. Log bids.**

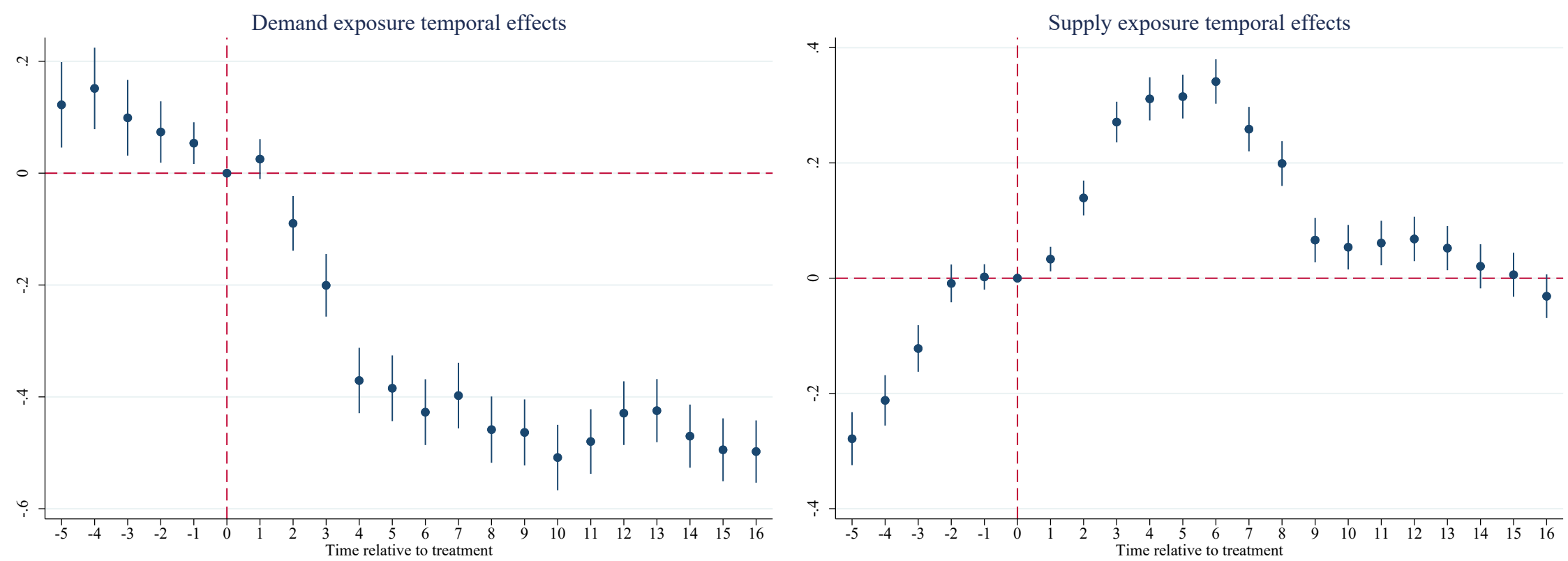

**Panel A2. Similarity to prior behavior.**

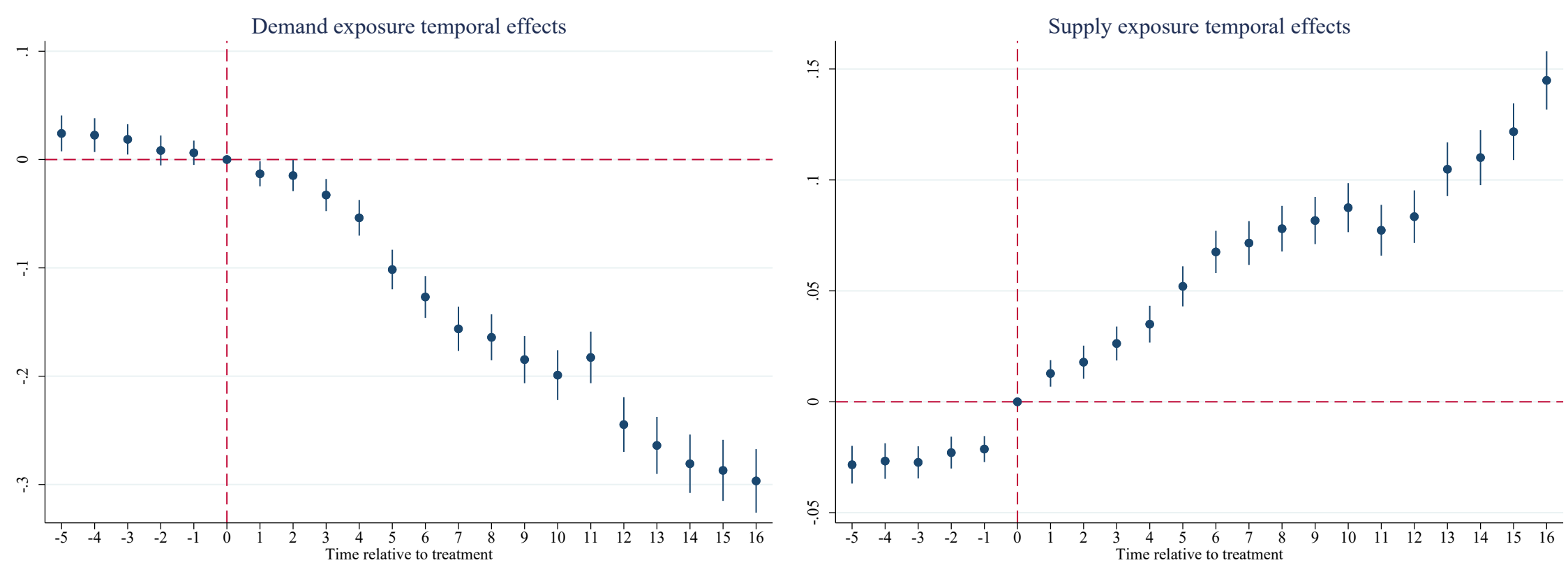

**Panel A3. Percent high-value bids.**

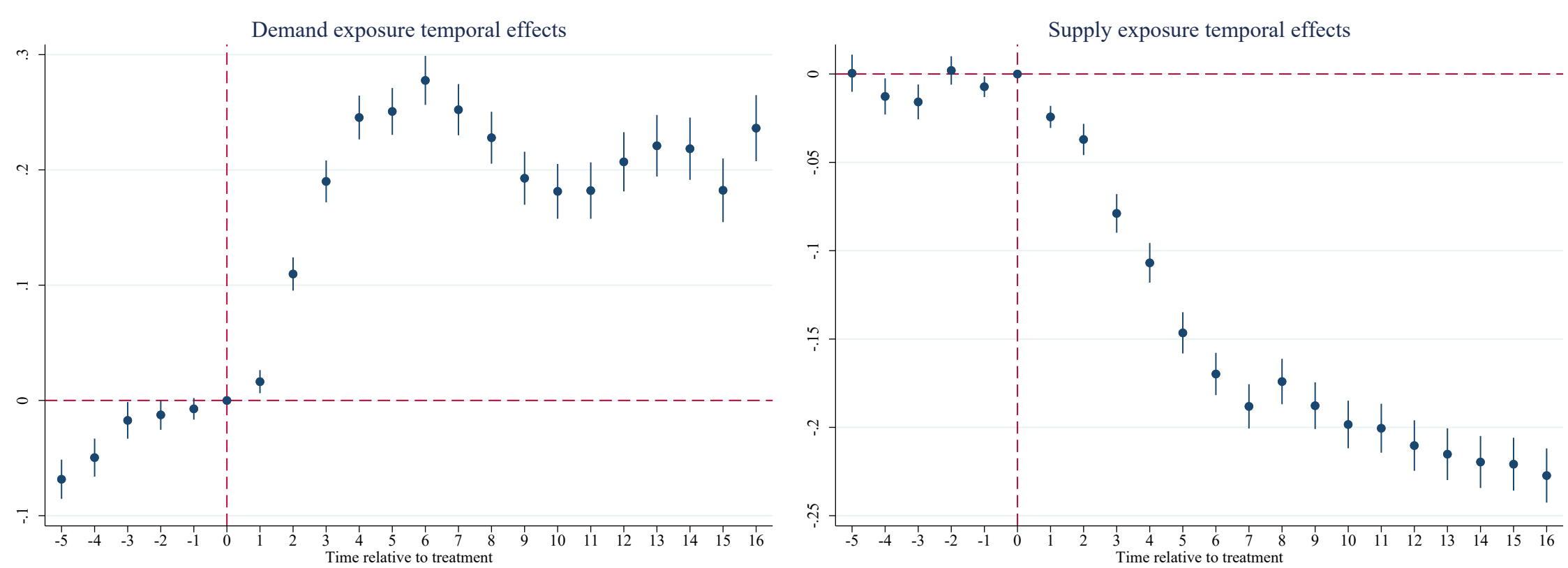

**Panel B. Coarsened Exact Matching (CEM) estimates.**
**Panel B1. Log bids.**

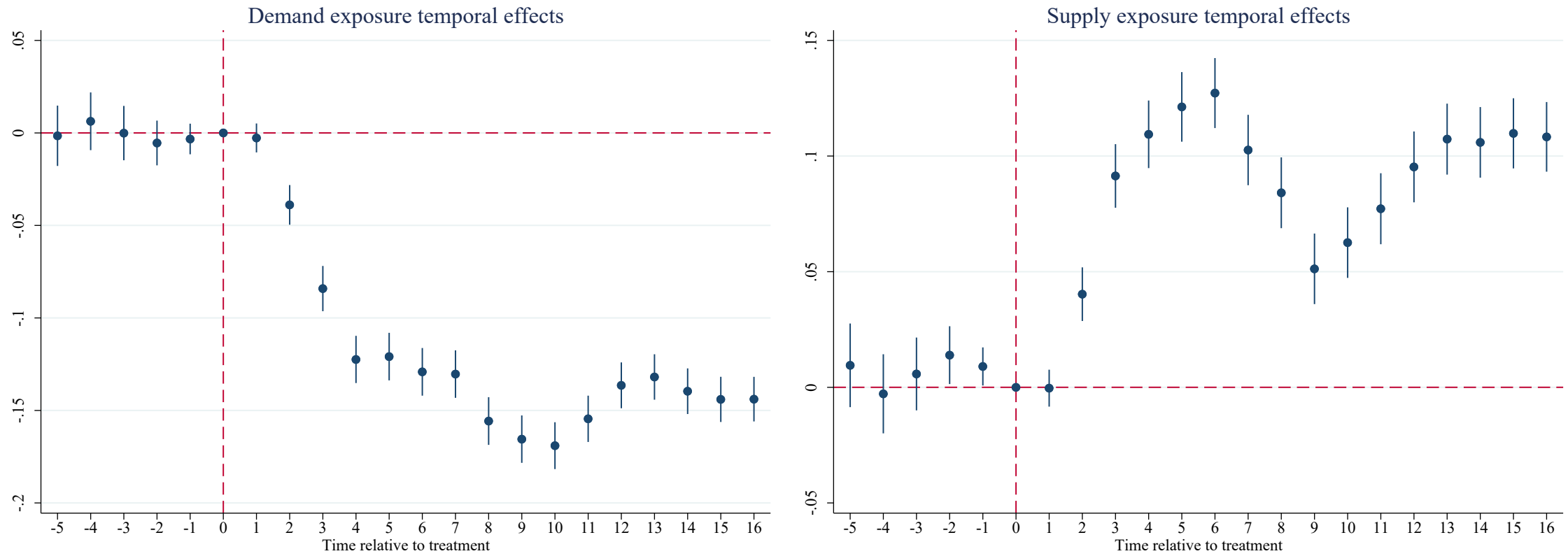

**Panel B2. Similarity to prior behavior.**

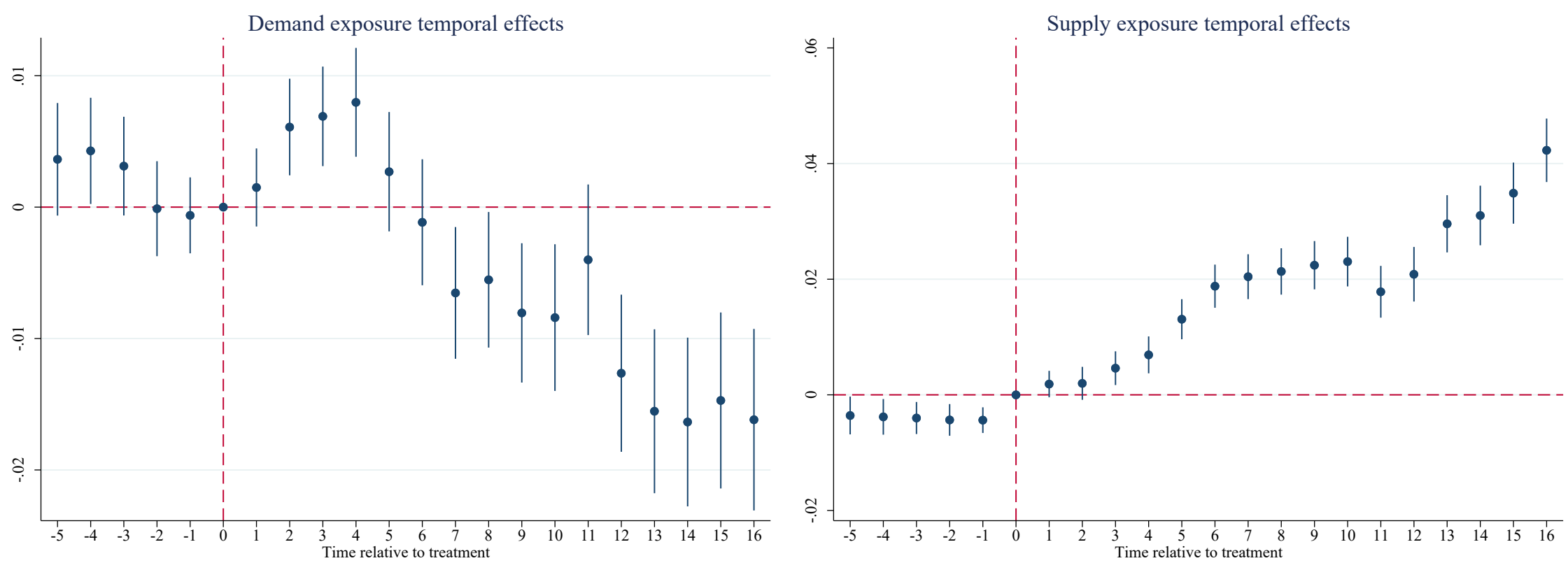

**Panel B3. Percent high-value bids.**

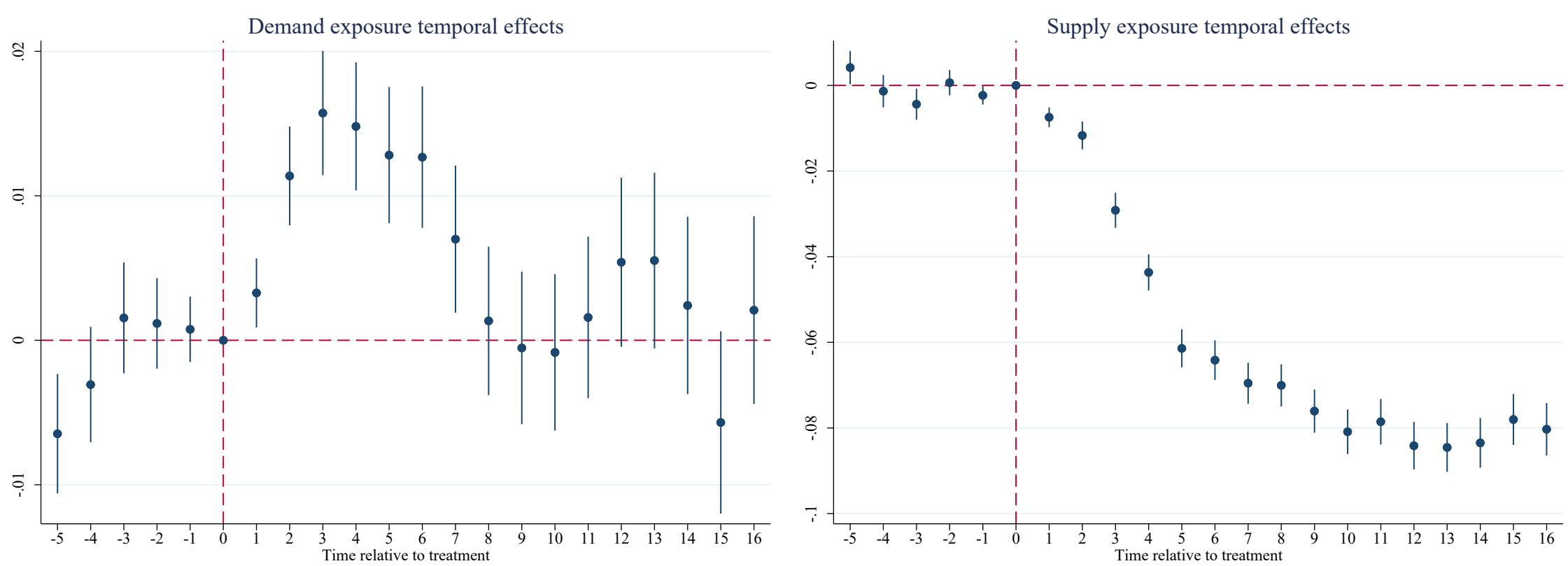

**Figure 6. Changes in freelancer strategic positioning pre- and post-ChatGPT by pre-period skill level.**

The following figures present the results of relative time models documenting changes in freelancer positioning from June 2022 through March 2024 by pre-period skill-level. The CEM procedure is described in the manuscript. The x-axis represents months relative to the launch of ChatGPT. Corresponding regression models are found in *Appendix Tables A9-A10*.

**Panel A. Full sample estimates.**

A1. Log bids.

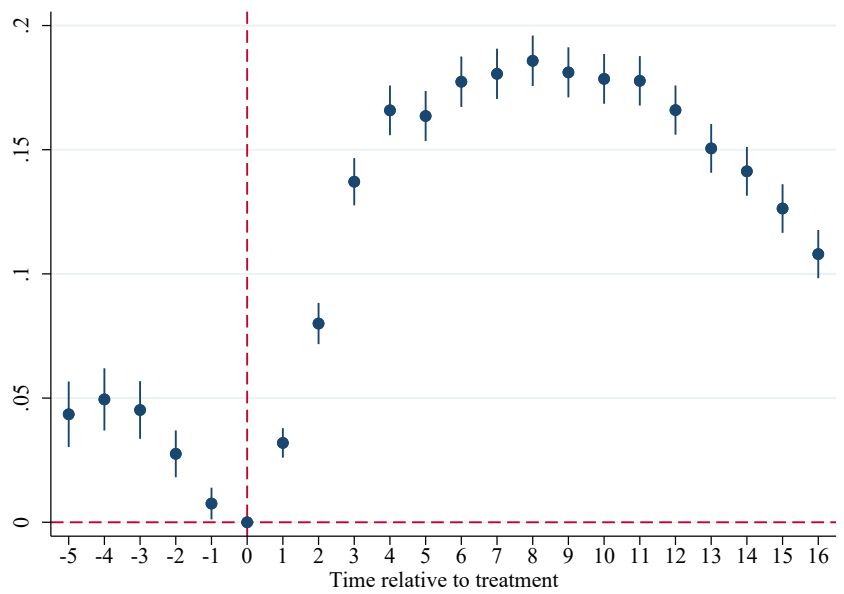

A2. Similarity to prior behavior.

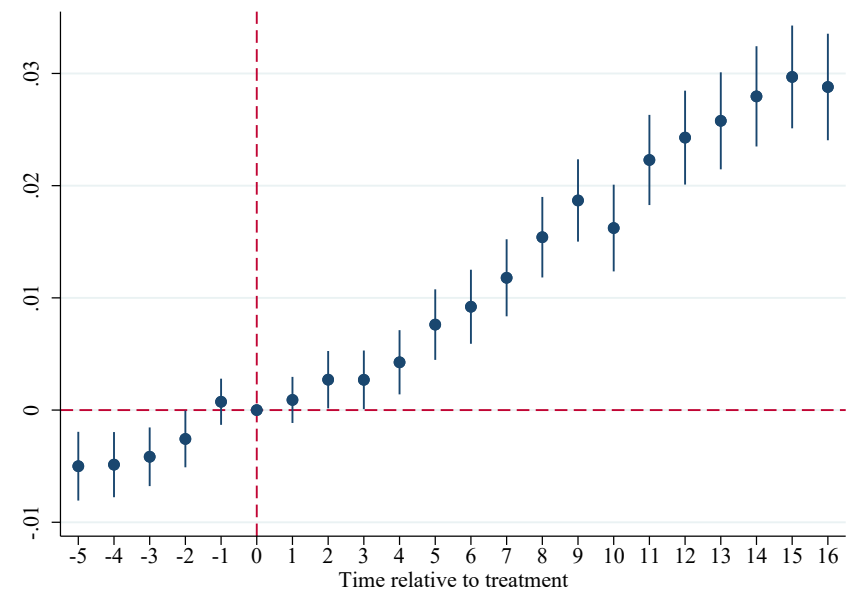

A3. Percent high value bids.

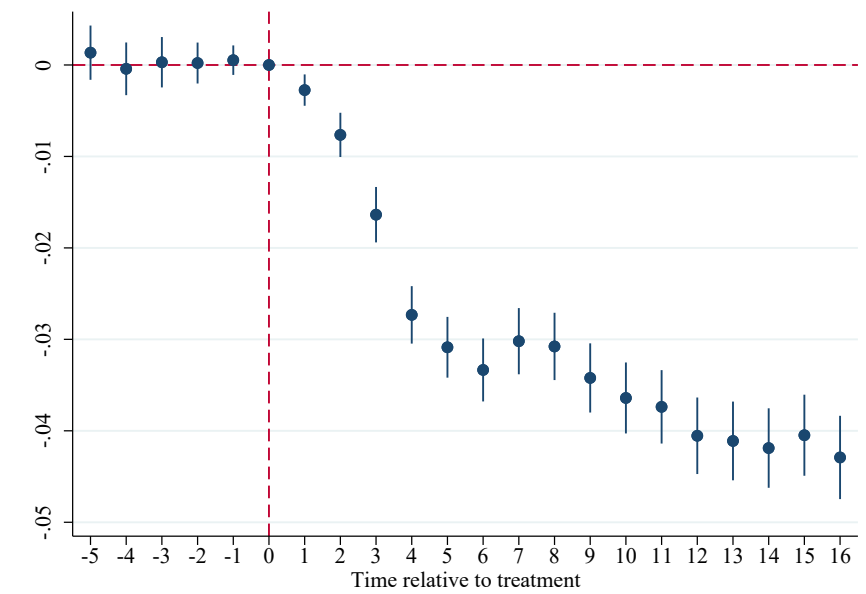

**Panel B. CEM estimates.**

B1. Log bids.

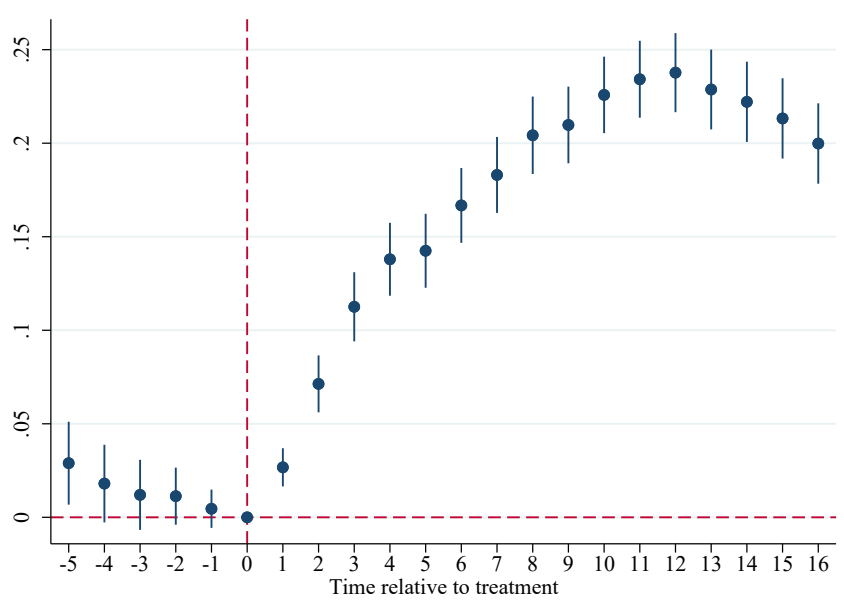

B2. Similarity to prior behavior.

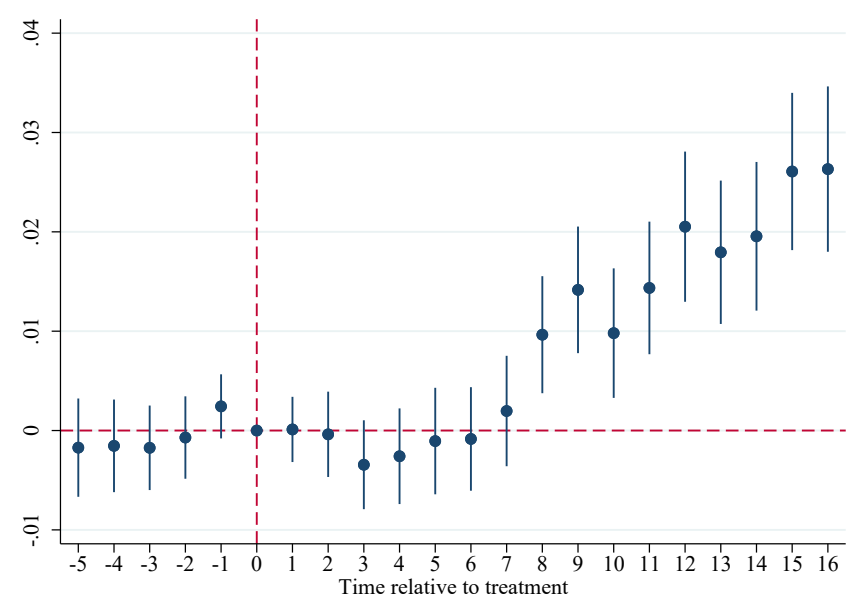

B3. Percent high value bids.

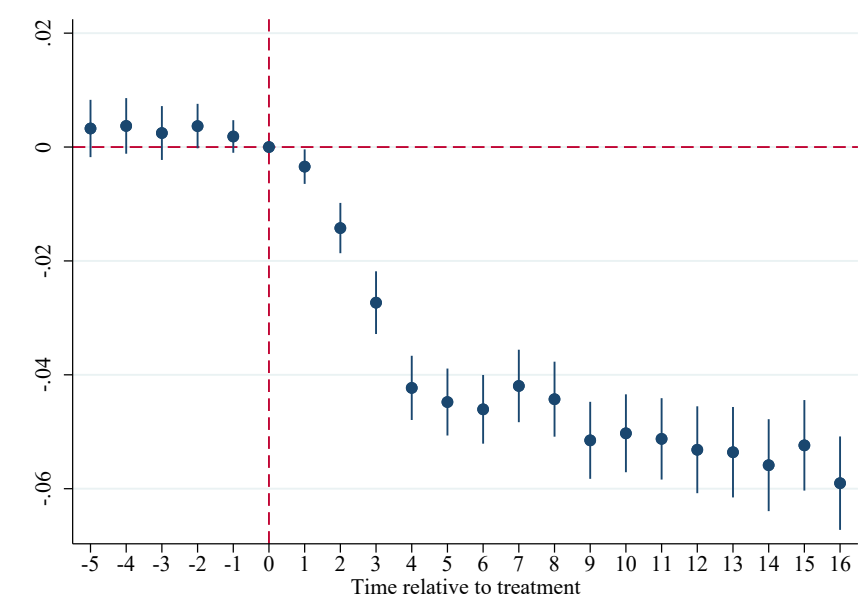

**Table 1. Summary statistics.**

The following table shows means and standard deviations for relevant variables used in the main analysis. All variables are at the freelancer-month level.

| | | | | |
|---|---|---|---|---|
| Freelancer-month observations | | | | 6,633,007 |
| Number of freelancers | | | | 312,143 |
| | | Full sample | Pre-ChatGPT | Post-ChatGPT |
| Number of bids | *mean* | 10.821 | 17.229 | 8.713 |
| | *std. dev* | *45.264* | *55.106* | *41.302* |
| Similarity to prior behavior | *mean* | 0.871 | 0.889 | 0.862 |
| | *std. dev* | *0.249* | *0.220* | *0.262* |
| Percent high-value bids | *mean* | 0.366 | 0.351 | 0.374 |
| | *std. dev* | *0.341* | *0.316* | *0.353* |
| Supply change exposure | *mean* | 0.200 | | |
| | *std. dev* | *0.116* | | |
| Demand change exposure | *mean* | -0.141 | | |
| | *std. dev* | *0.070* | | |
| Average skill pre-treatment | *mean* | 2.008 | | |
| | *std. dev* | *0.521* | | |

**Table 2. Estimates of the effect of ChatGPT on freelancer strategic positioning.**

The following table presents regression estimates of the effect of ChatGPT on freelancer bidding activity, following equation (2). Model 1-3 uses the baseline sample. Model 4-6 implements the panel differences approach (for more details, see the draft), which includes month (but not month-by-year) fixed effects to account for cyclical patterns in bidding behavior across a calendar year. Robust standard errors clustered at the freelancer level in parentheses. ***, **, and * denote significance at the 1%, 5%, and 10% level, respectively.

| | (1) | (2) | (3) | (4) | (5) | (6) |
|---|---|---|---|---|---|---|
| | Baseline sample | | | Panel differences sample | | |
| Dependent variable: | Log bids | Similarity to prior behavior | Percent high-value bids | Log bids | Similarity to prior behavior | Percent high-value bids |
| Post-ChatGPT | -0.963*** | -0.019*** | 0.006*** | | | |
| | (0.002) | (0.000) | (0.000) | | | |
| Post x treated | | | | -0.717*** | -0.018*** | -0.037*** |
| | | | | (0.003) | (0.001) | (0.001) |
| Log bids | | 0.050*** | -0.017*** | | 0.041*** | -0.013*** |
| | | (0.000) | (0.000) | | (0.000) | (0.000) |
| Freelancer fixed effects | Yes | Yes | Yes | Yes | Yes | Yes |
| Month fixed effects | No | No | No | Yes | Yes | Yes |
| Mean DV | 1.367 | 0.871 | 0.366 | 1.850 | 0.891 | 0.354 |
| No. of freelancers | 312,143 | 298,926 | 310,189 | 312,143 | 299,944 | 309,885 |
| R-squared | 0.605 | 0.458 | 0.583 | 0.537 | 0.446 | 0.633 |
| Observations | 6,633,007 | 3,790,149 | 4,430,937 | 5,702,926 | 3,999,792 | 4,602,481 |

## Table 3. Matrices documenting changes in freelancer modal work category pre- and post-ChatGPT.

Matrices are constructed by constructing a modal L1 for each freelancer in the eleven months prior to the launch of ChatGPT (January 2022 through November 2022) and in the year post-ChatGPT (December 2022 through November 2023) and then aggregating across work categories to document the propensity of freelancers to shift across work categories. To contextualize the changes from the pre- to post-period, we present normalized matrices that use the prior year as a control. For control matrices, the pre-period is January 2021 through November 2021 and the post-period is December 2021 through November 2022. The matrices underlying the normalized matrix presented in Panel A are constructed using all freelancers that submitted at least one bid during the relevant time period. The matrices underlying the normalized matrix presented in Panel B are constructed using only incumbent freelancers that meet the main sample criteria outlined in the draft for each relevant time period. Because of this sample criteria, Panel B does not include any freelancers that were inactive in the pre-period.

### Panel A. All freelancers.

| | | Post-period modal L1 | | | | | | | | | | | | |
|---|---|---|---|---|---|---|---|---|---|---|---|---|---|---|
| | L1 Name | 1 | 2 | 3 | 4 | 5 | 6 | 7 | 8 | 9 | 10 | 11 | 12 | 13 |
| Pre-period modal L1 | 1 Writing | -3.1% | 0.1% | 0.0% | 0.0% | 0.0% | 0.0% | 0.3% | 0.0% | 0.4% | 0.0% | 0.3% | 0.2% | 1.7% |
| | 2 Translation | 0.9% | -1.2% | 0.1% | 0.0% | 0.1% | 0.2% | 0.6% | 0.0% | 1.1% | 0.0% | 0.2% | 0.2% | -2.3% |
| | 3 Customer Service | 0.2% | 0.0% | -3.0% | 0.1% | 0.1% | -0.1% | 0.3% | -0.5% | 0.3% | 0.1% | 0.5% | 0.2% | 1.9% |
| | 4 Engineering & Architecture | 0.0% | 0.0% | 0.0% | -2.5% | 0.1% | -0.1% | -0.3% | 0.0% | 0.2% | 0.0% | 0.1% | 0.2% | 2.4% |
| | 5 Accounting & Consulting | -0.2% | 0.0% | 0.0% | 0.0% | -3.0% | -0.1% | 0.1% | 0.0% | 0.4% | 0.0% | 0.1% | 0.1% | 2.6% |
| | 6 Web, Mobile & Software Dev | 0.1% | 0.0% | 0.0% | 0.0% | 0.0% | -7.1% | 0.0% | 0.0% | 0.0% | 0.0% | -0.1% | 0.1% | 7.2% |
| | 7 Design & Creative | 0.3% | 0.0% | 0.0% | 0.1% | 0.0% | -0.2% | -3.8% | 0.0% | 0.3% | 0.0% | 0.2% | 0.1% | 2.9% |
| | 8 IT & Networking | 0.1% | 0.0% | 0.1% | 0.0% | 0.1% | -1.4% | 0.0% | -2.8% | 0.4% | 0.0% | 0.1% | 0.1% | 3.1% |
| | 9 Admin Support | 0.1% | 0.0% | -0.4% | 0.0% | -0.2% | -0.1% | 0.2% | 0.0% | -2.8% | 0.0% | -0.2% | 0.2% | 3.3% |
| | 10 Legal | -0.8% | -0.2% | 0.0% | 0.1% | -0.2% | -0.1% | 0.1% | -0.1% | -0.5% | -5.6% | -0.1% | 0.1% | 7.3% |
| | 11 Sales & Marketing | 0.2% | 0.1% | -0.4% | 0.0% | 0.0% | -0.2% | 0.4% | 0.0% | -0.1% | 0.0% | -4.7% | 0.2% | 4.4% |
| | 12 Data Science & Analytics | 0.1% | 0.0% | 0.0% | -0.1% | -0.1% | -1.4% | 0.0% | 0.0% | 0.3% | 0.0% | -0.1% | -1.7% | 3.0% |
| | 13 Not active | 0.2% | -1.1% | -0.7% | -0.4% | -0.5% | 7.1% | -2.2% | 0.3% | -4.1% | 0.0% | -0.2% | 1.6% | 0.0% |

### Panel B. Incumbent freelancers.

| | | Post-period modal L1 | | | | | | | | | | | | |
|---|---|---|---|---|---|---|---|---|---|---|---|---|---|---|
| | L1 Name | 1 | 2 | 3 | 4 | 5 | 6 | 7 | 8 | 9 | 10 | 11 | 12 | 13 |
| Pre-period modal L1 | 1 Writing | -10.2% | 0.6% | 0.0% | 0.2% | 0.2% | 0.1% | 0.9% | 0.0% | 1.8% | 0.0% | 1.3% | 0.6% | 4.4% |
| | 2 Translation | 1.1% | -12.0% | 0.1% | 0.0% | 0.1% | 0.3% | 1.2% | 0.0% | 2.8% | -0.1% | 0.2% | 0.4% | 5.9% |
| | 3 Customer Service | -0.1% | -0.4% | -10.5% | 0.2% | 0.0% | -0.1% | 0.6% | -0.6% | 3.3% | 0.1% | 1.9% | 0.3% | 5.3% |
| | 4 Engineering & Architecture | -0.1% | 0.1% | 0.1% | -4.8% | 0.1% | -0.2% | -0.1% | 0.0% | 0.3% | 0.0% | 0.2% | 0.2% | 4.3% |
| | 5 Accounting & Consulting | -0.7% | 0.1% | 0.0% | 0.1% | -6.9% | -0.1% | 0.0% | 0.0% | 2.4% | -0.2% | 0.0% | 0.4% | 4.9% |
| | 6 Web, Mobile & Software Dev | 0.1% | 0.1% | 0.0% | 0.0% | 0.0% | -4.6% | 0.5% | 0.1% | 0.4% | 0.0% | 0.2% | 0.4% | 2.7% |
| | 7 Design & Creative | 0.3% | 0.2% | 0.0% | 0.4% | 0.0% | -0.1% | -8.5% | 0.0% | 0.9% | 0.0% | 0.4% | 0.2% | 6.2% |
| | 8 IT & Networking | 0.0% | 0.0% | 0.5% | 0.0% | 0.1% | -1.6% | -0.1% | -1.9% | 0.3% | 0.0% | 0.2% | 0.3% | 2.3% |
| | 9 Admin Support | 0.4% | 0.2% | -0.4% | 0.1% | -0.3% | -0.1% | 0.8% | 0.0% | -8.9% | 0.0% | -0.9% | 0.5% | 8.7% |
| | 10 Legal | -0.2% | 0.1% | 0.1% | 0.3% | -1.0% | 0.0% | 0.1% | -0.1% | -0.1% | -1.4% | 0.2% | 0.0% | 2.1% |
| | 11 Sales & Marketing | 0.1% | 0.1% | -0.1% | 0.0% | 0.0% | 0.3% | 1.0% | 0.0% | 1.5% | 0.0% | -9.0% | 0.4% | 5.6% |
| | 12 Data Science & Analytics | 0.1% | 0.0% | 0.0% | 0.0% | -0.3% | -2.1% | 0.2% | 0.0% | 1.2% | 0.0% | 0.1% | -2.7% | 3.5% |

**Table 4. Estimates of the heterogeneous effect of ChatGPT on freelancer strategic positioning by freelancer ex-ante demand and supply change exposure.**

The following table presents regression estimates of the heterogeneous effect of ChatGPT on freelancer bidding activity by demand change exposure and supply change exposure, following equation (3). Panel A shows results with the baseline sample, Panel B shows results with the coarsened exact matching (CEM) sample. In Panel B, controls include *log bids* in cols. 2-3 and 5-6, Post-ChatGPT x supply expansion exposure in cols. 1-3, and post-ChatGPT x demand contraction exposure in cols. 4-6. Robust standard errors clustered at the freelancer level in parentheses. ***, **, and * denote significance at the 1%, 5%, and 10% level, respectively.

**Panel A. Baseline sample**

| | (1) | (2) | (3) |
|---|---|---|---|
| Dependent variable: | Log bids | Similarity to prior behavior | Percent high-value bids |
| Post-ChatGPT x … | | | |
| Demand contraction exposure | -0.458*** | -0.137*** | 0.210*** |
| | (0.022) | (0.005) | (0.006) |
| Supply expansion exposure | 0.232*** | 0.081*** | -0.141*** |
| | (0.015) | (0.003) | (0.003) |
| Log number of bids | | 0.045*** | -0.014*** |
| | | (0.000) | (0.000) |
| Freelancer fixed effects | Yes | Yes | Yes |
| Month-by-year fixed effects | Yes | Yes | Yes |
| Mean DV | 1.367 | 0.871 | 0.366 |
| No. of freelancers | 312,143 | 298,926 | 310,189 |
| R-squared | 0.654 | 0.461 | 0.584 |
| Observations | 6,633,007 | 3,790,149 | 4,430,937 |

**Panel B. Coarsened exact matching sample.**

| | (1) | (2) | (3) | (4) | (5) | (6) |
|---|---|---|---|---|---|---|
| Dependent variable: | Log bids | Similarity to prior behavior | Percent high-value bids | Log bids | Similarity to prior behavior | Percent high-value bids |
| Post-ChatGPT x … | | | | | | |
| High demand contraction exposure | -0.122*** | -0.004*** | 0.007*** | | | |
| | (0.005) | (0.001) | (0.001) | | | |
| High supply expansion exposure | | | | 0.081*** | 0.019*** | -0.056*** |
| | | | | (0.006) | (0.001) | (0.001) |
| Controls | Yes | Yes | Yes | Yes | Yes | Yes |
| Freelancer fixed effects | Yes | Yes | Yes | Yes | Yes | Yes |
| Month-by-year fixed effects | Yes | Yes | Yes | Yes | Yes | Yes |
| Mean DV | 1.244 | 0.817 | 0.347 | 1.313 | 0.873 | 0.442 |
| No. of freelancers | 150662 | 143611 | 149637 | 147707 | 142126 | 146927 |
| R-squared | 0.613 | 0.431 | 0.531 | 0.641 | 0.454 | 0.596 |
| Observations | 3,189,325 | 1,753,343 | 2,070,483 | 3,148,930 | 1,801,072 | 2,109,938 |

**Table 5. Estimates of the heterogeneous effect of ChatGPT on freelancer strategic positioning by freelancer skill.**

The following table presents regression estimates of the heterogeneous effect of ChatGPT on freelancer bidding activity by freelancer skill, following equation (3). Model 1-3 shows results with the baseline sample, Model 4-6 shows results with the coarsened exact matching sample. Robust standard errors clustered at the freelancer level in parentheses. ***, **, and * denote significance at the 1%, 5%, and 10% level, respectively.

| | (1) | (2) | (3) | (4) | (5) | (6) |
|---|---|---|---|---|---|---|
| | | Baseline sample | | | CEM sample | |
| Dependent variable: | Log bids | Similarity to prior behavior | Percent high-value bids | Log bids | Similarity to prior behavior | Percent high-value bids |
| Post-ChatGPT x average skill pre-treatment | 0.120*** | 0.014*** | -0.028*** | | | |
| | (0.004) | (0.001) | (0.001) | | | |
| Post-ChatGPT x high average skill pre-treatment | | | | 0.164*** | 0.007*** | -0.042*** |
| | | | | (0.008) | (0.002) | (0.002) |
| Log number of bids | | 0.045*** | -0.015*** | | 0.051*** | -0.015*** |
| | | (0.000) | (0.000) | | (0.001) | (0.001) |
| Freelancer fixed effects | Yes | Yes | Yes | Yes | Yes | Yes |
| Modal pre-period work category x date fixed effects | Yes | Yes | Yes | Yes | Yes | Yes |
| Mean DV | 1.570 | 0.880 | 0.389 | 1.401 | 0.859 | 0.392 |
| No. of freelancers | 223,384 | 219,112 | 222,913 | 101,194 | 98,504 | 100,842 |
| R-squared | 0.658 | 0.455 | 0.584 | 0.639 | 0.456 | 0.591 |
| Observations | 4,800,894 | 3,107,253 | 3,515,942 | 2,159,230 | 1,308,865 | 1,511,164 |

# Appendix

## Contents

## Appendix Figure A1. Changes in freelancer bids to fixed vs. hourly contract jobs post-ChatGPT.

The following figures present the results of relative time models documenting changes in freelancer positioning from June 2022 through March 2024, following equation (1). The dependent variables used are the log-transformed count of bids to hourly contract jobs, fixed contract jobs, and the percent of bids to hourly contract jobs. The x-axis represents months relative to the launch of ChatGPT. For a description of the PD estimation approach, please see the draft. Panel A contains simple time trends with freelancer fixed effects for sample freelancers. Panel B documents binned scatterplots with freelancer fixed effects using the PD approach discussed in the draft.

**Panel A. Sample time trends.**

A1. Bids to hourly contract jobs.

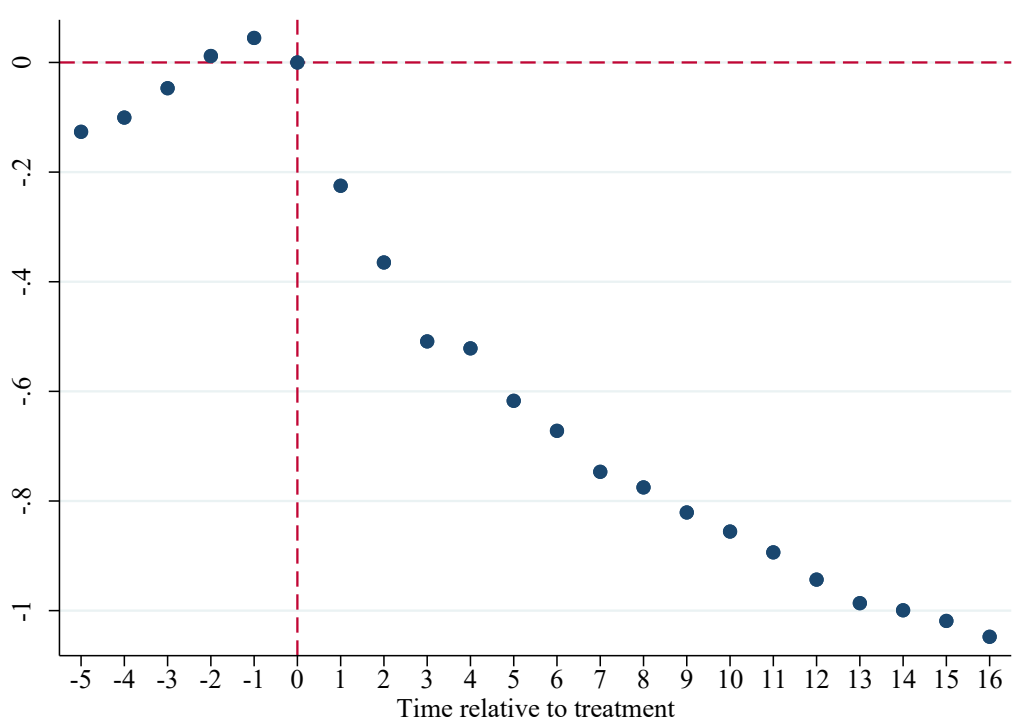

A2. Bids to fixed contract jobs.

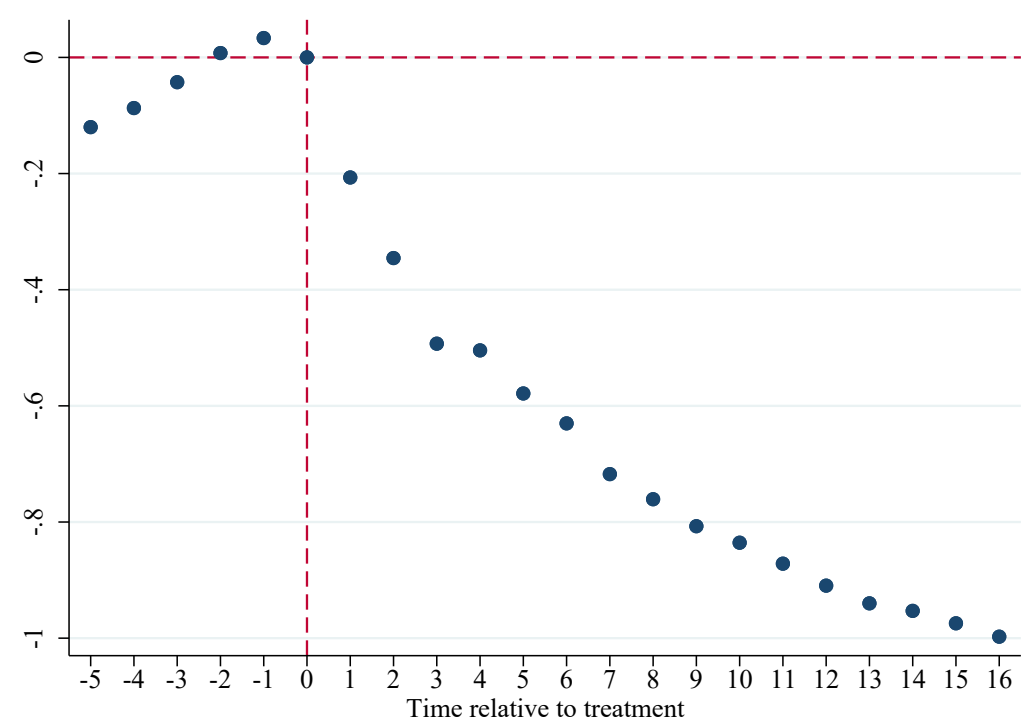

A3. Percent hourly contract bids.

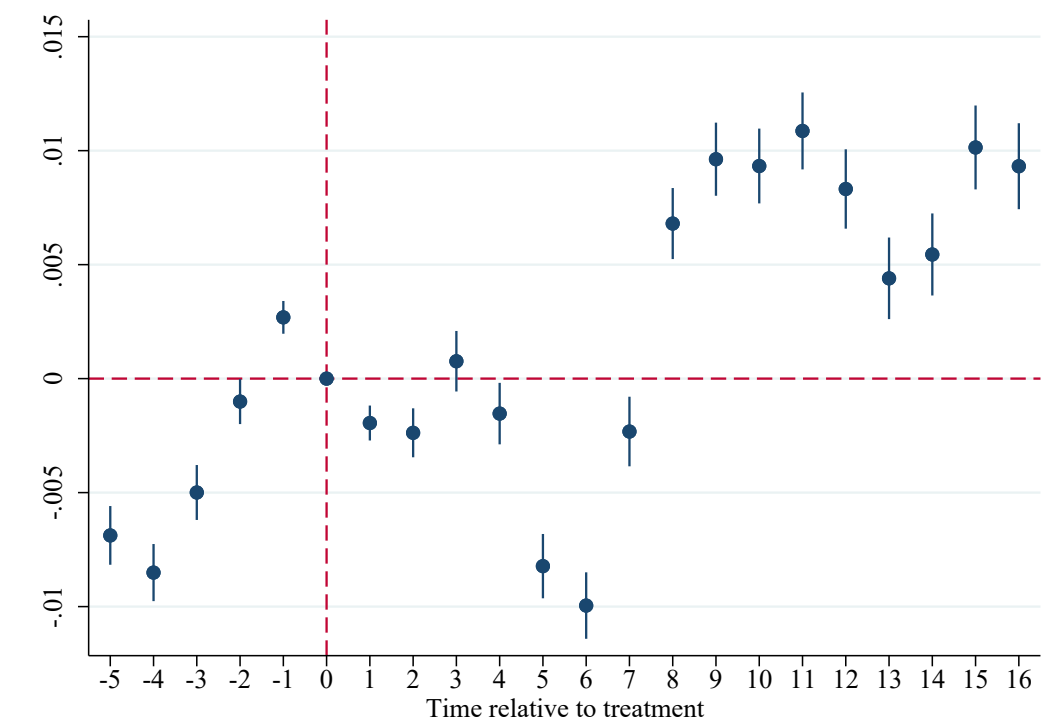

**Panel B. Panel differences binned scatterplots.**

B1. Bids to hourly contract jobs.

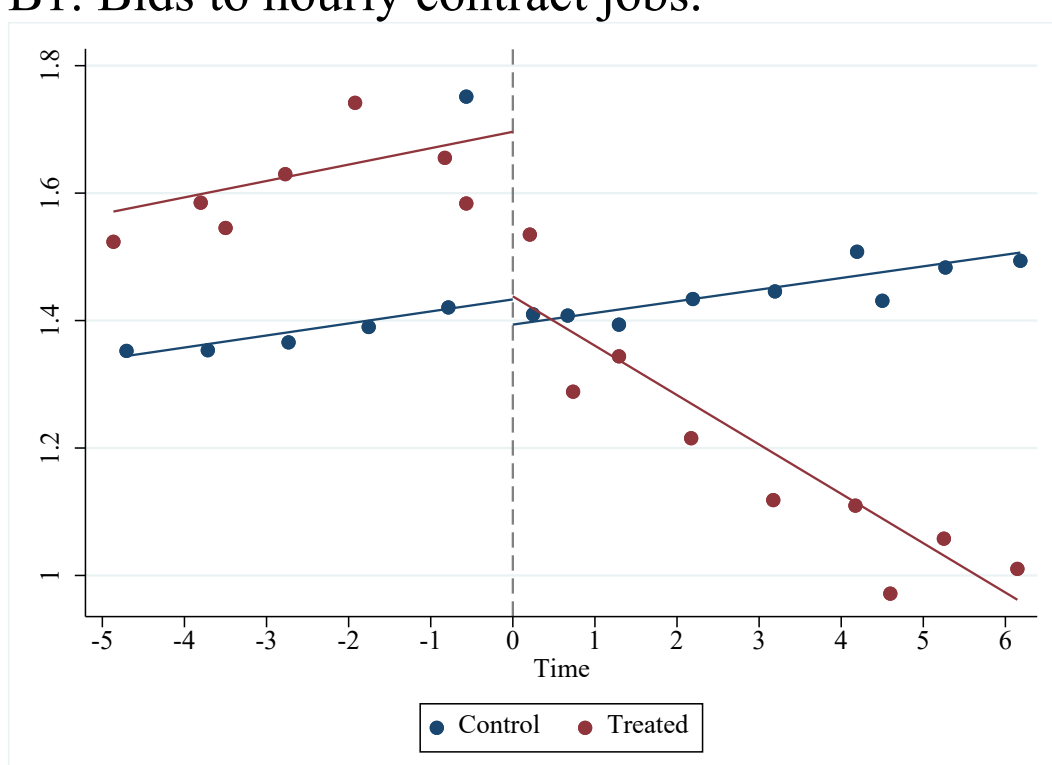

B2. Bids to fixed contract jobs.

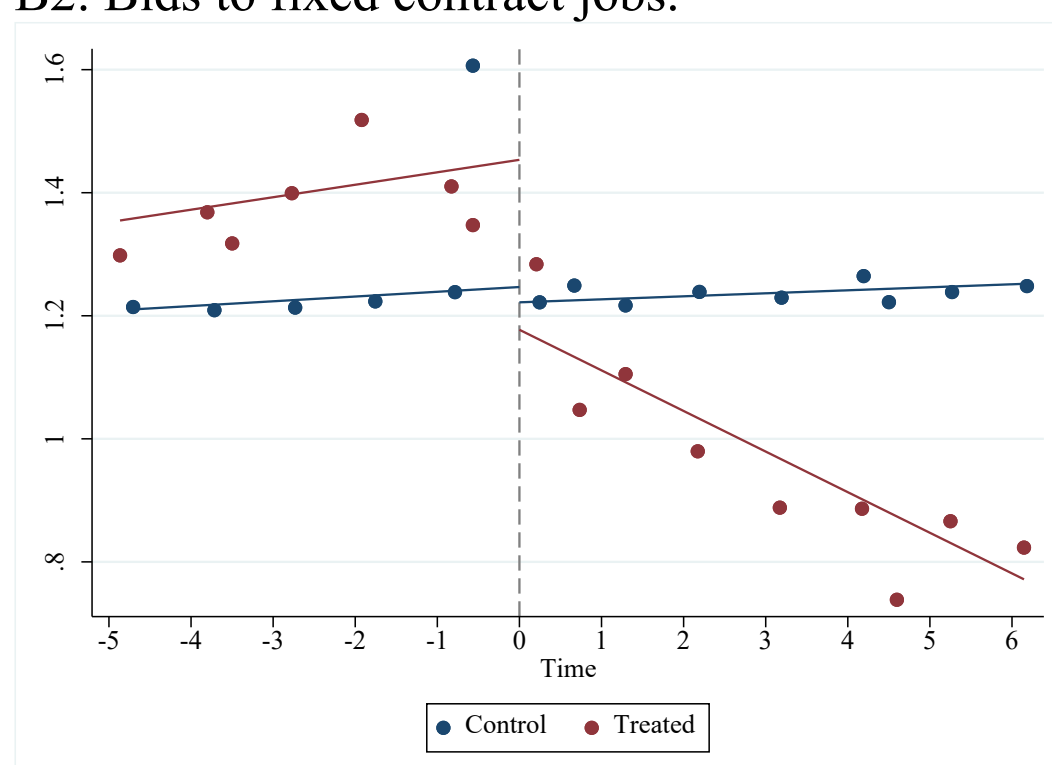

B3. Percent hourly contract bids.

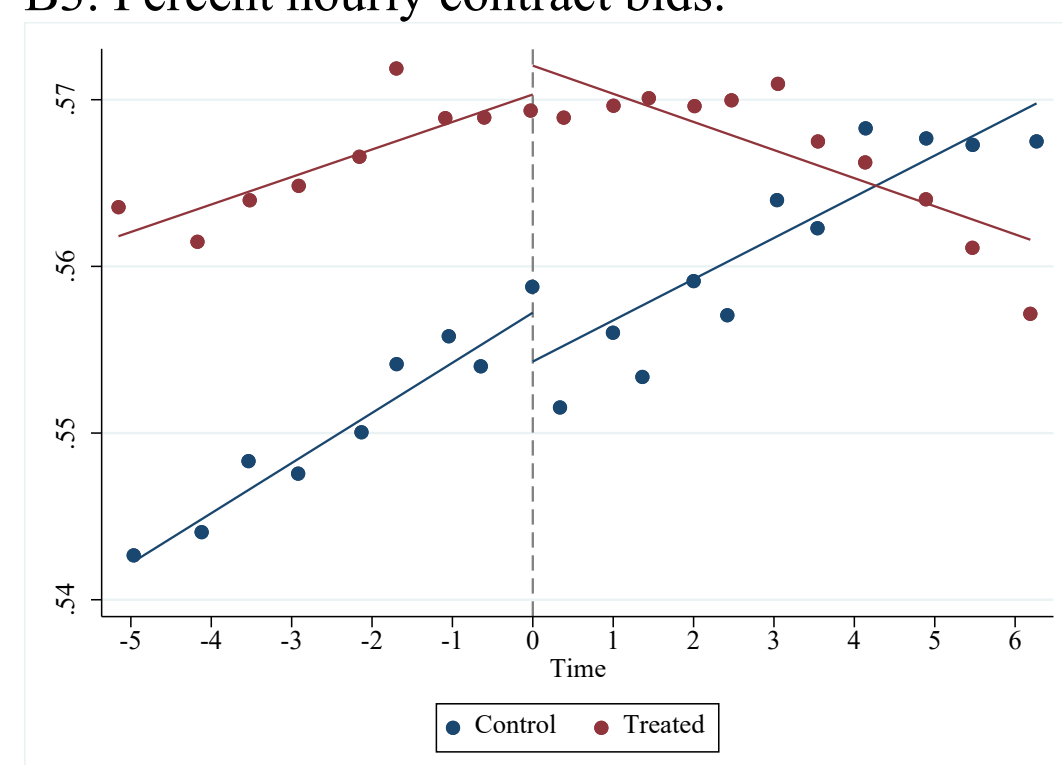

**Appendix Figure A2. Binned scatterplot of supply expansion exposure and demand contraction exposure.**

This figure presents a binned scatterplot displaying the relationship between demand contraction exposure and supply expansion exposure. Demand contraction exposure and supply expansion exposure are defined in the manuscript.

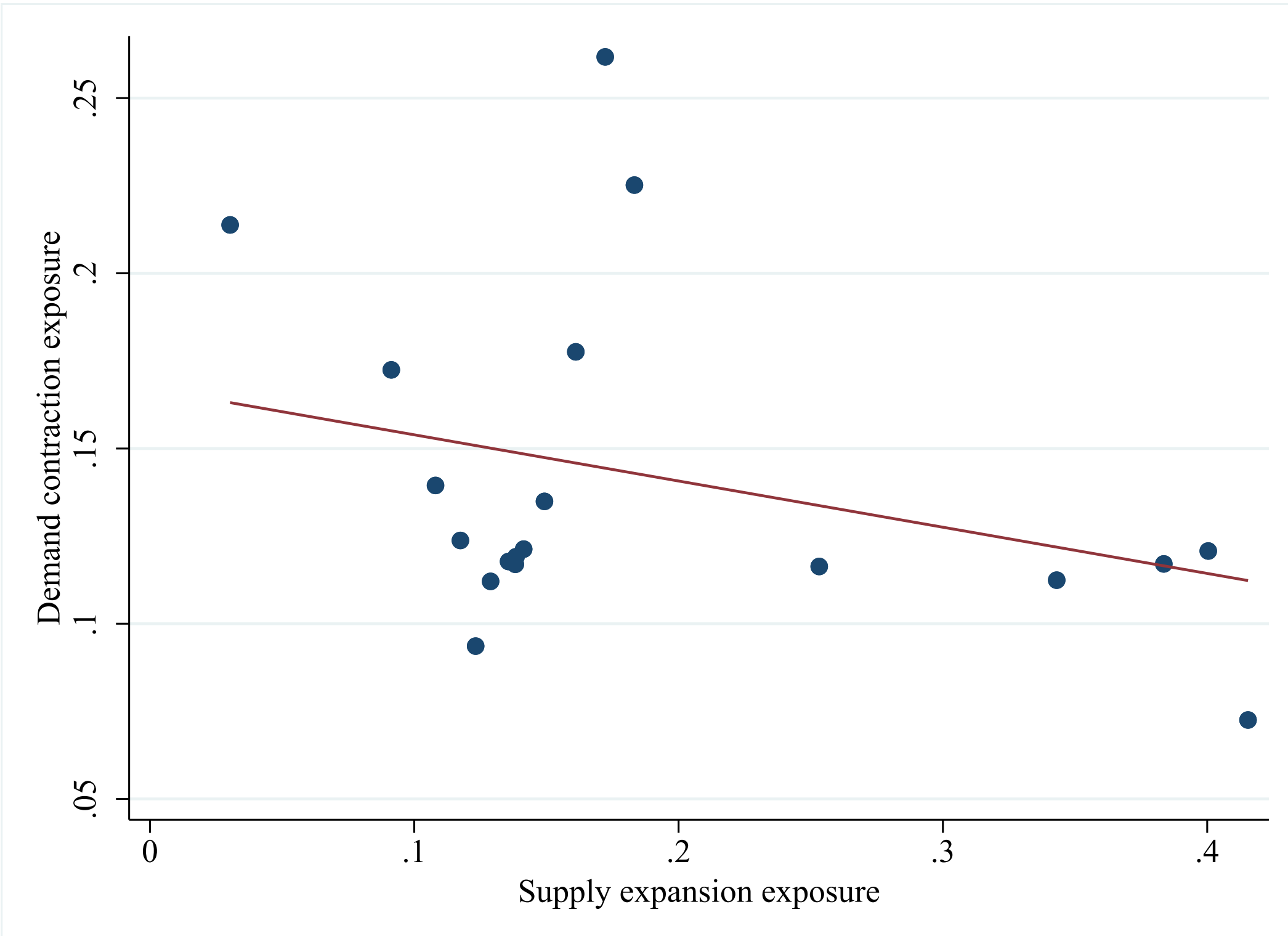

## Appendix Figure A3. Changes in freelancer strategic positioning pre- and post-ChatGPT with alternative dependent variables.

The following figures present the results of relative time models documenting changes in freelancer positioning from June 2022 through March 2024, following equation (1). The dependent variables used are described in the manuscript. The x-axis represents months relative to the launch of ChatGPT. For a description of the PD estimation approach, please see the draft. Panel A contains simple time trends with freelancer fixed effects for sample freelancers. Corresponding regression models are found in *Appendix Table A3*. Panel B documents binned scatterplots with freelancer fixed effects using the PD approach discussed in the draft. In both panels, we control for *log bids* when *cosine similarity to prior behavior* or *percent high value bids* are dependent variables.

**Panel A. Sample time trends.**

A1. Bid for job.

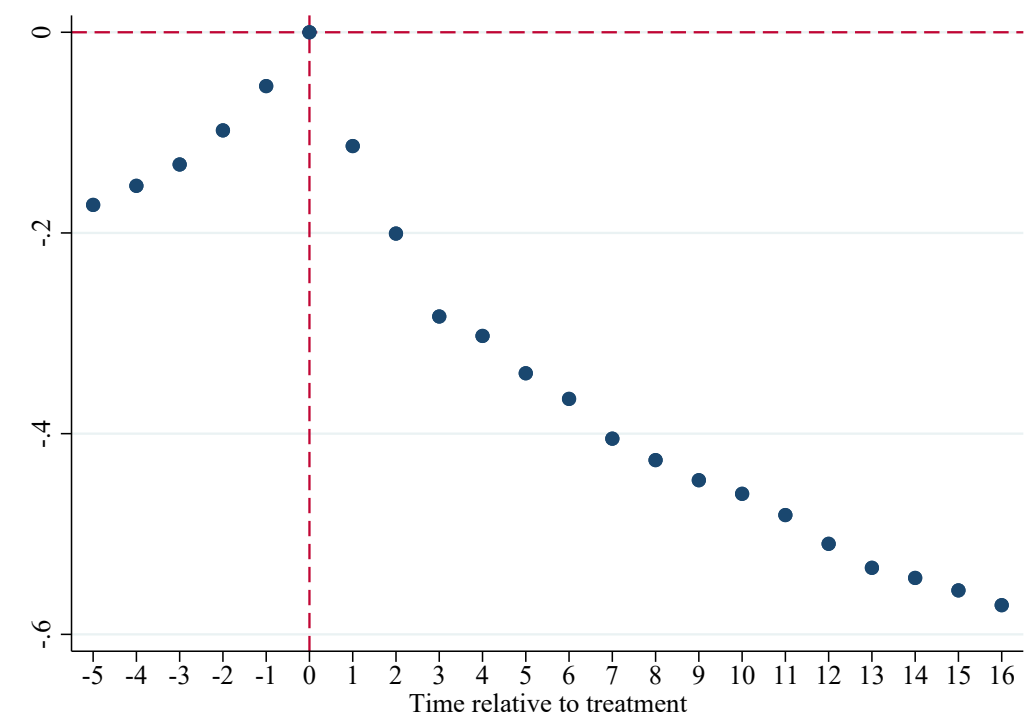

A2. Bid concentration.

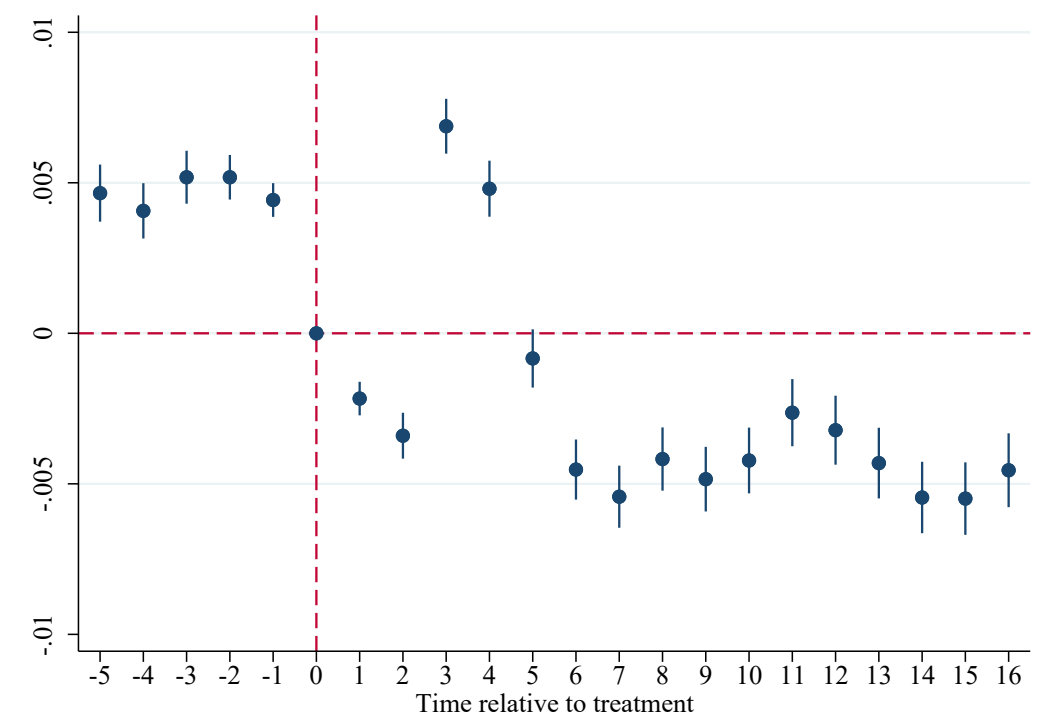

A3. New modal work category.

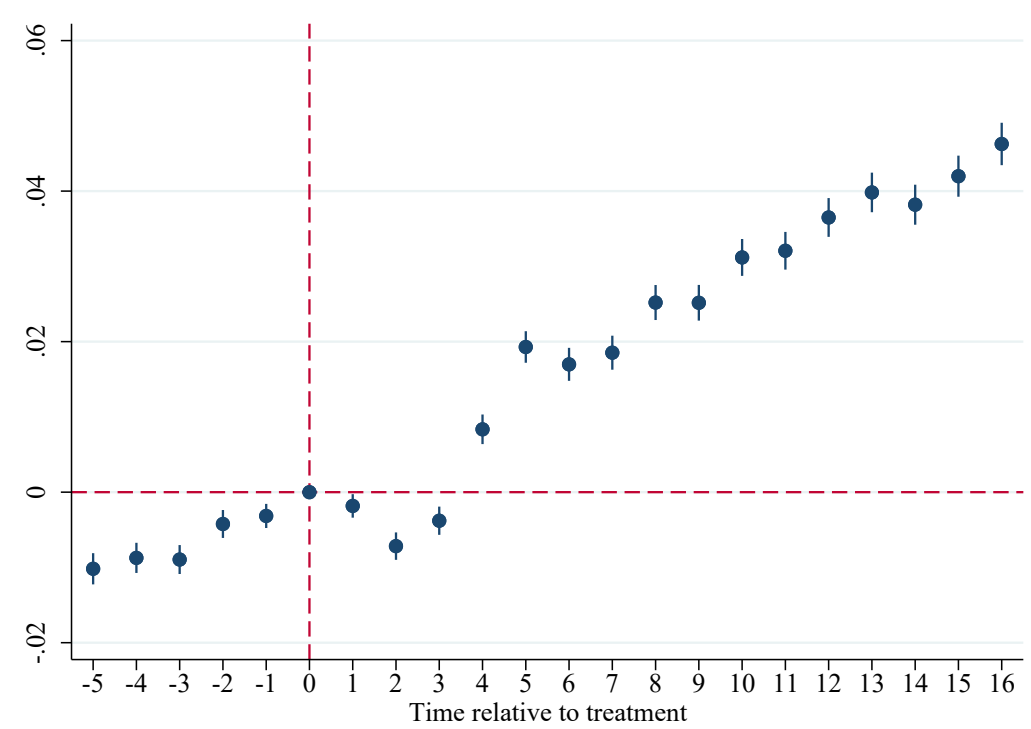

A4. AI keyword in profile.

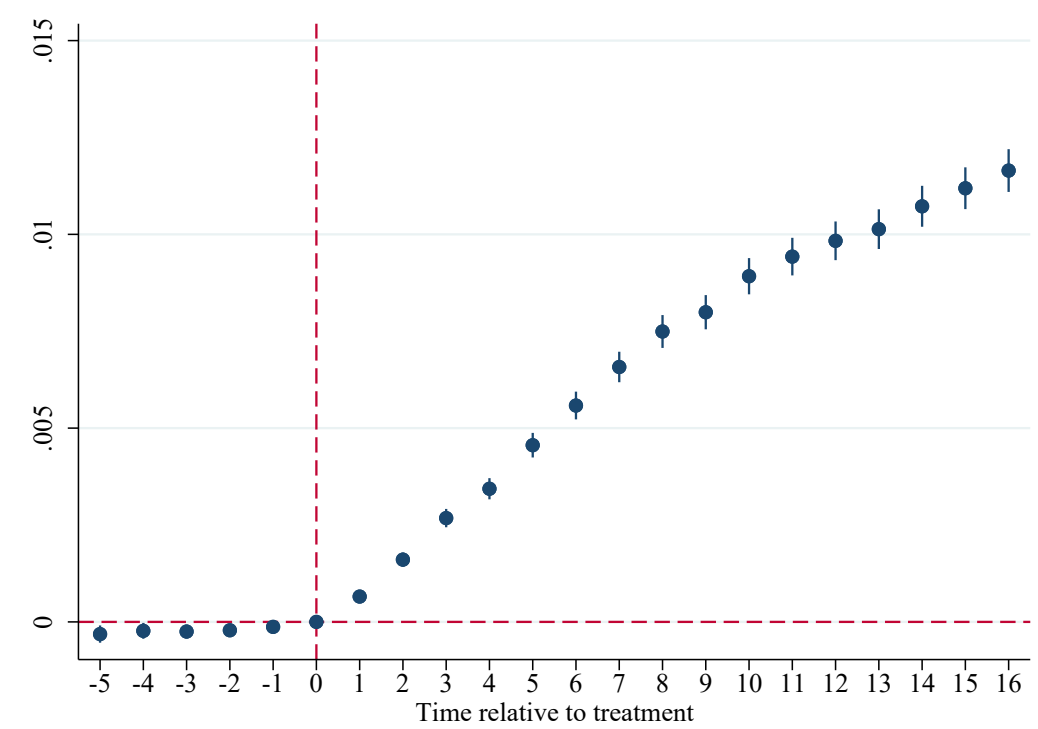

**Panel B. Panel differences binned scatterplots.**

B1. Bid for job.

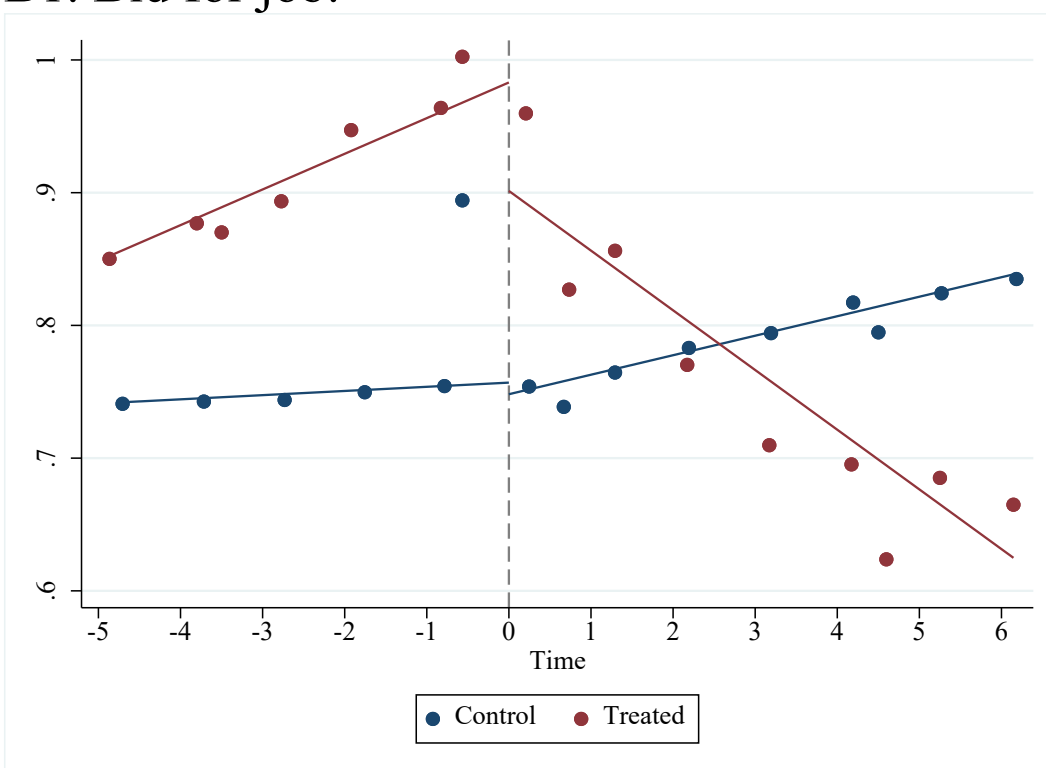

B2. Bid concentration.

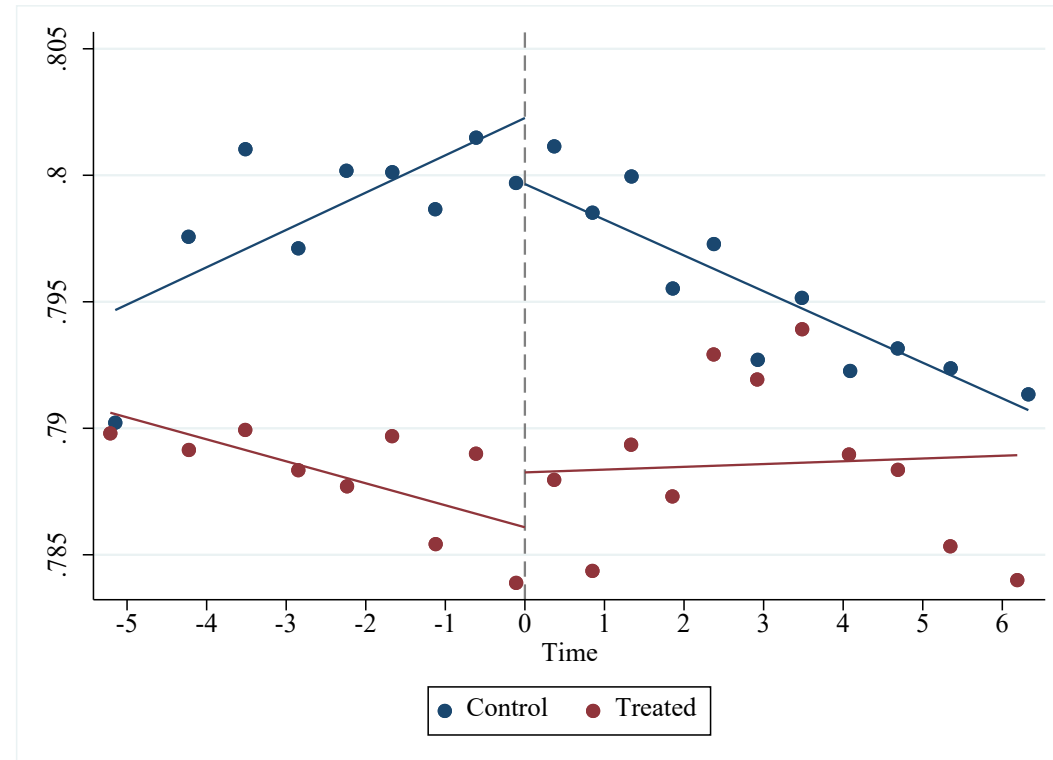

B3. New modal work category.

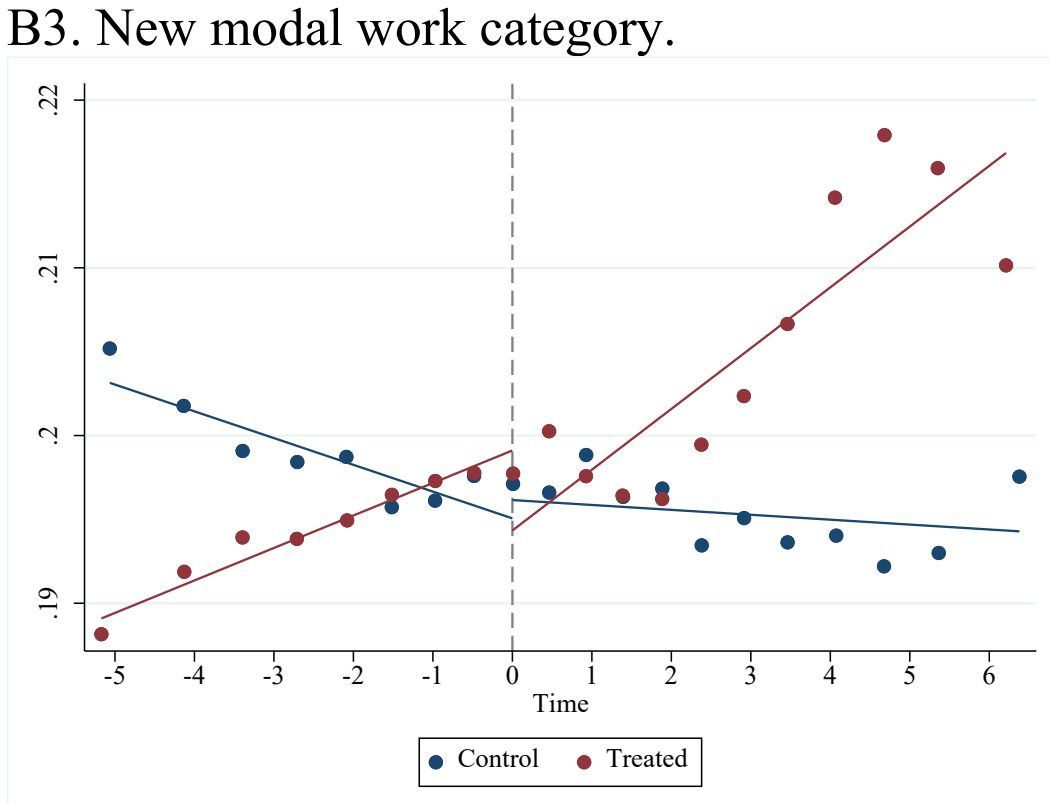

B4. AI keyword in profile.

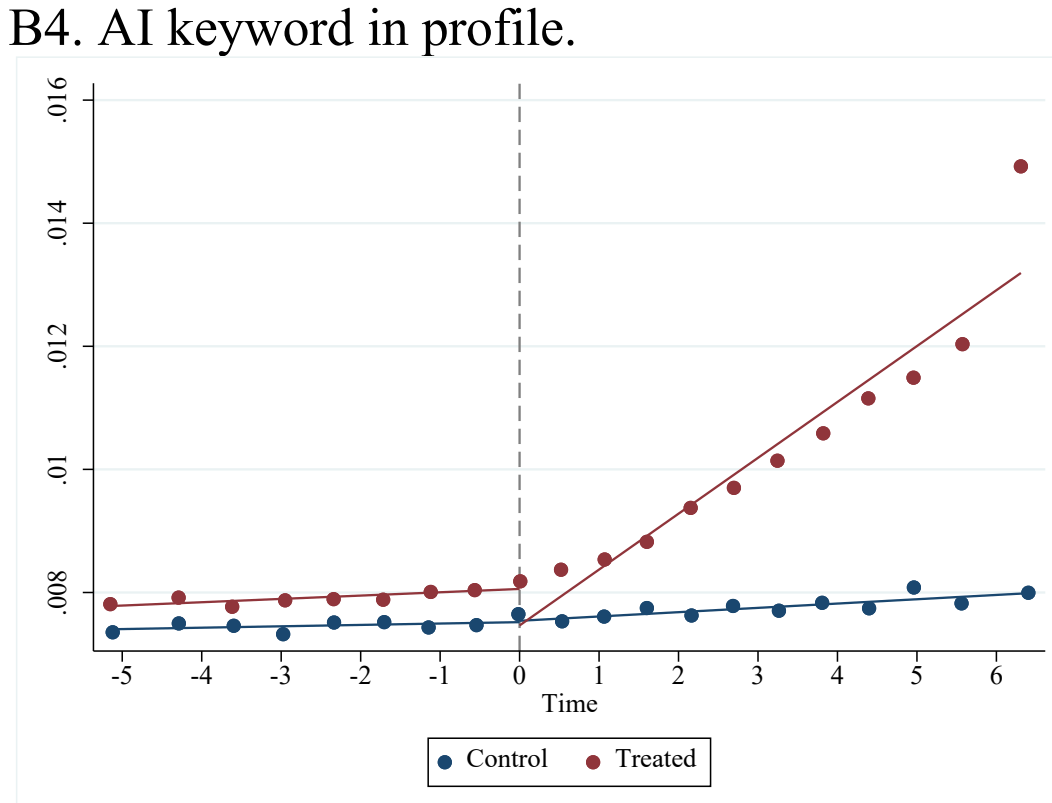

## Appendix Figure A4. Changes in freelancer strategic positioning pre- and post-ChatGPT by demand contraction and supply expansion exposure with alternative dependent variables.

The following figures present the results of relative time models documenting changes in freelancer positioning from June 2022 through March 2024 by demand contraction exposure and supply expansion exposure. The dependent variables used are described in the manuscript. The CEM procedure is described in the manuscript. The x-axis represents months relative to the launch of ChatGPT.

**Panel A. Full sample estimates.**

**Panel A1. Bid for job.**

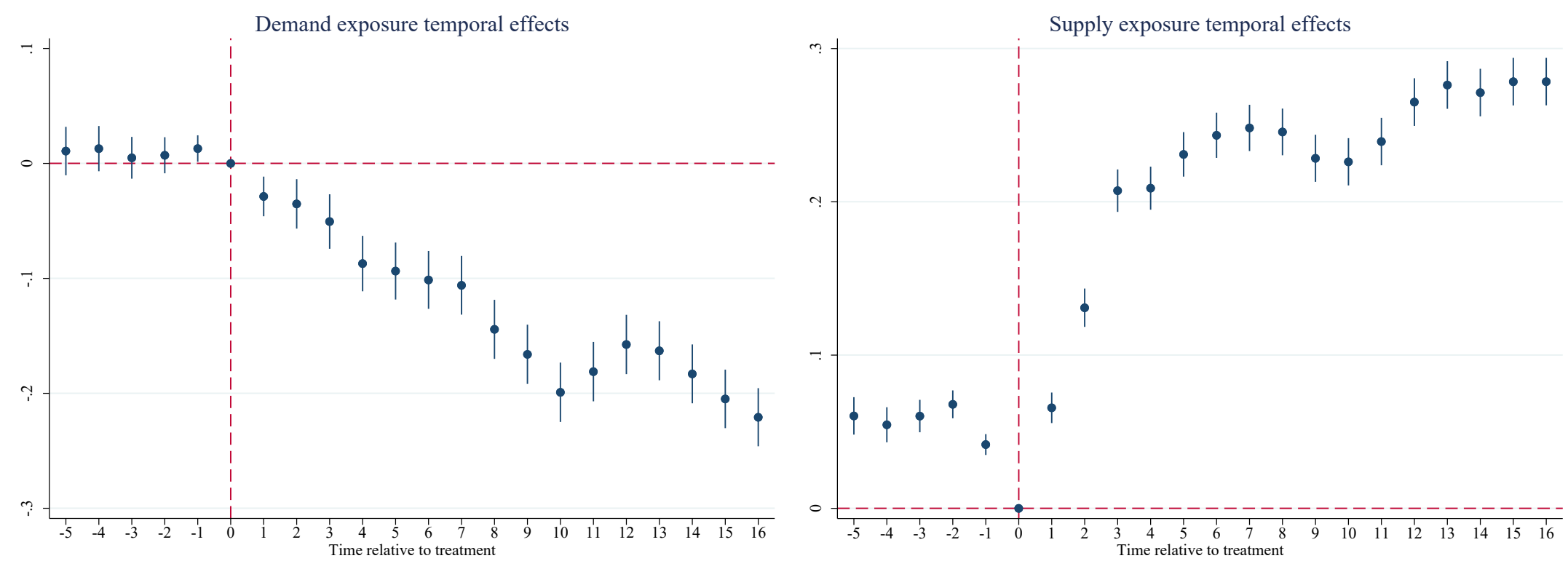

**Panel A2. Bid concentration.**

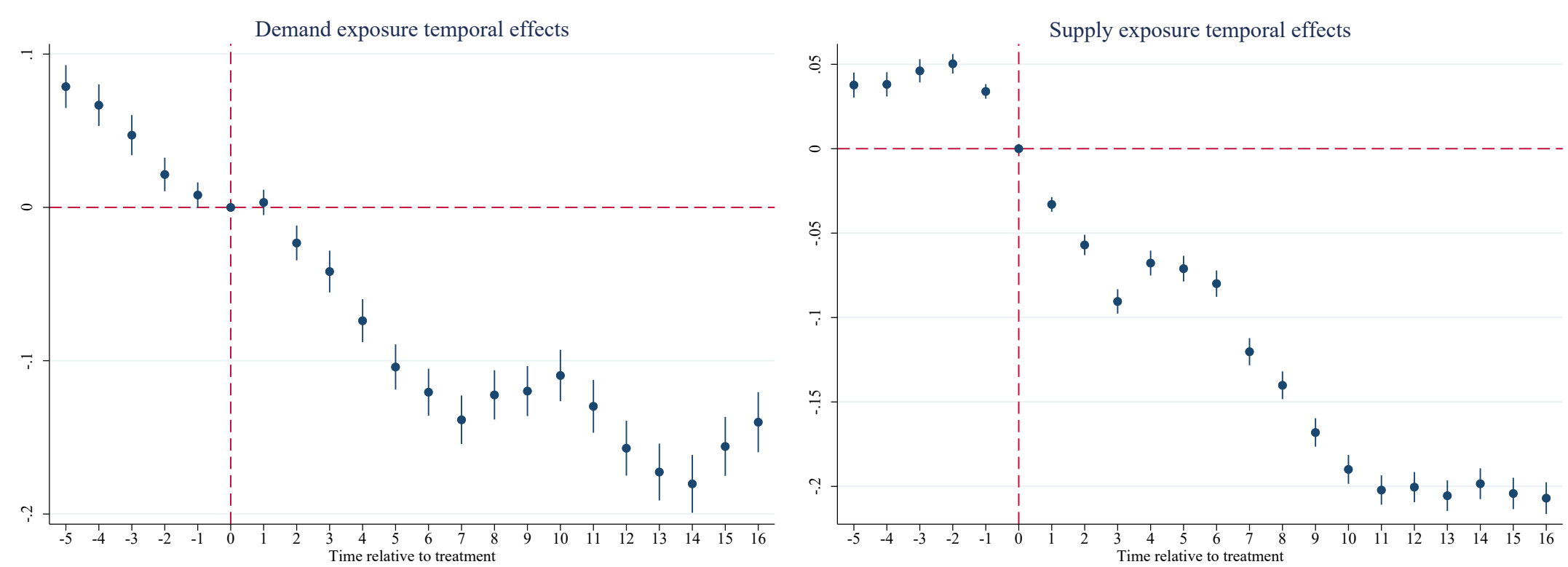

**Panel A3. New modal work category.**

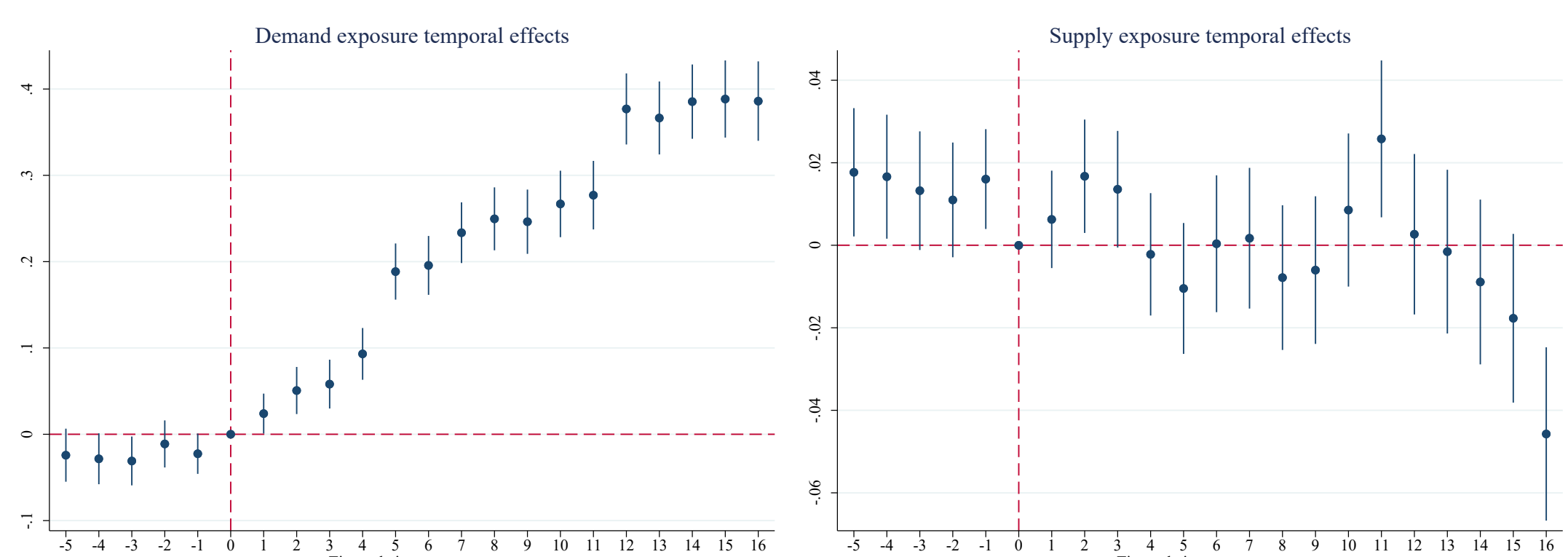

**Panel A3. AI keyword in profile.**

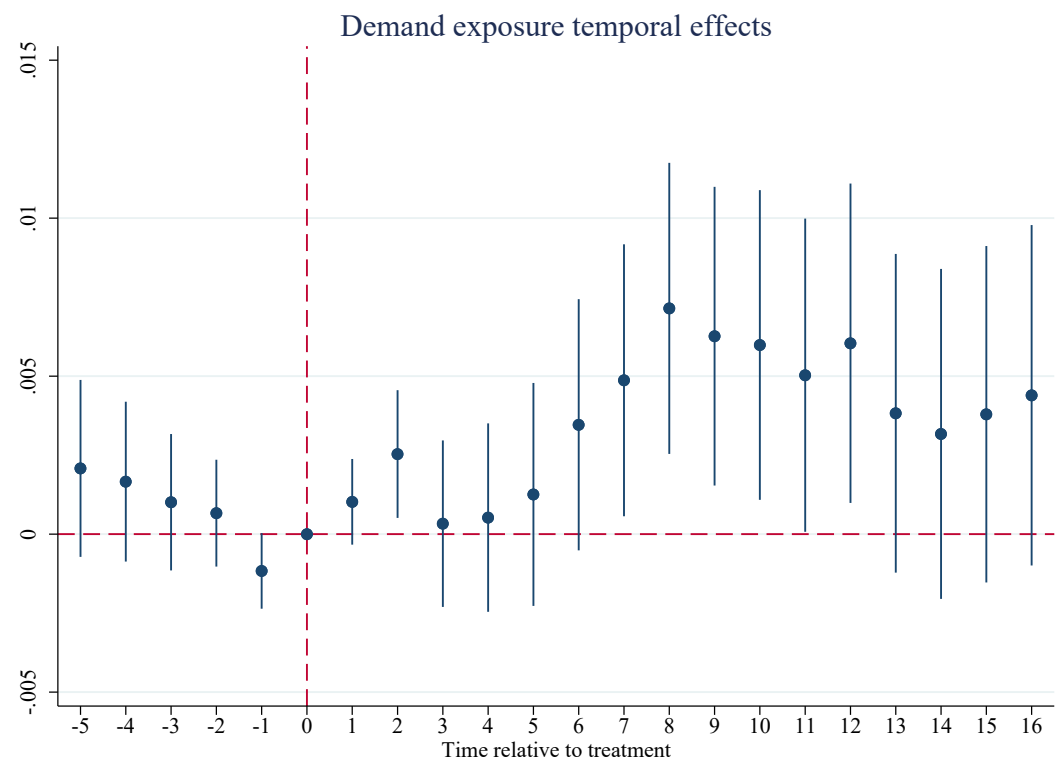

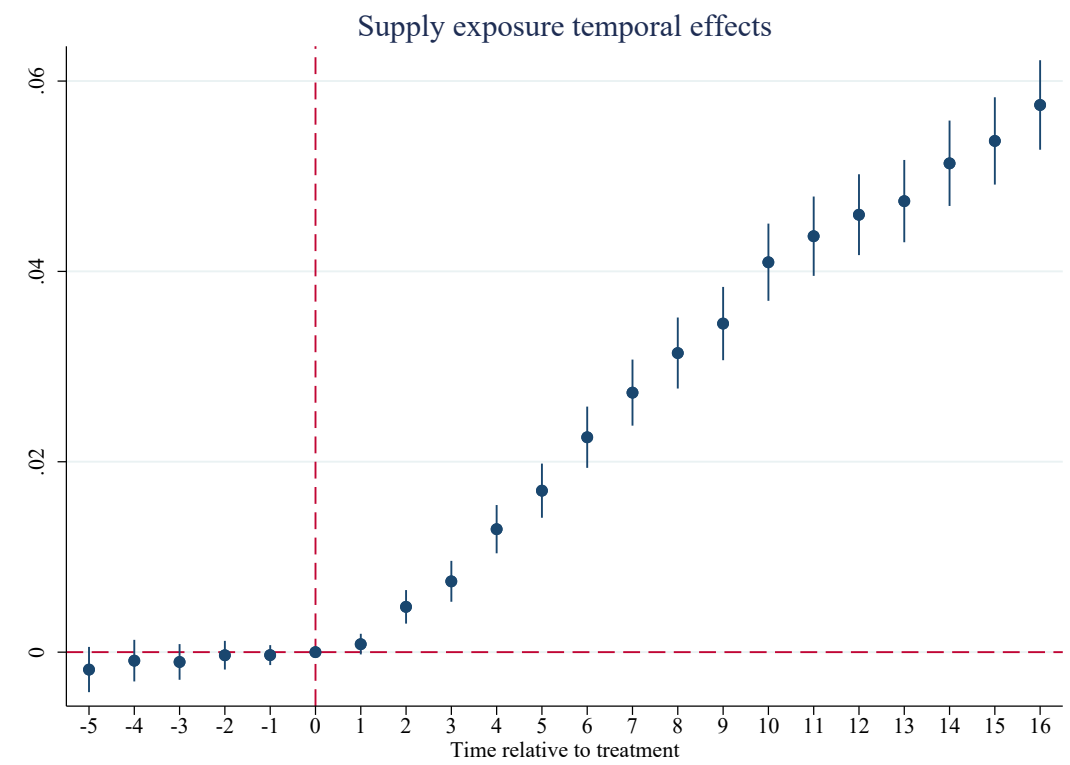

**Panel B. CEM estimates.**
**Panel B1. Bid for job.**

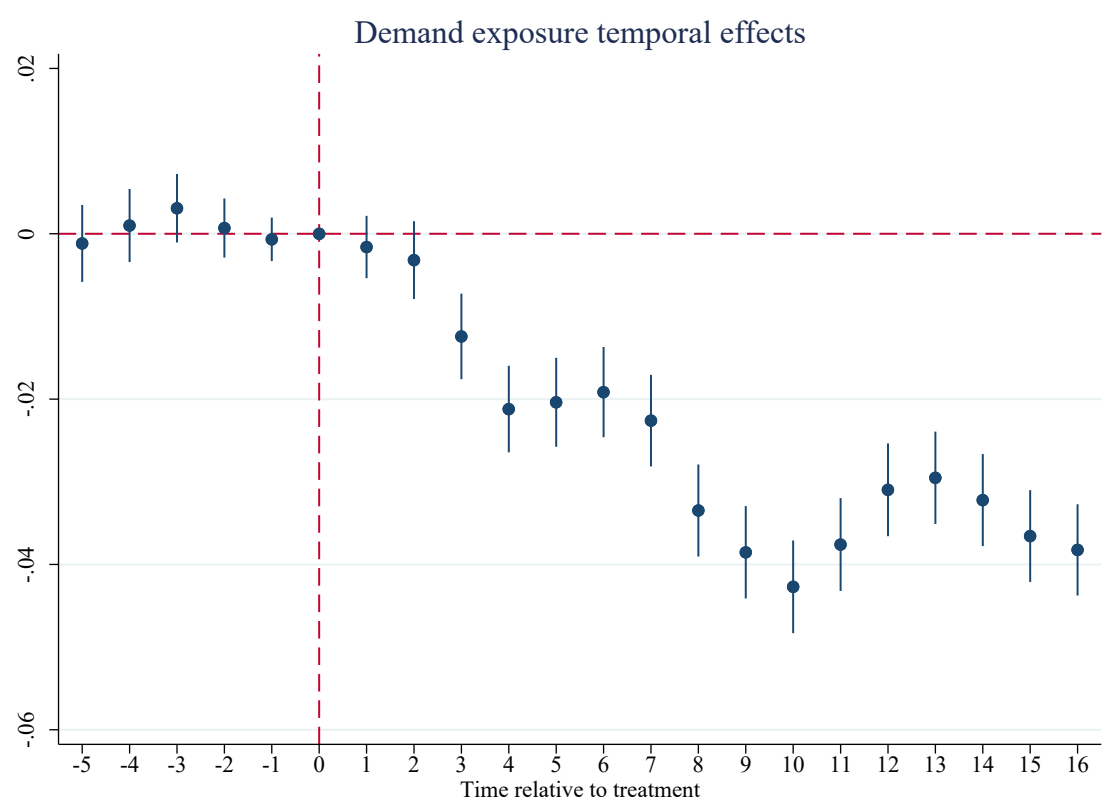

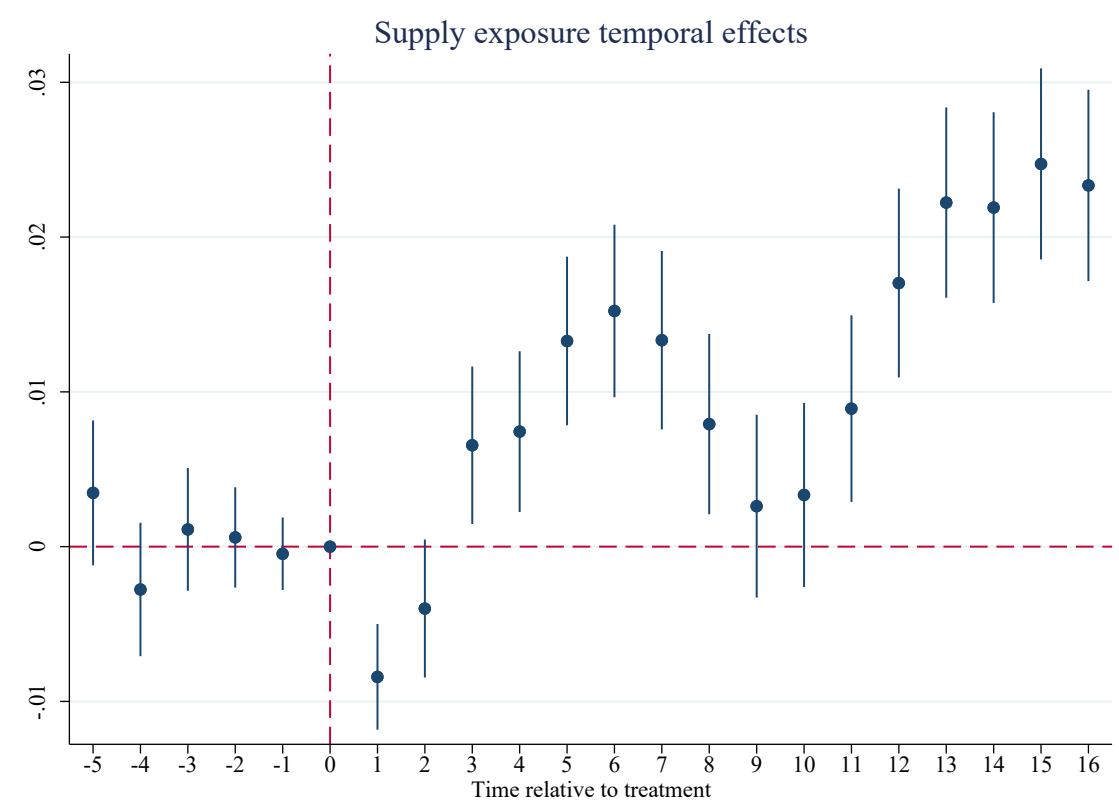

**Panel B2. Bid concentration.**

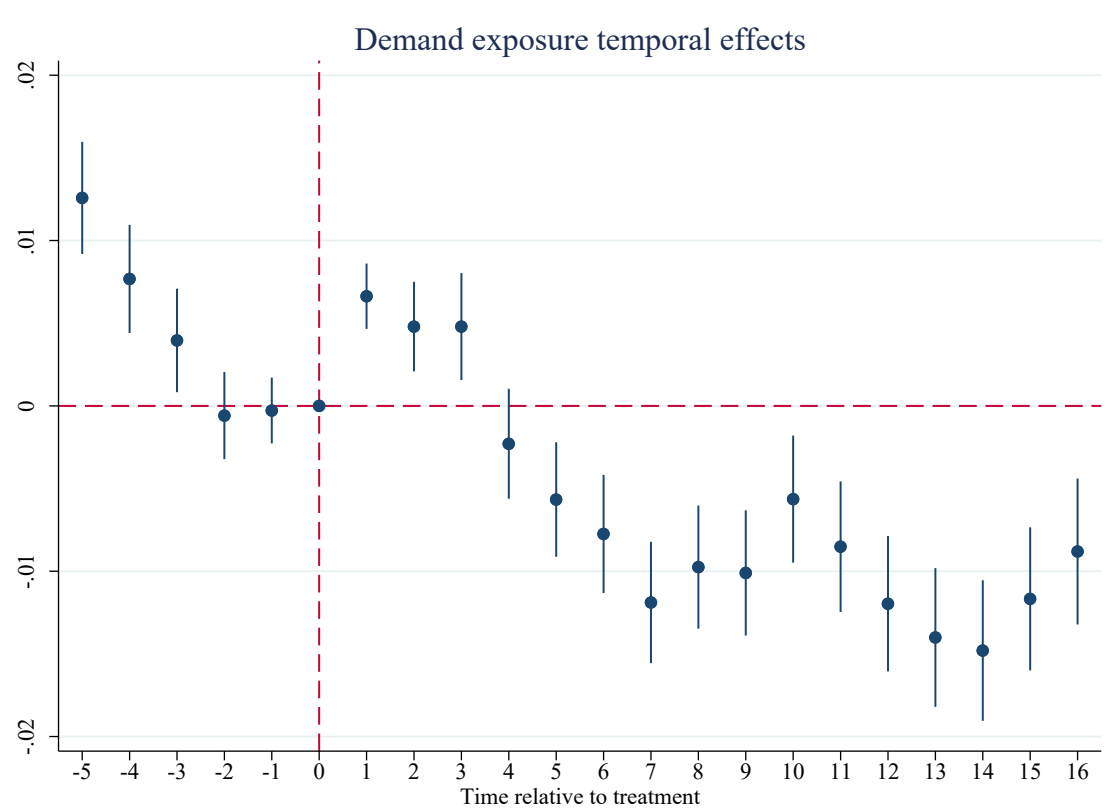

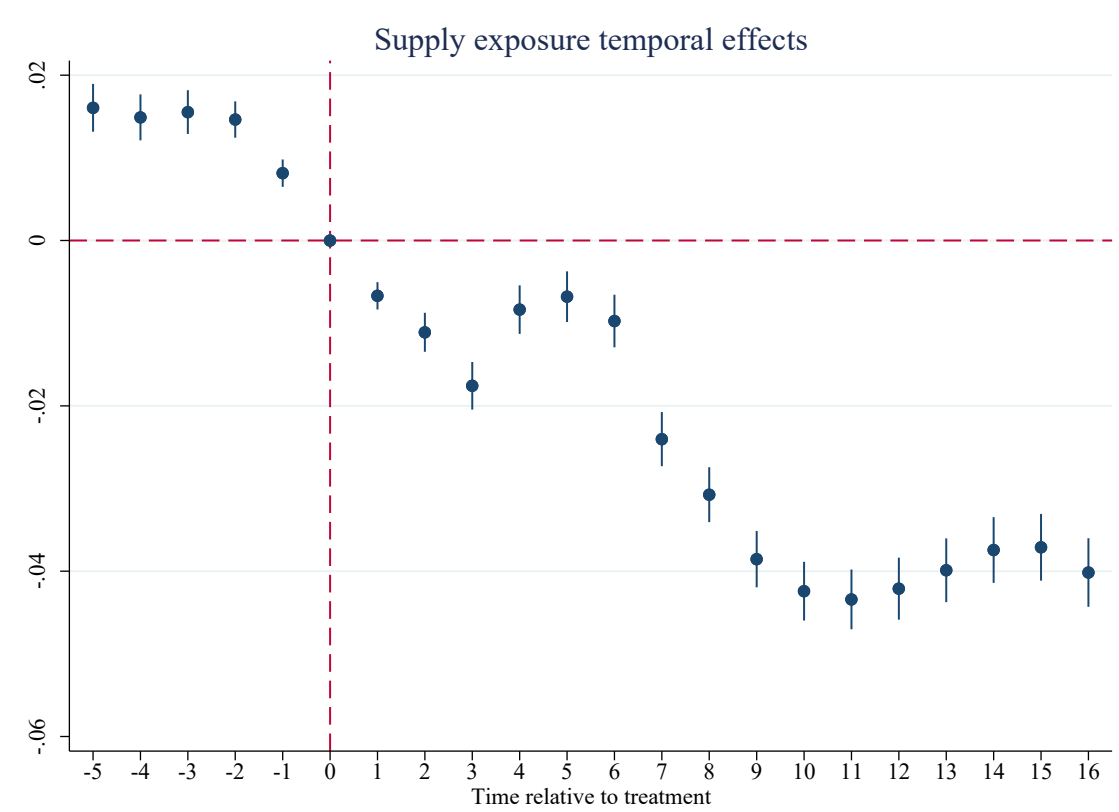

**Panel B3. New modal work category.**

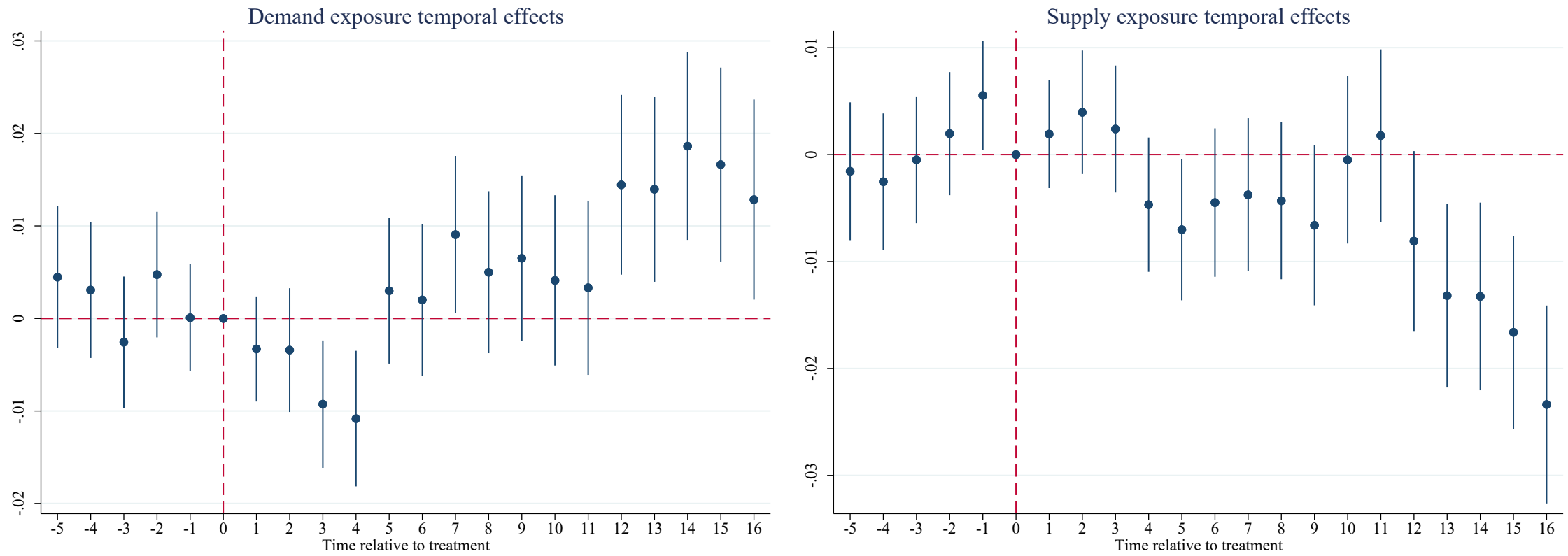

**Panel B4. AI keyword in profile.**

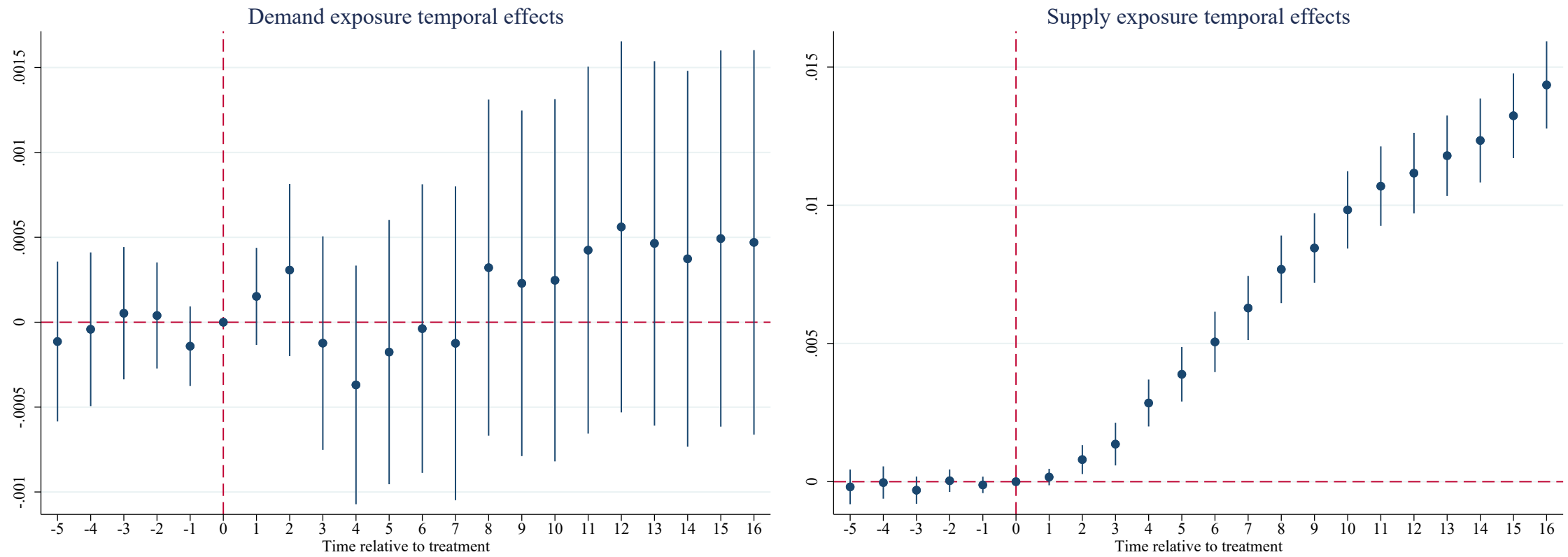

## Appendix Figure A5. Changes in freelancer strategic positioning pre- and post-ChatGPT by pre-period skill level with alternative dependent variables.

The following figures present the results of relative time models documenting changes in freelancer positioning from June 2022 through March 2024 by pre-period skill-level. The dependent variables used are described in the manuscript. The CEM procedure is described in the manuscript. The x-axis represents months relative to the launch of ChatGPT.

**Panel A. Full sample estimates.**

A1. Bid for job.

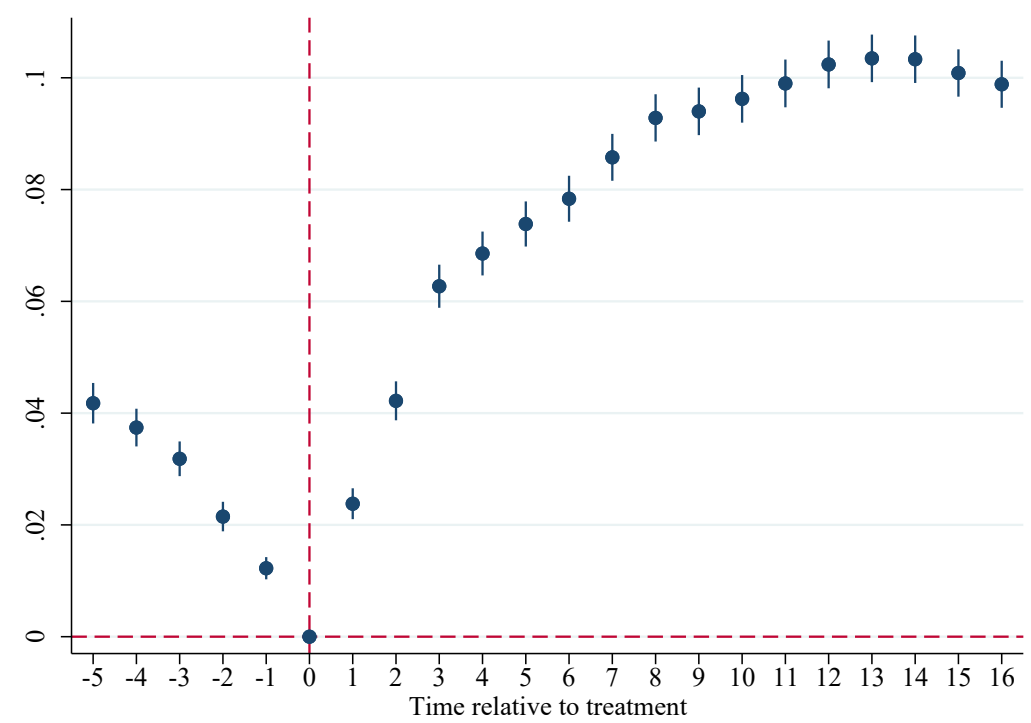

A2. Bid concentration.

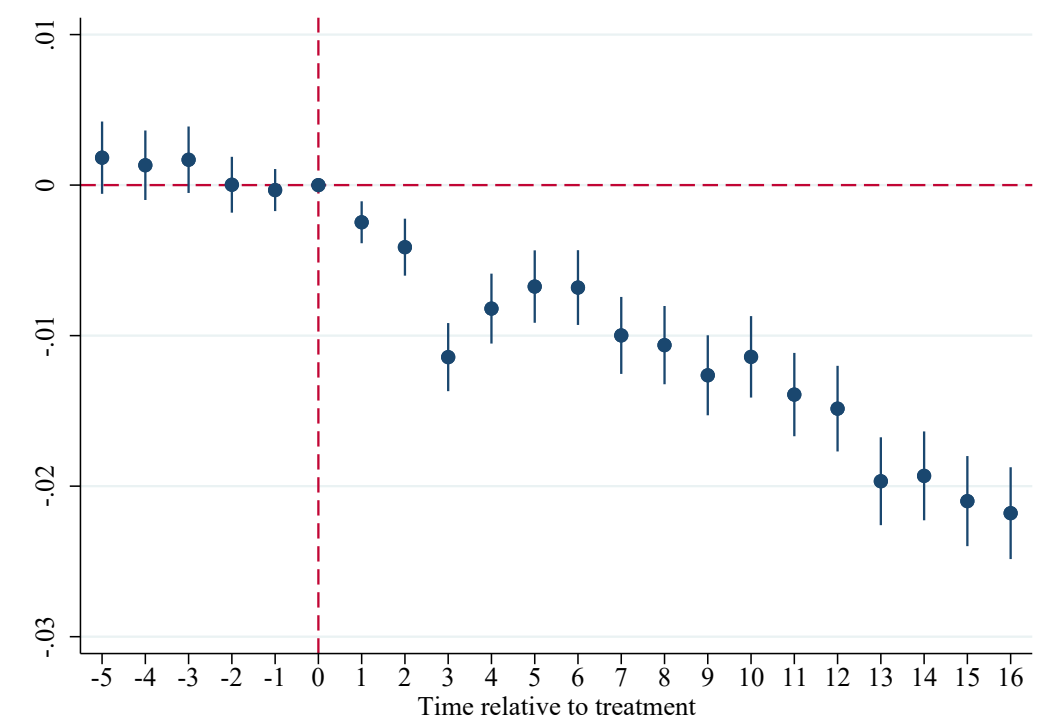

A3. New modal work category.

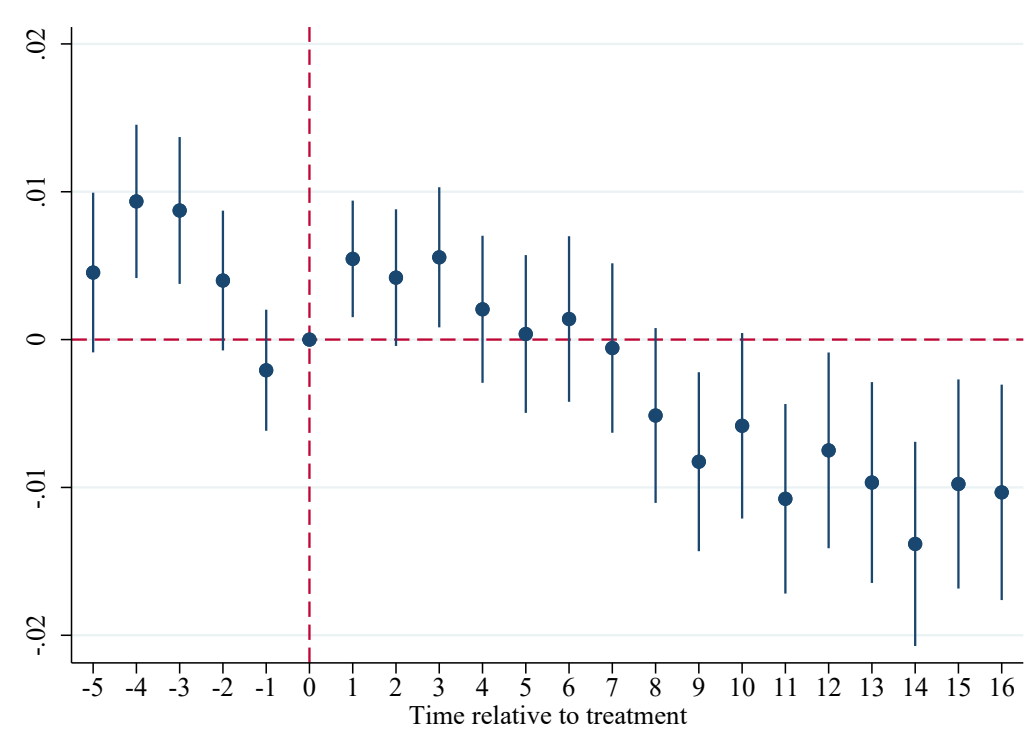

A4. AI keyword in profile.

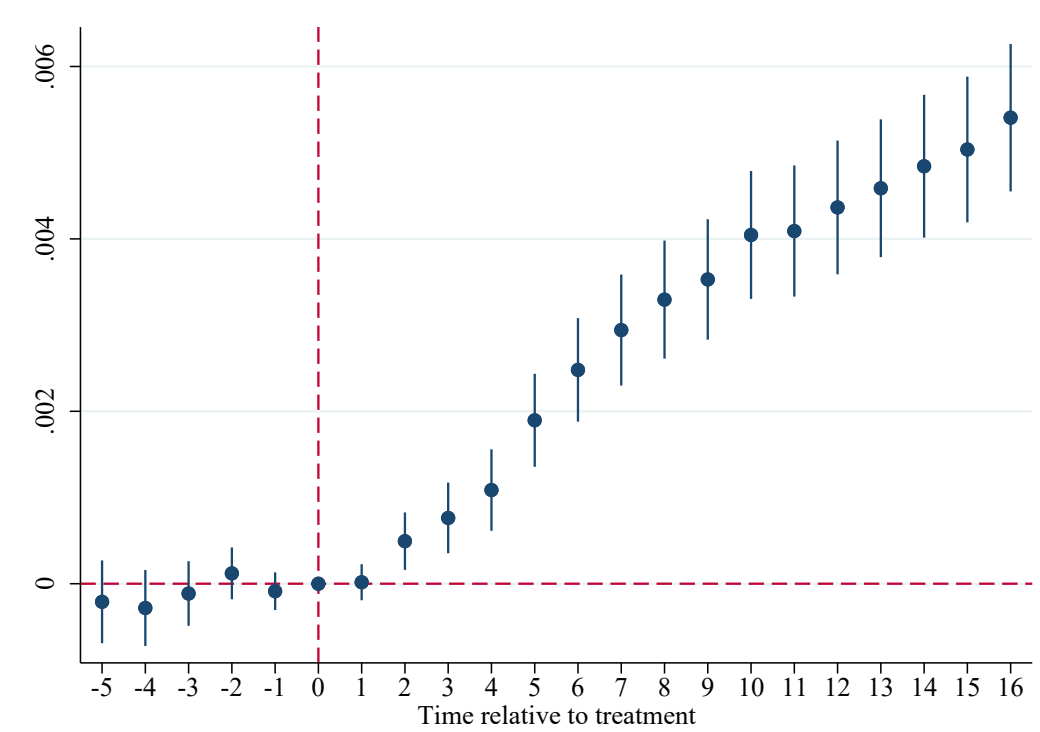

**Panel B. CEM estimates.**

B1. Bid for job.

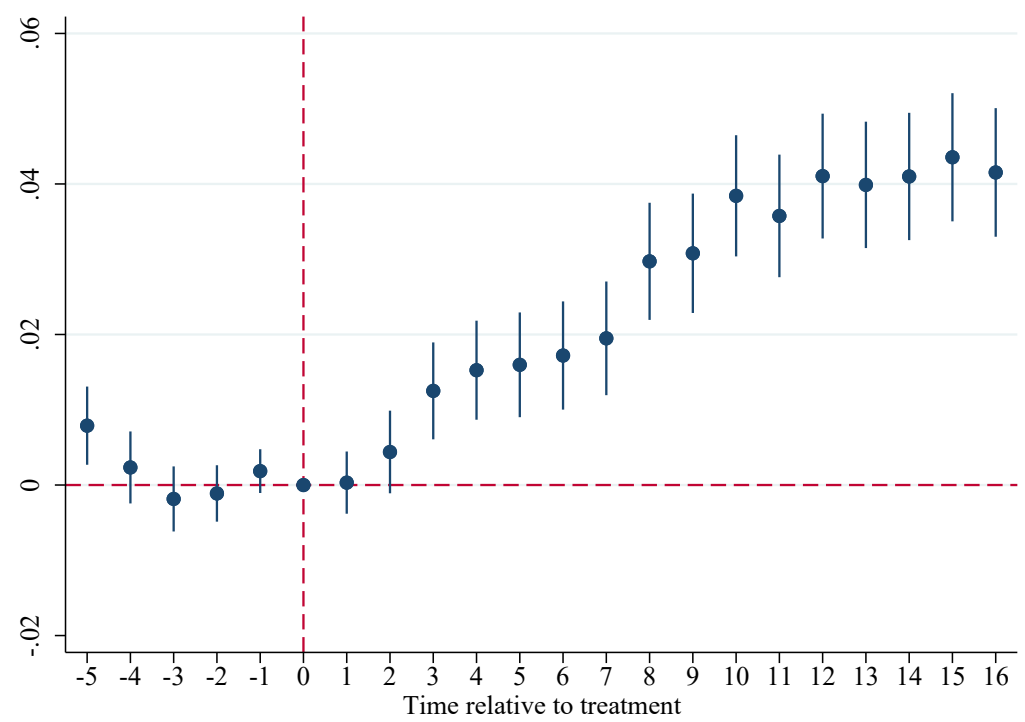

B2. Bid concentration.

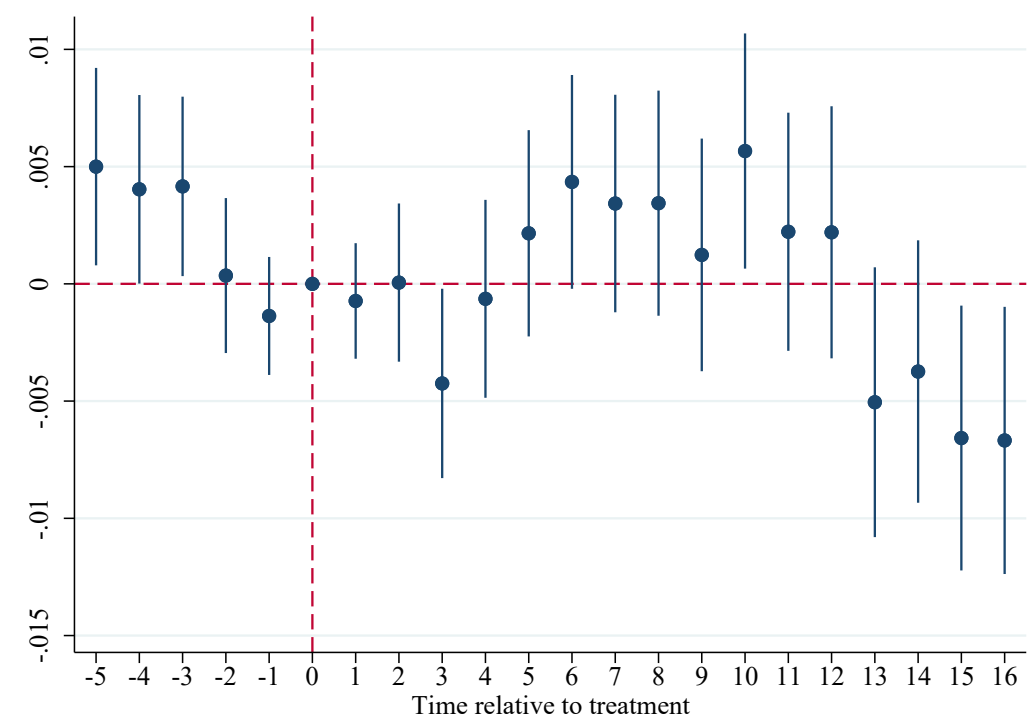

B3. New modal work category.

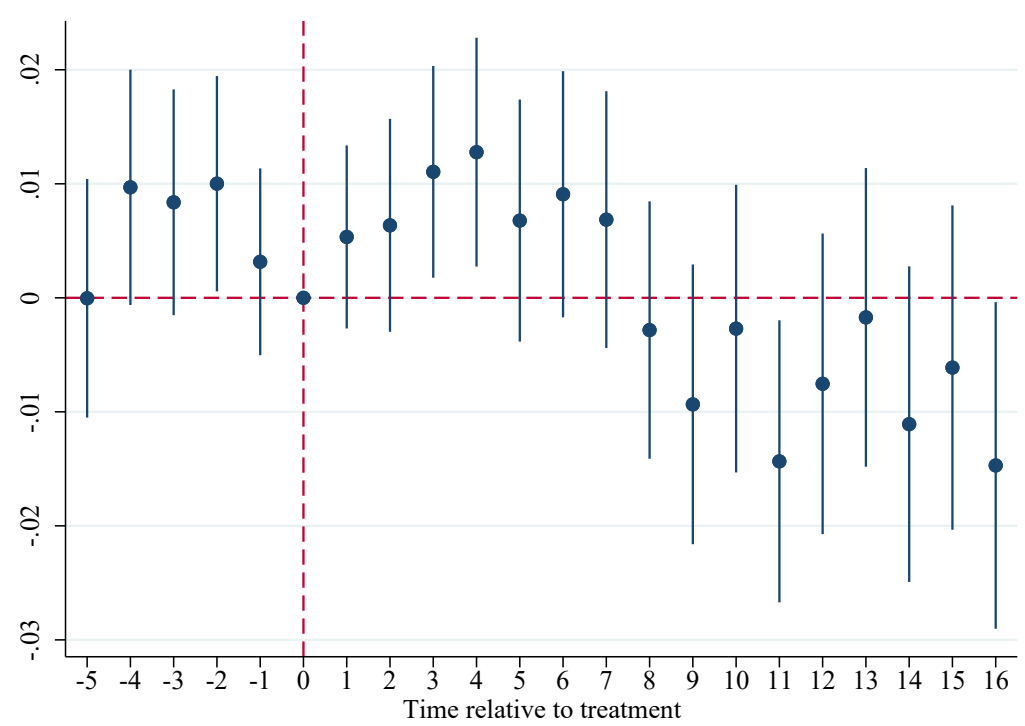

B4. AI keyword in profile.

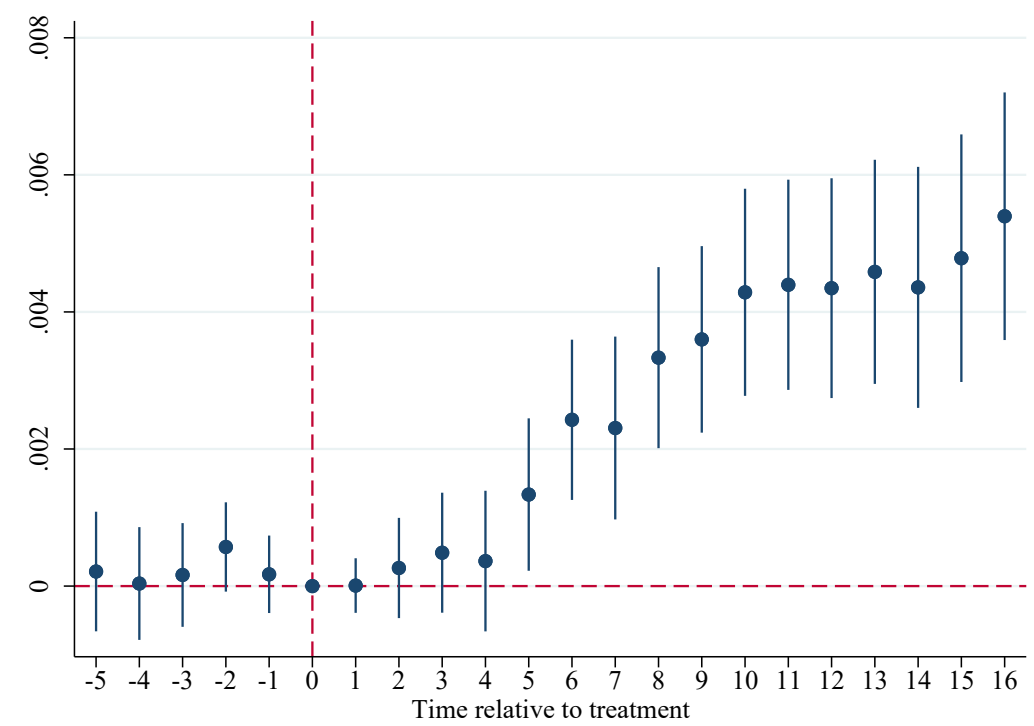

**Appendix Figure A6. Changes in freelancer strategic positioning pre- and post-ChatGPT by pre-period experience.**

The following figures present the results of relative time models documenting changes in freelancer positioning from June 2022 through March 2024 by pre-period experience, as measured by the number of jobs completed. The x-axis represents months relative to the launch of ChatGPT.

**Panel A. Log bids.**

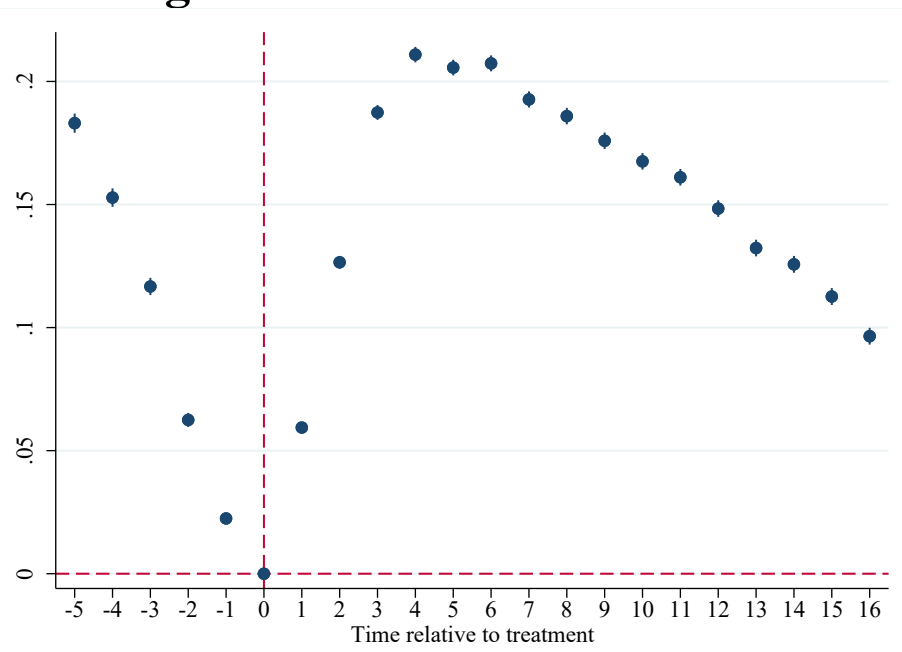

**Panel B. Cosine similarity to prior behavior.**

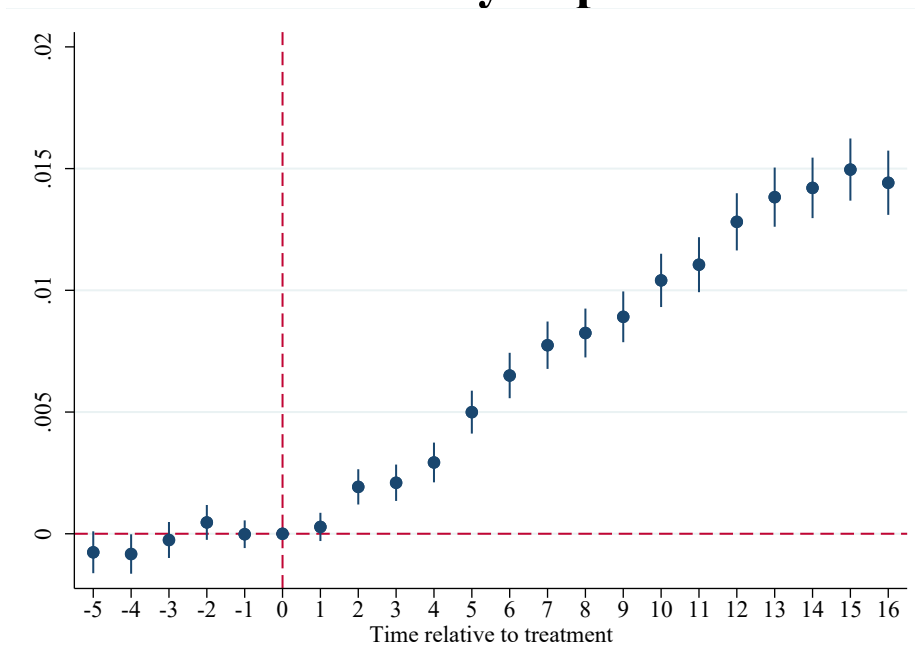

**Panel C. Percent high value bids.**

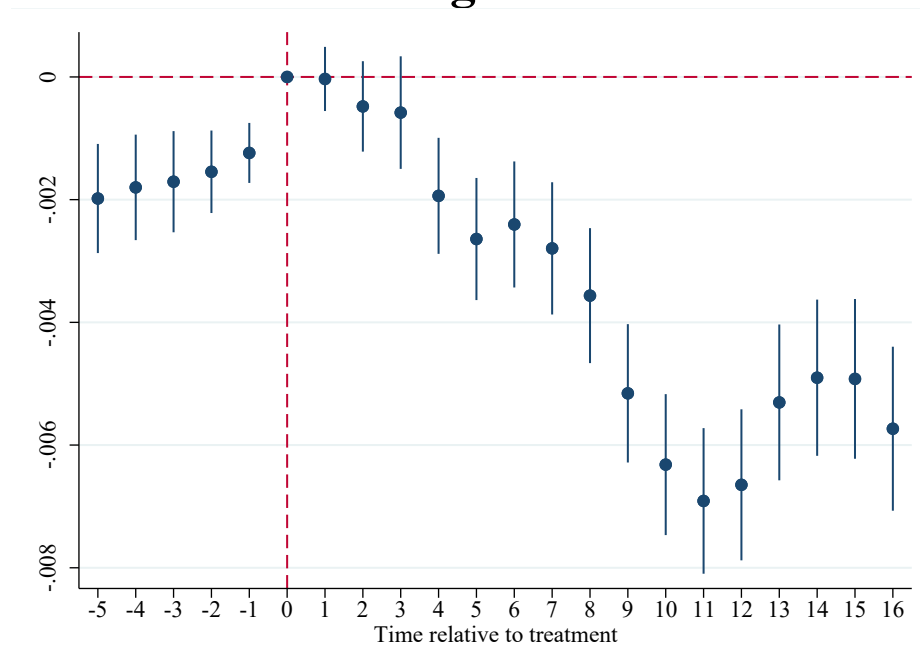

## Appendix Figure A7. Changes in freelancer strategic positioning pre- and post-ChatGPT with alternative sample selection.

This figure is analogous to Figure 3 in the main draft, except that it use an alternative sample of freelancer-year-month observations with at least 3 bids made in a rolling three-month window.

**Panel A. Sample time trends.**

A1. Log bids.

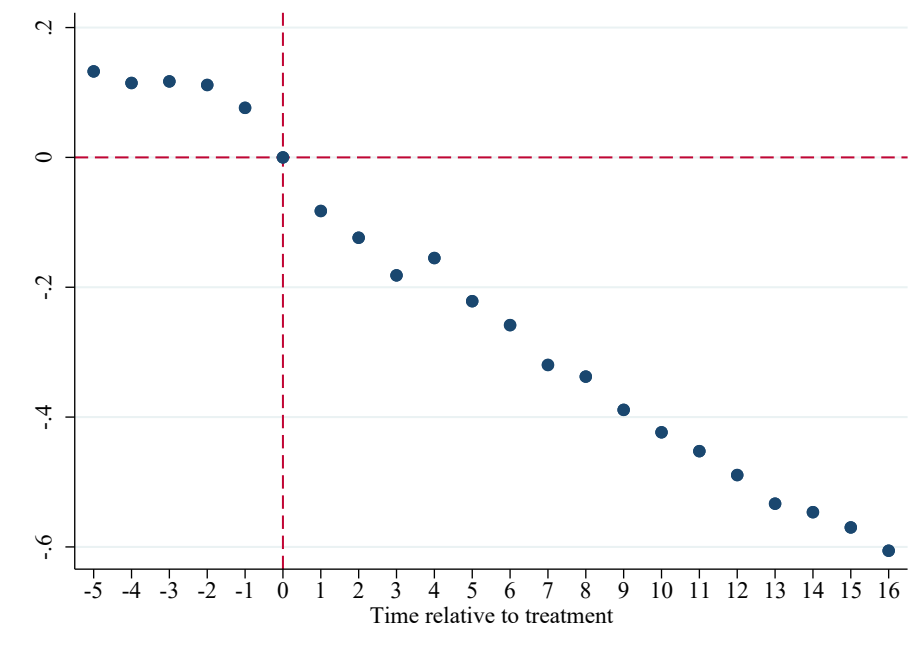

A2. Cosine similarity to prior behavior.

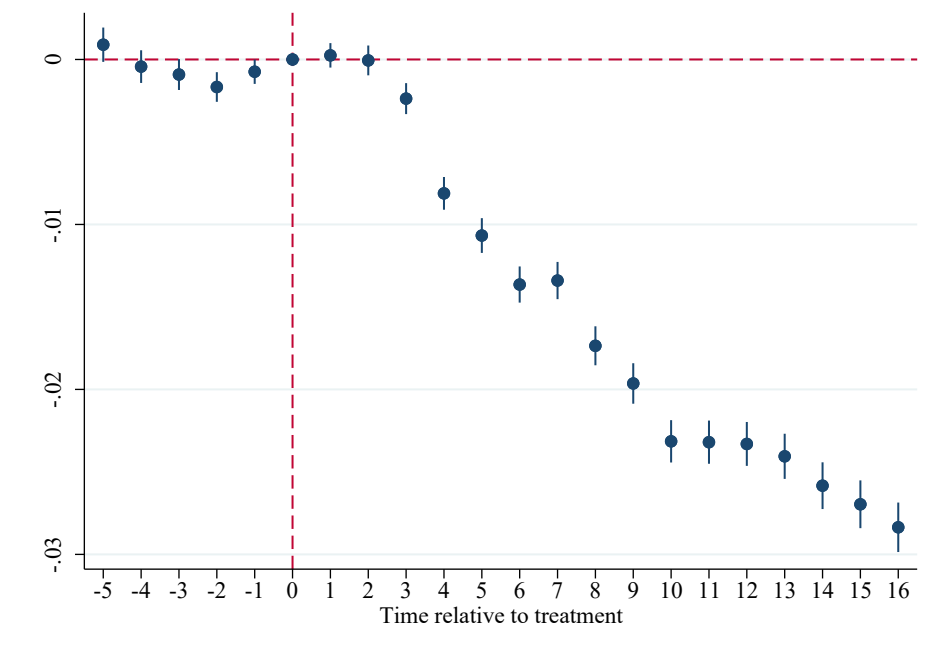

A3. Percent high value bids.

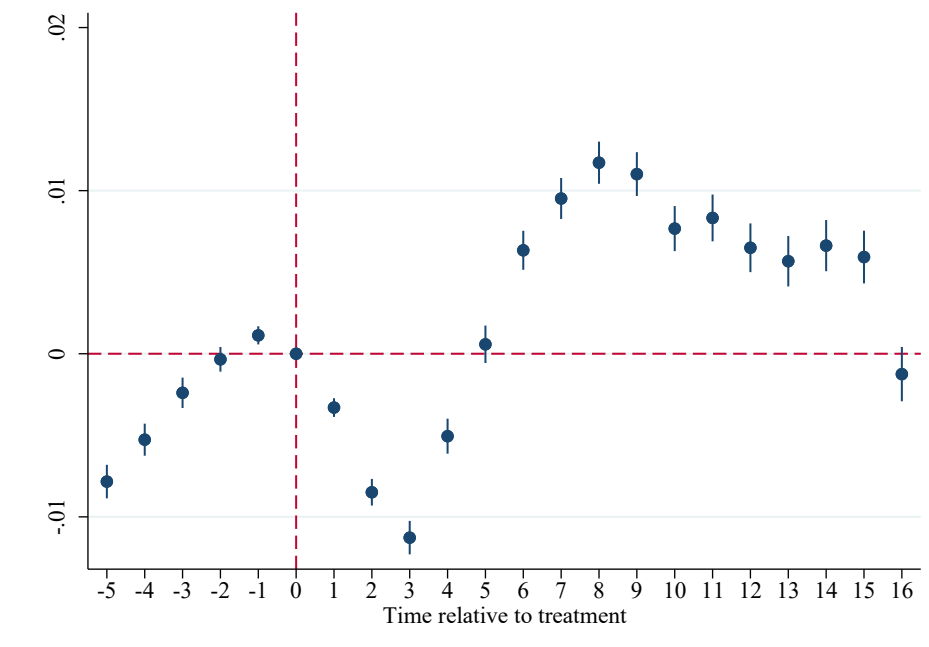

**Panel B. Panel differences binned scatterplots.**

B1. Log bids.

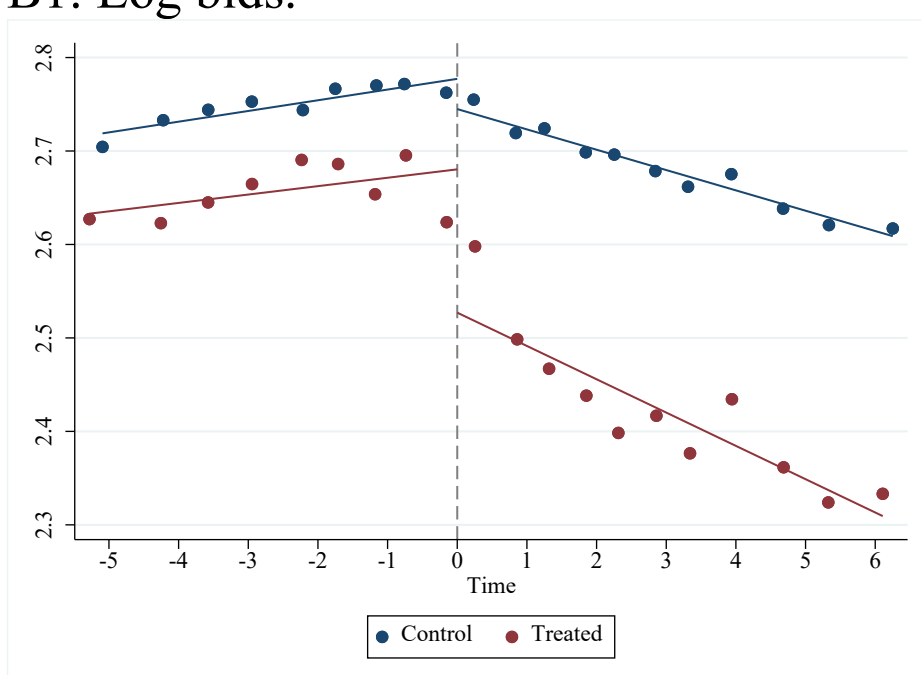

B2. Similarity to prior behavior.

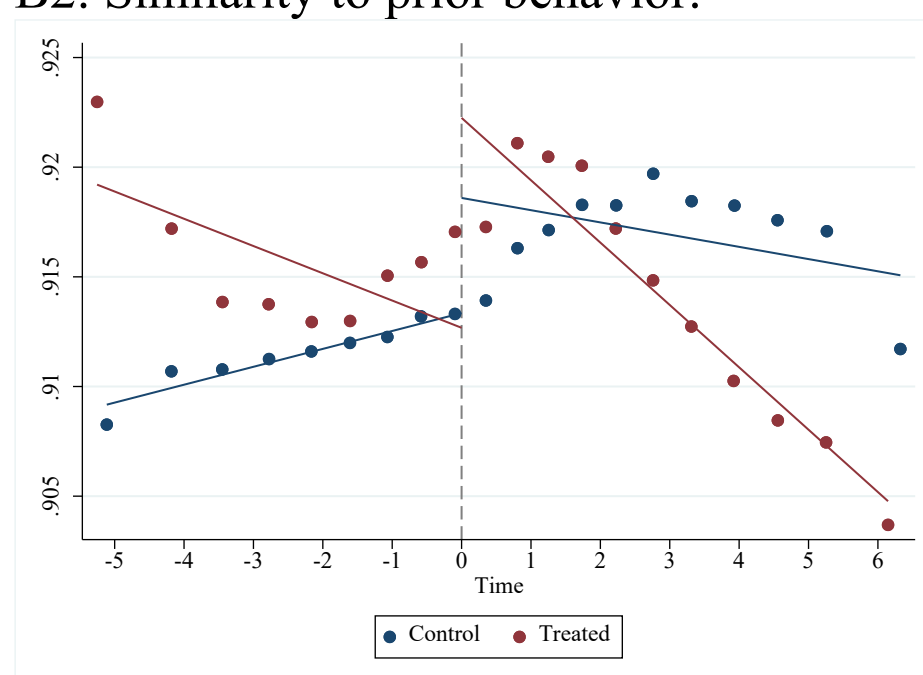

B3. Percent high value bids.

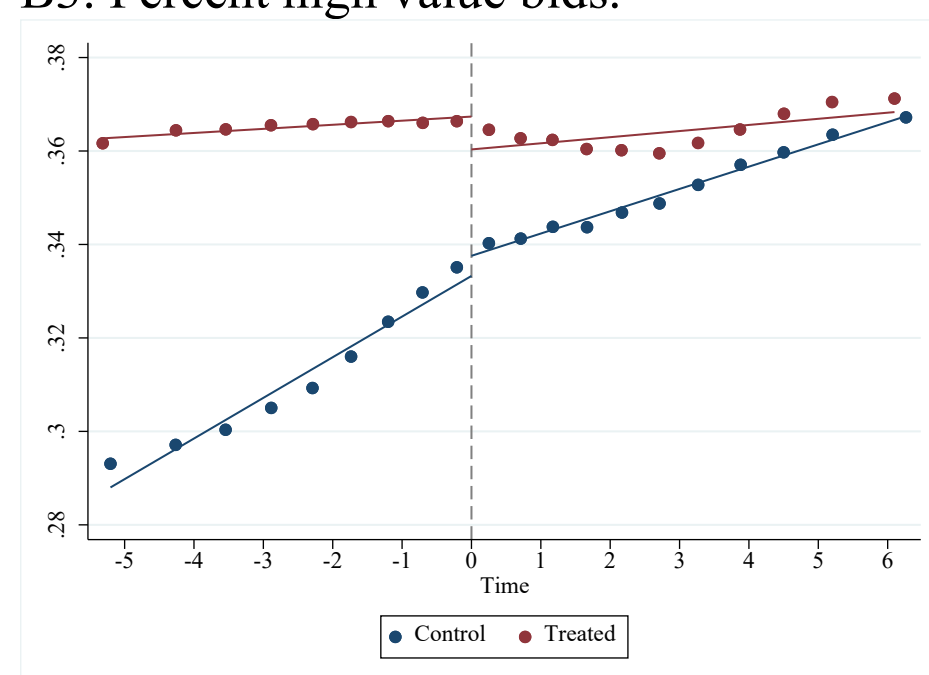

## Appendix Figure A8. Changes in freelancer strategic positioning pre- and post-ChatGPT by demand contraction and supply expansion exposure with alternative sample selection.

This figure is analogous to Figure 5 in the main draft, except that it use an alternative sample of freelancer-year-month observations with at least 3 bids made in a rolling three-month window.

**Panel A. Full sample estimates.**

**Panel A1. Log bids.**

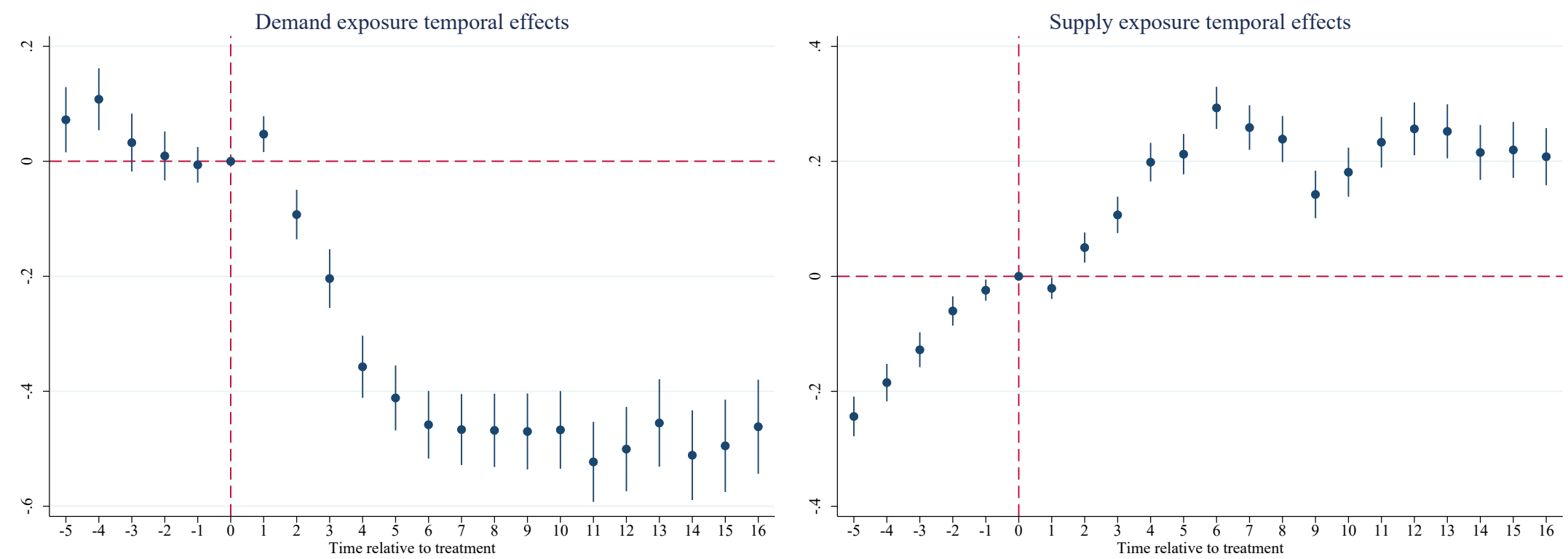

**Panel A2. Similarity to prior behavior.**

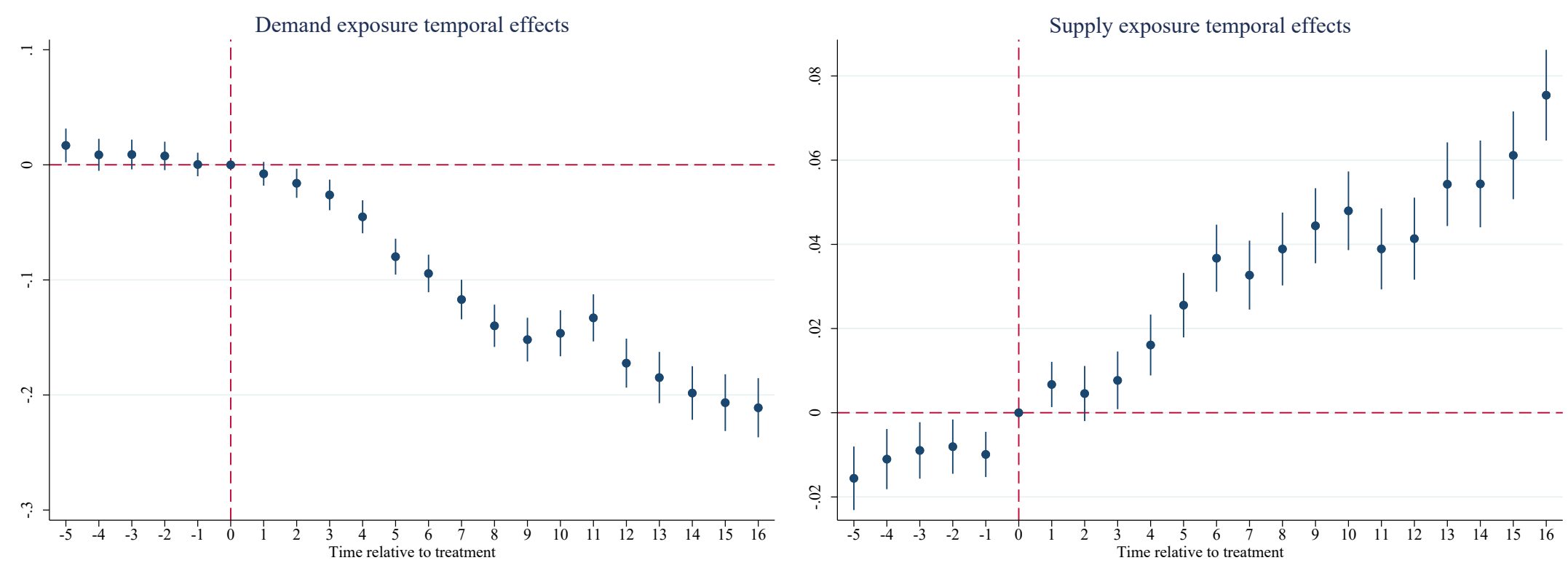

**Panel A3. Percent high-value bids.**

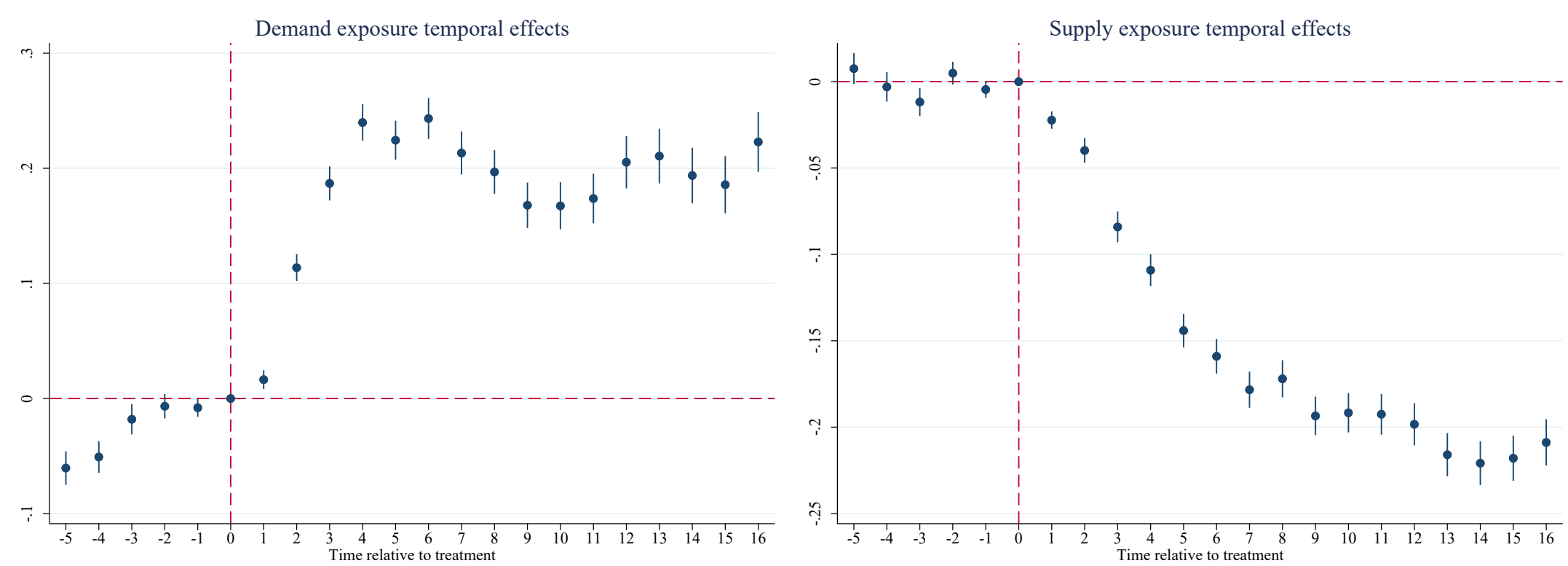

**Panel B. CEM estimates.**
**Panel B1. Log bids.**

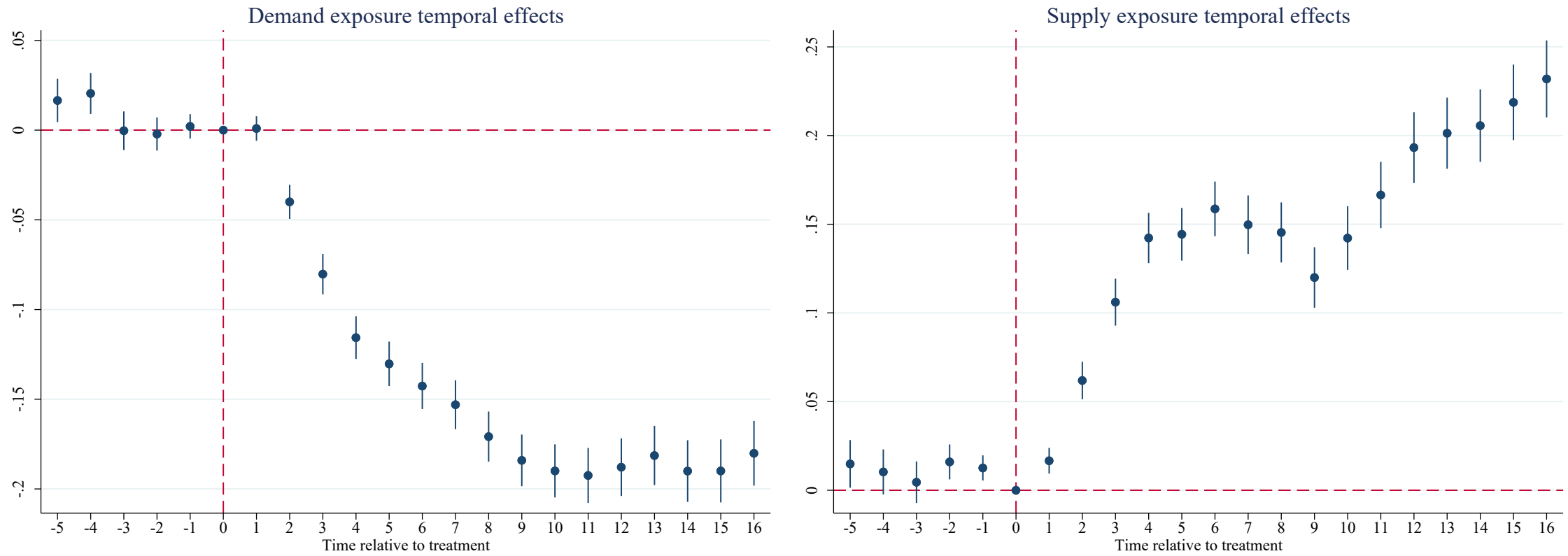

**Panel B2. Similarity to prior behavior.**

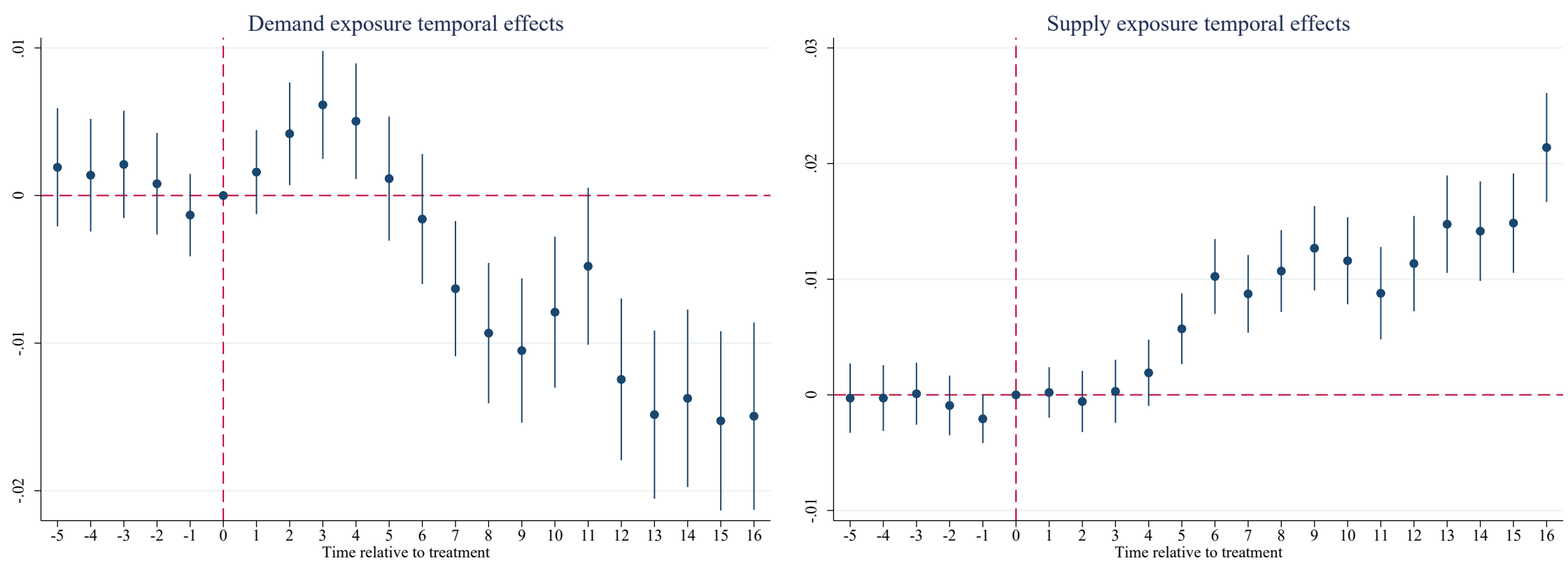

**Panel B3. Percent high-value bids.**

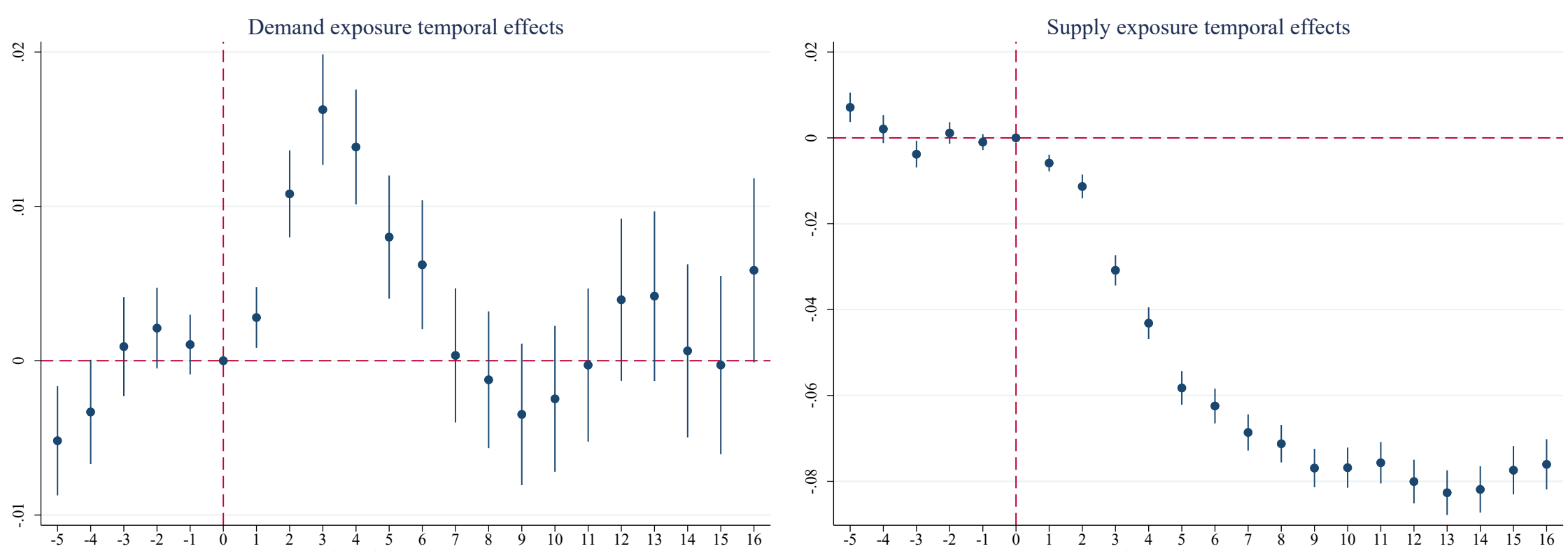

**Appendix Figure A9. Changes in freelancer strategic positioning pre- and post-ChatGPT by pre-period skill level with alternative sample selection.**

This figure is analogous to Figure 6 in the main draft, except that it uses an alternative sample of freelancer-year-month observations with at least 3 bids made in a rolling three-month window.

**Panel A. Full sample estimates.**

A1. Log bids.

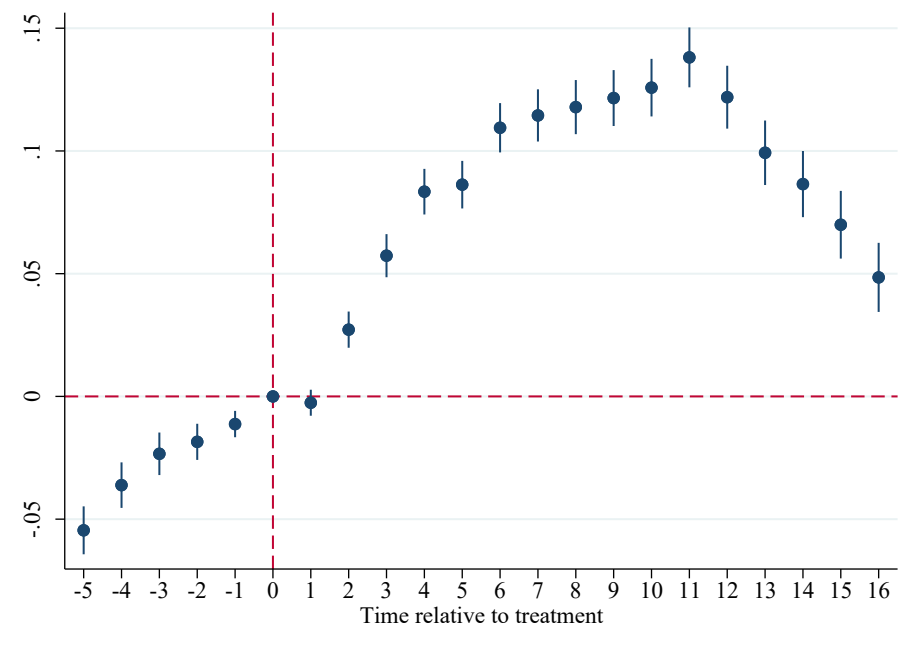

A2. Cosine similarity to prior behavior.

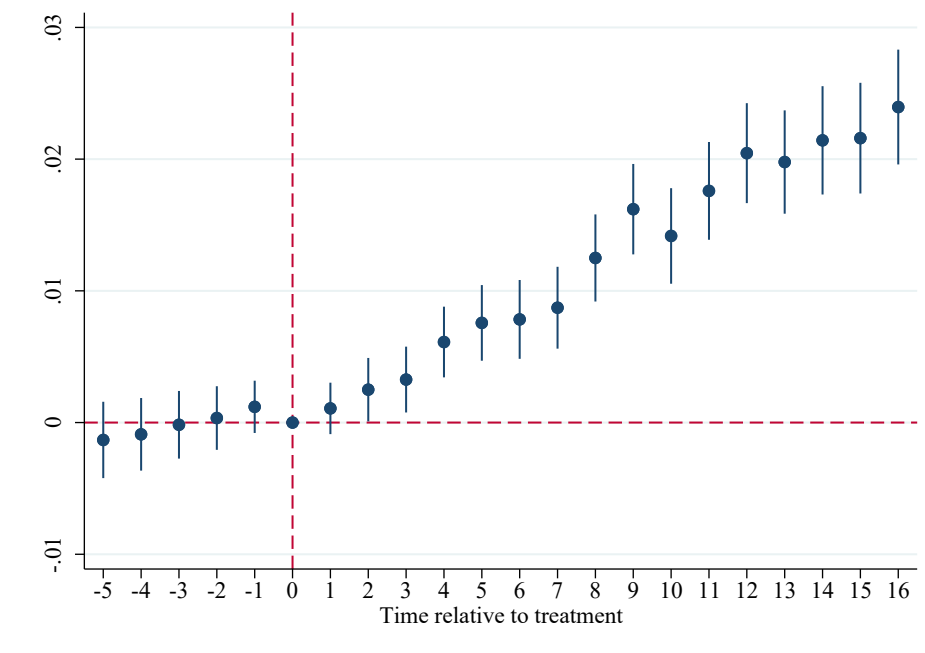

A3. Percent high value bids.

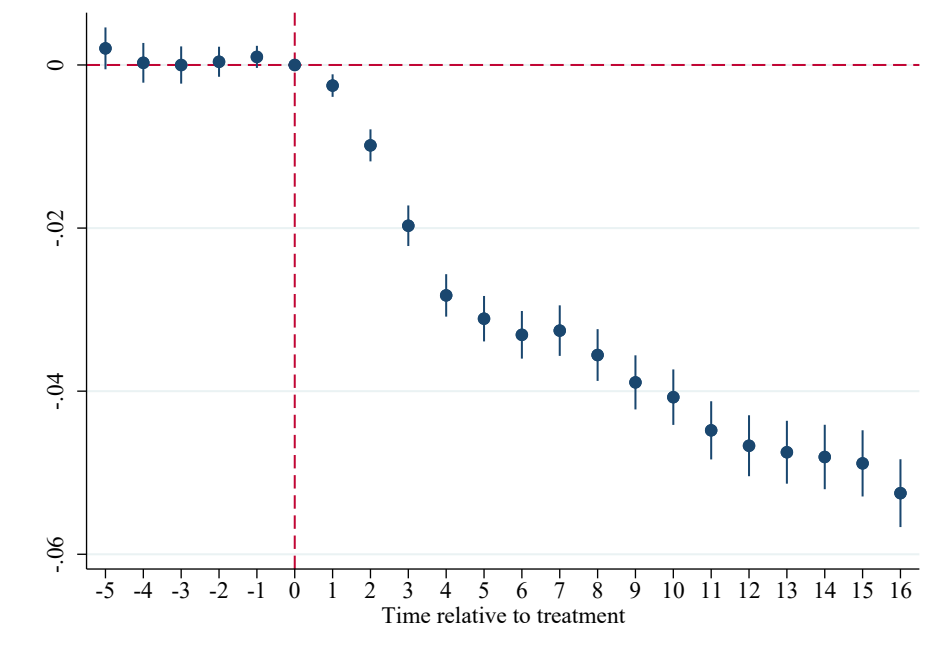

**Panel B. CEM estimates.**

B1. Log bids.

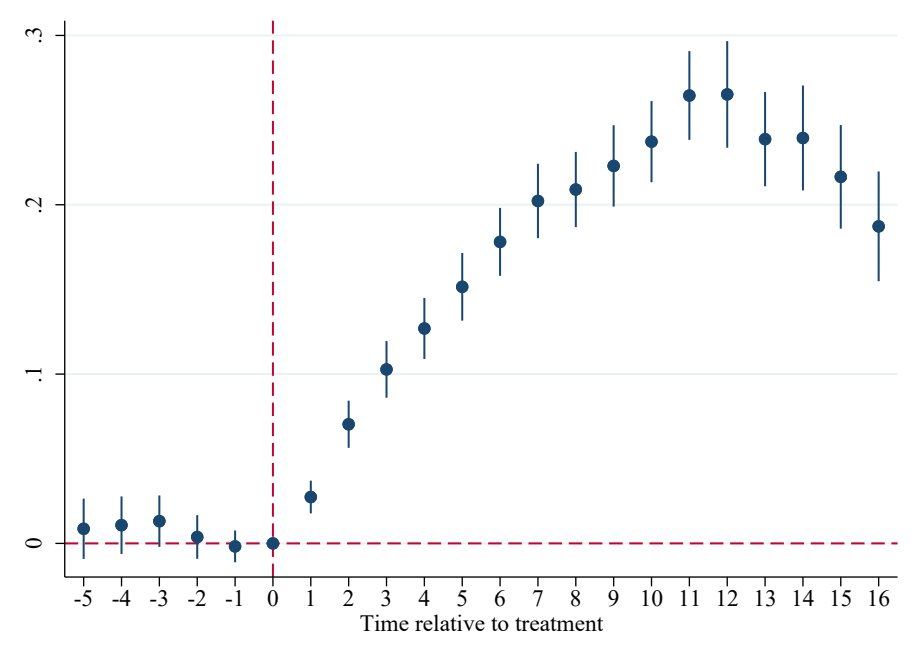

B2. Similarity to prior behavior.

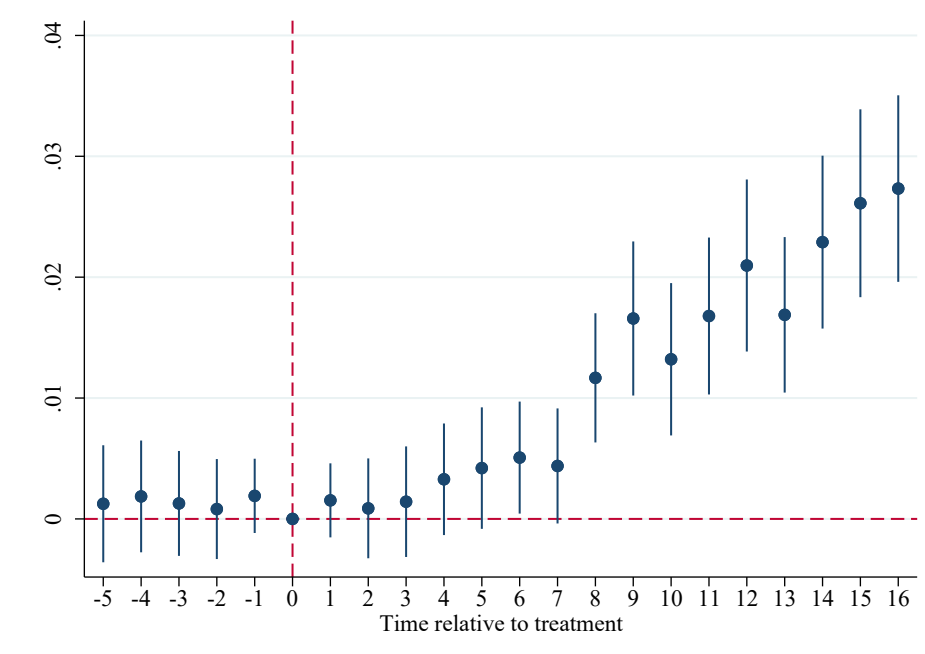

B3. Percent high value bids.

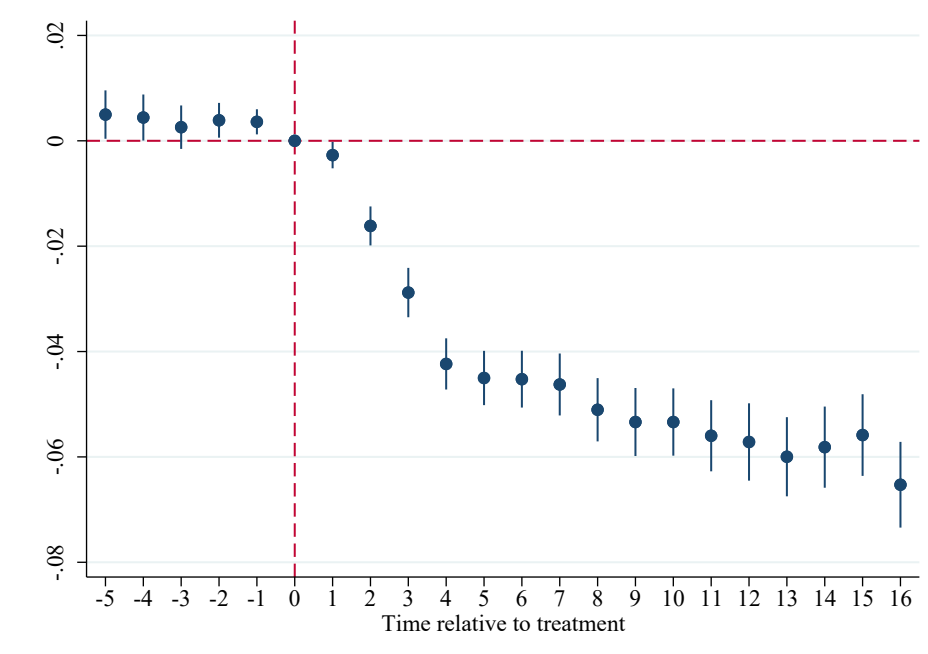

## Appendix Figure A10. Changes in freelancer strategic positioning pre- and post-ChatGPT by alternate measure of demand contraction and supply expansion exposure.

The following figures present the results of relative time models documenting changes in freelancer positioning from June 2022 through March 2024 by demand contraction exposure and supply expansion exposure. This figure is analogous to Figure 5 in the main manuscript, except that the measure of demand contraction and supply expansion exposure are constructed at the more granular work specialization level (rather than the broader work category level).

**Panel A. Full sample estimates.**

**Panel A1. Log bids.**

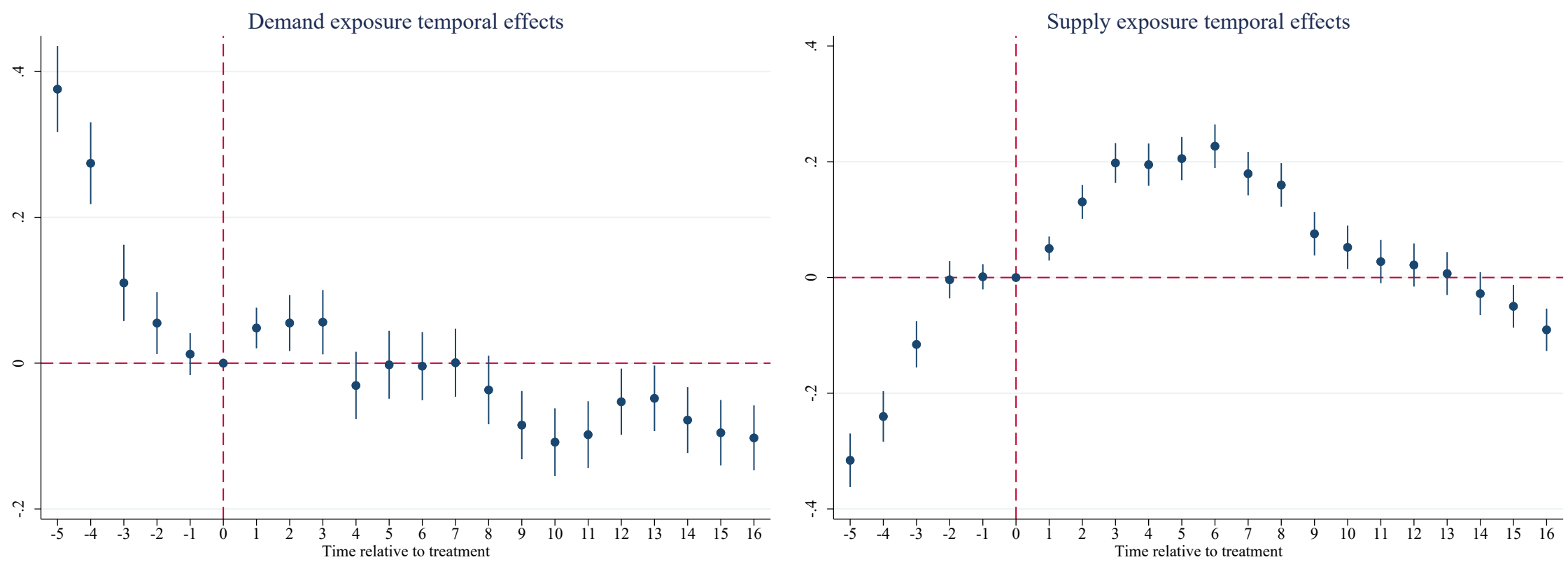

**Panel A2. Similarity to prior behavior.**

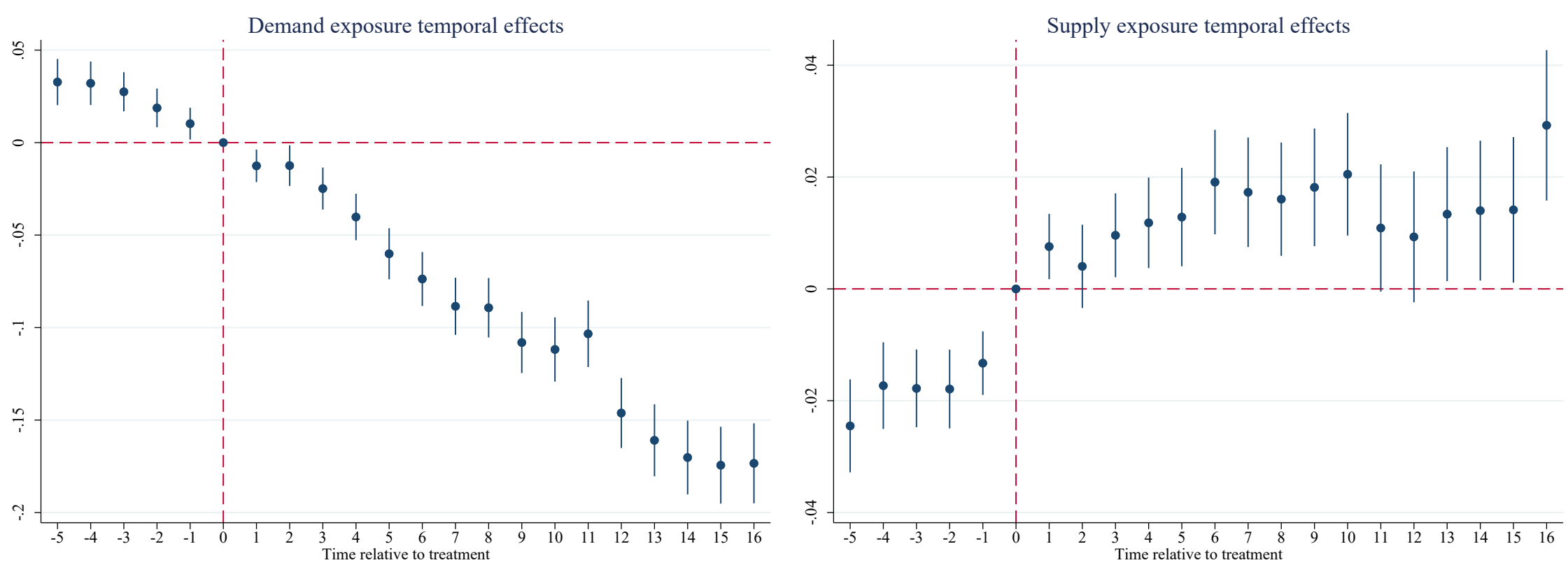

**Panel A3. Percent high-value bids.**

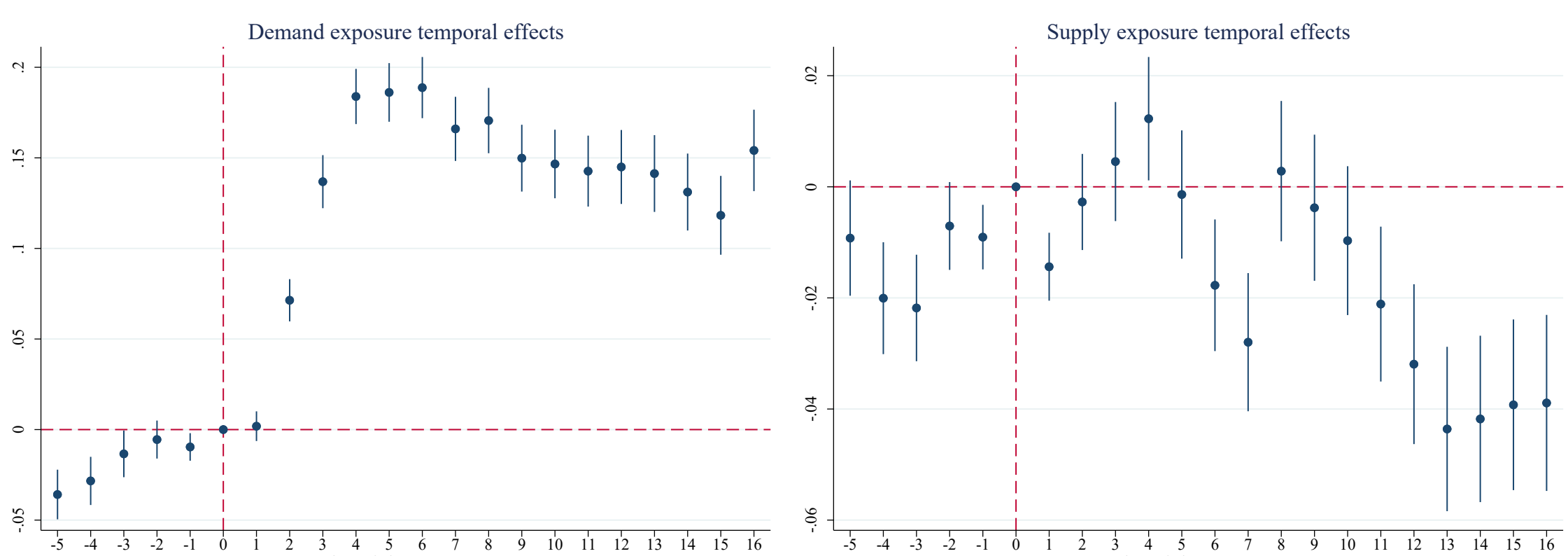

**Panel B. Coarsened Exact Matching (CEM) estimates.**
**Panel B1. Log bids.**

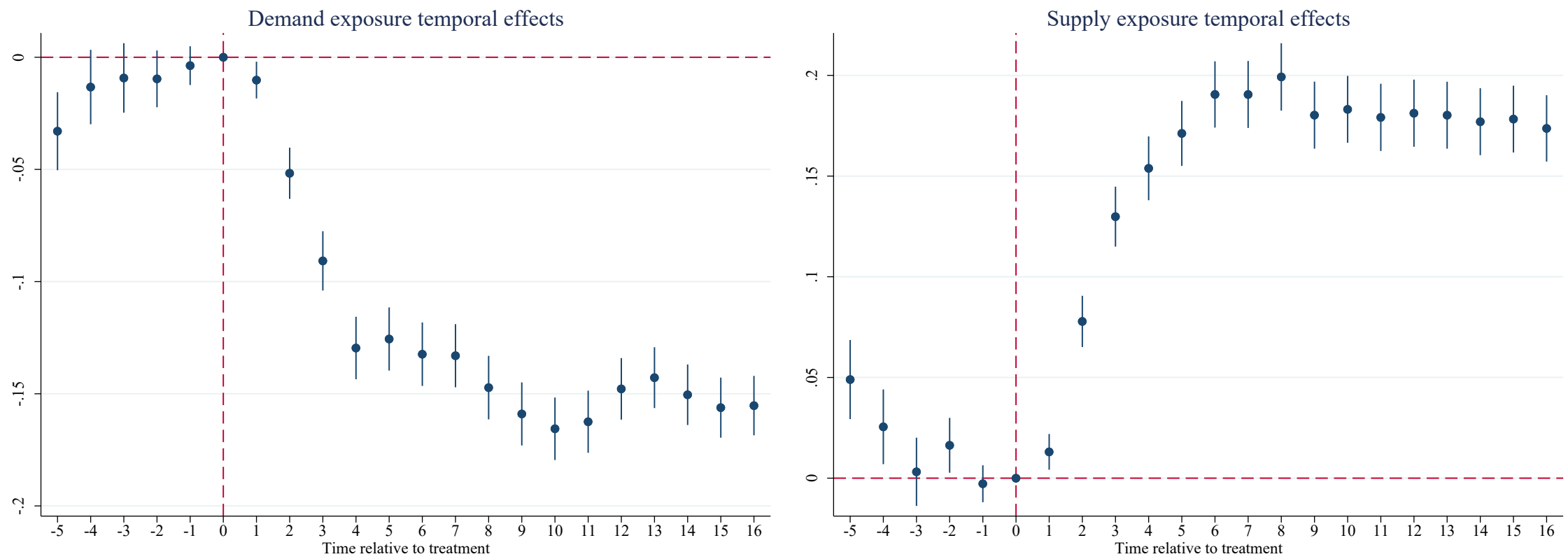

**Panel B2. Similarity to prior behavior.**

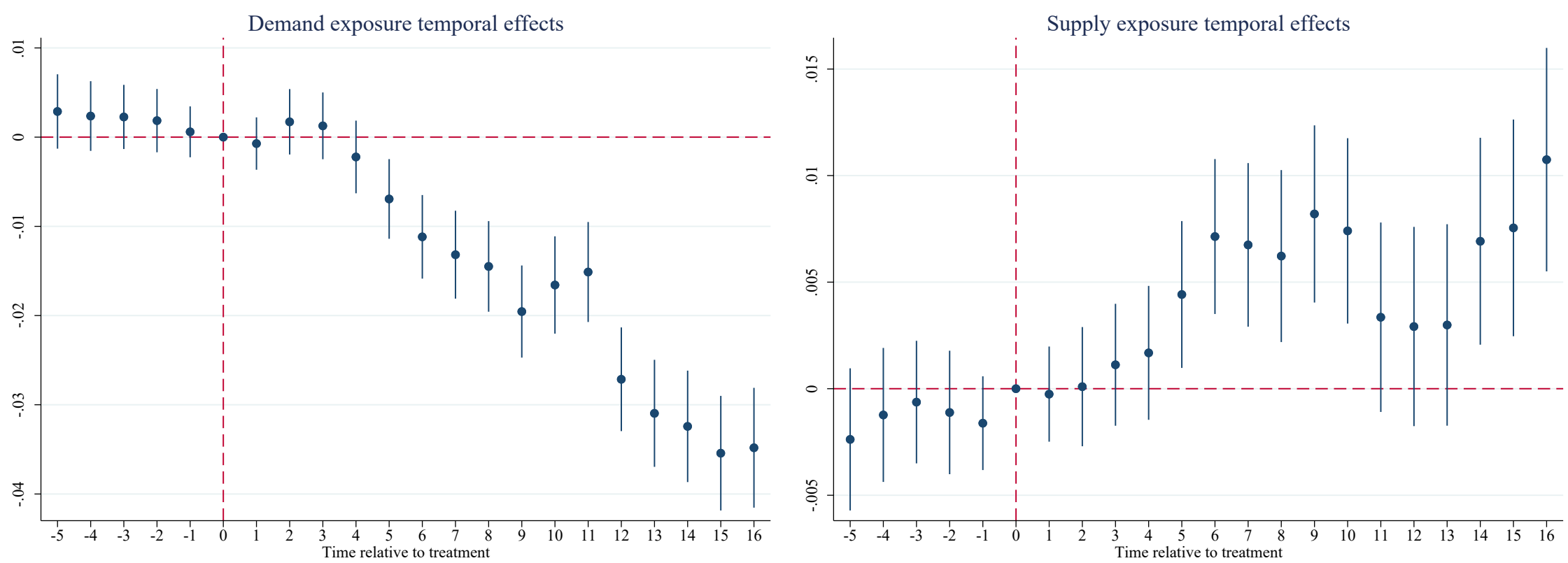

**Panel B3. Percent high-value bids.**

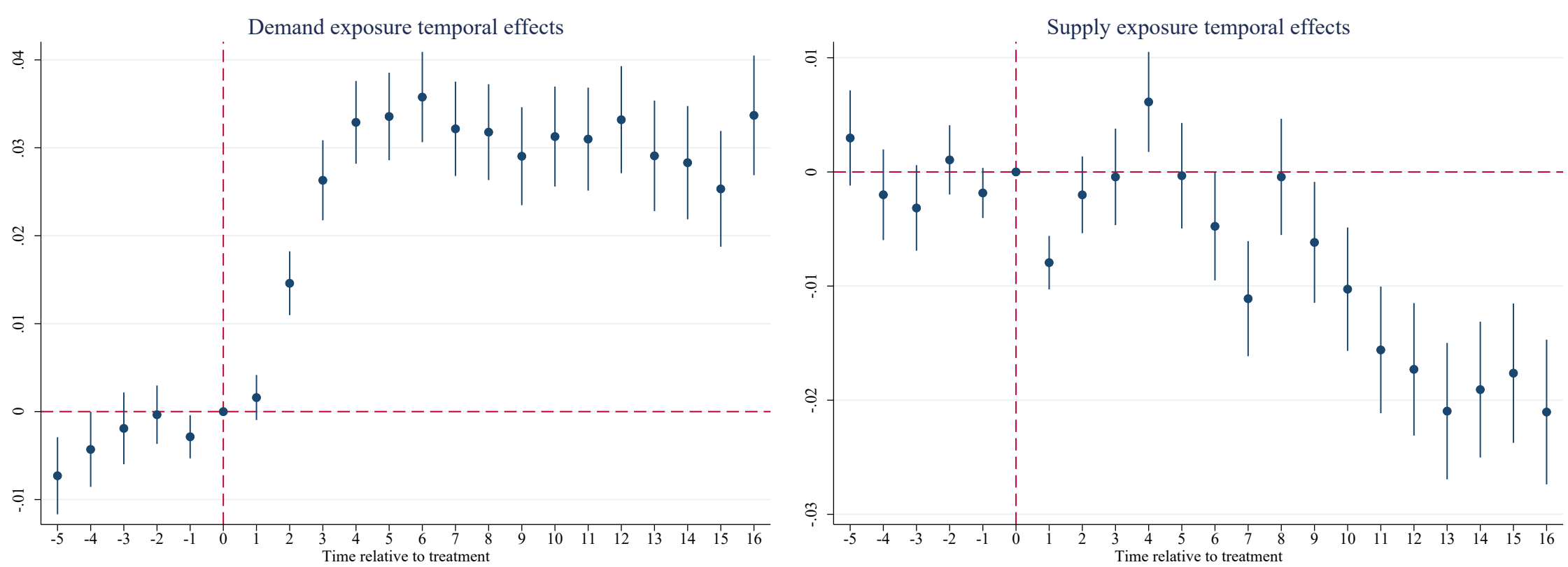

## Appendix Table A1. Pairwise correlations

The following tables shows pairwise correlations for relevant variables used in the main analysis. All variables are at the freelancer-month level. All correlations are significant at $p < 0.01$.

| | Variable | 1 | 2 | 3 | 4 | 5 | 6 | 7 |
|---|---|---|---|---|---|---|---|---|
| 1 | Number of bids | 1.000 | | | | | | |
| 2 | Similarity to prior behavior | 0.100 | 1.000 | | | | | |
| 3 | Percent high-value bids | 0.050 | 0.015 | 1.000 | | | | |
| 4 | Demand contraction exposure | 0.074 | 0.165 | 0.151 | 1.000 | | | |
| 5 | Supply expansion exposure | 0.038 | 0.062 | 0.102 | 0.219 | 1.000 | | |
| 6 | Total jobs pre-treatment | 0.407 | 0.090 | 0.013 | 0.014 | -0.032 | 1.000 | |
| 7 | Average skill pre-treatment | 0.084 | 0.153 | 0.209 | 0.219 | 0.055 | 0.105 | 1.000 |

**Appendix Table A2. Estimates of the effect of ChatGPT on freelancer strategic positioning.**

This table is analogous to Table 2 in the manuscript except that it displays robust standard errors clustered at the month-by-year level (rather than the freelancer-level) as in the main manuscript.

| | (1) | (2) | (3) | (4) | (5) | (6) |
|---|---|---|---|---|---|---|
| | Baseline sample | | | Panel differences sample | | |
| Dependent variable: | Log bids | Similarity to prior behavior | Percent high-value bids | Log bids | Similarity to prior behavior | Percent high-value bids |
| Post-ChatGPT | -0.963*** | -0.019*** | 0.006*** | | | |
| | (0.002) | (0.000) | (0.000) | | | |
| Post x treated | | | | -0.717*** | -0.018*** | -0.037*** |
| | | | | (0.003) | (0.001) | (0.001) |
| Log bids | | 0.050*** | -0.017*** | | 0.041*** | -0.013*** |
| | | (0.000) | (0.000) | | (0.000) | (0.000) |
| Freelancer fixed effects | Yes | Yes | Yes | Yes | Yes | Yes |
| Month fixed effects | No | No | No | Yes | Yes | Yes |
| Mean DV | 1.367 | 0.871 | 0.366 | 1.850 | 0.891 | 0.354 |
| No. of freelancers | 312,143 | 298,926 | 310,189 | 312,143 | 299,944 | 309,885 |
| R-squared | 0.605 | 0.458 | 0.583 | 0.537 | 0.446 | 0.633 |
| Observations | 6,633,007 | 3,790,149 | 4,430,937 | 5,702,926 | 3,999,792 | 4,602,481 |

**Appendix Table A3. Estimates of the heterogeneous effect of ChatGPT on freelancer strategic positioning by freelancer ex-ante demand and supply change exposure.**

This table is analogous to Table 4 in the manuscript except that it displays robust standard errors clustered at the month-by-year level (rather than the freelancer-level) as in the main manuscript.

**Panel A. Baseline sample**

| | (1) | (2) | (3) |
|---|---|---|---|
| Dependent variable: | Log bids | Similarity to prior behavior | Percent high-value bids |
| Post-ChatGPT x … | | | |
| Demand contraction exposure | -0.458*** | -0.137*** | 0.210*** |
| | (0.022) | (0.005) | (0.006) |
| Supply expansion exposure | 0.232*** | 0.081*** | -0.141*** |
| | (0.015) | (0.003) | (0.003) |
| Log number of bids | | 0.045*** | -0.014*** |
| | | (0.000) | (0.000) |
| Freelancer fixed effects | Yes | Yes | Yes |
| Month-by-year fixed effects | Yes | Yes | Yes |
| Mean DV | 1.367 | 0.871 | 0.366 |
| No. of freelancers | 312,143 | 298,926 | 310,189 |
| R-squared | 0.654 | 0.461 | 0.584 |
| Observations | 6,633,007 | 3,790,149 | 4,430,937 |

**Panel B. Coarsened exact matching sample.**

| | (1) | (2) | (3) | (4) | (5) | (6) |
|---|---|---|---|---|---|---|
| Dependent variable: | Log bids | Similarity to prior behavior | Percent high-value bids | Log bids | Similarity to prior behavior | Percent high-value bids |
| Post-ChatGPT x … | | | | | | |
| High demand contraction exposure | -0.122*** | -0.004* | 0.007*** | | | |
| | (0.012) | (0.002) | (0.002) | | | |
| High supply expansion exposure | | | | 0.081*** | 0.019*** | -0.056*** |
| | | | | (0.009) | (0.002) | (0.008) |
| Controls | Yes | Yes | Yes | Yes | Yes | Yes |
| Freelancer fixed effects | Yes | Yes | Yes | Yes | Yes | Yes |
| Month-by-year fixed effects | Yes | Yes | Yes | Yes | Yes | Yes |
| Mean DV | 1.244 | 0.817 | 0.347 | 1.313 | 0.873 | 0.442 |
| No. of freelancers | 150,662 | 143,611 | 149,637 | 147,707 | 142,126 | 146,927 |
| R-squared | 0.613 | 0.431 | 0.531 | 0.641 | 0.454 | 0.596 |
| Observations | 3,189,325 | 1,753,343 | 2,070,483 | 3,148,930 | 1,801,072 | 2,109,938 |

**Appendix Table A4. Estimates of the heterogeneous effect of ChatGPT on freelancer strategic positioning by freelancer skill.**

This table is analogous to Table 5 in the manuscript except that it displays robust standard errors clustered at the month-by-year level (rather than the freelancer-level) as in the main manuscript.

| | (1) | (2) | (3) | (4) | (5) | (6) |
|---|---|---|---|---|---|---|
| | | Baseline sample | | | CEM sample | |
| Dependent variable: | Log bids | Similarity to prior behavior | Percent high-value bids | Log bids | Similarity to prior behavior | Percent high-value bids |
| Post-ChatGPT x average skill pre-treatment | 0.120*** | 0.014*** | -0.028*** | | | |
| | (0.013) | (0.002) | (0.004) | | | |
| Post-ChatGPT x high average skill pre-treatment | | | | 0.164*** | 0.007*** | -0.042*** |
| | | | | (0.017) | (0.002) | (0.005) |
| Log number of bids | | 0.045*** | -0.015*** | | 0.051*** | -0.015*** |
| | | (0.001) | (0.001) | | (0.002) | (0.001) |
| Freelancer fixed effects | Yes | Yes | Yes | Yes | Yes | Yes |
| Modal pre-period work category x date fixed effects | Yes | Yes | Yes | Yes | Yes | Yes |
| Mean DV | 1.570 | 0.880 | 0.389 | 1.401 | 0.859 | 0.392 |
| No. of freelancers | 223,384 | 219,112 | 222,913 | 101,194 | 98,504 | 100,842 |
| R-squared | 0.658 | 0.455 | 0.584 | 0.639 | 0.456 | 0.591 |
| Observations | 4,800,894 | 3,107,253 | 3,515,942 | 2,159,230 | 1,308,865 | 1,511,164 |

**Appendix Table A5. Relative time estimates of the effect of ChatGPT on freelancer strategic positioning.**

The following table presents relative time regression estimates, following equation (1), the effect of ChatGPT on freelancer bidding activity. Robust standard errors clustered at the freelancer level in parentheses. ***, **, and * denote significance at the 1%, 5%, and 10% level, respectively.

| | (1) | (2) | (3) |
|---|---|---|---|
| Dependent variable: | Log bids | Similarity to prior behavior | Percent high-value bids |
| June 2022 | -0.214*** | 0.003*** | -0.007*** |
| | (0.003) | (0.001) | (0.001) |
| July 2022 | -0.173*** | 0.002*** | -0.005*** |
| | (0.003) | (0.001) | (0.001) |
| August 2022 | -0.107*** | 0.002*** | -0.001* |
| | (0.002) | (0.001) | (0.001) |
| September 2022 | -0.029*** | -0.000 | 0.001** |
| | (0.002) | (0.000) | (0.000) |
| October 2022 | 0.026*** | -0.000 | 0.002*** |
| | (0.001) | (0.000) | (0.000) |
| December 2022 | -0.299*** | 0.001* | -0.003*** |
| | (0.001) | (0.000) | (0.000) |
| January 2023 | -0.500*** | -0.002*** | -0.008*** |
| | (0.002) | (0.001) | (0.001) |
| February 2023 | -0.708*** | -0.005*** | -0.010*** |
| | (0.002) | (0.001) | (0.001) |
| March 2023 | -0.732*** | -0.012*** | -0.001** |
| | (0.002) | (0.001) | (0.001) |
| April 2023 | -0.848*** | -0.017*** | 0.006*** |
| | (0.002) | (0.001) | (0.001) |
| May 2023 | -0.923*** | -0.021*** | 0.013*** |
| | (0.002) | (0.001) | (0.001) |
| June 2023 | -1.035*** | -0.023*** | 0.015*** |
| | (0.002) | (0.001) | (0.001) |
| July 2023 | -1.086*** | -0.027*** | 0.015*** |
| | (0.002) | (0.001) | (0.001) |
| August 2023 | -1.148*** | -0.028*** | 0.015*** |
| | (0.002) | (0.001) | (0.001) |
| September 2023 | -1.192*** | -0.033*** | 0.013*** |
| | (0.002) | (0.001) | (0.001) |
| October 2023 | -1.244*** | -0.034*** | 0.013*** |
| | (0.002) | (0.001) | (0.001) |
| November 2023 | -1.308*** | -0.038*** | 0.014*** |
| | (0.002) | (0.001) | (0.001) |
| December 2023 | -1.361*** | -0.040*** | 0.013*** |
| | (0.002) | (0.001) | (0.001) |
| January 2024 | -1.379*** | -0.043*** | 0.013*** |
| | (0.002) | (0.001) | (0.001) |
| February 2024 | -1.408*** | -0.045*** | 0.012*** |
| | (0.002) | (0.001) | (0.001) |
| March 2024 | -1.445*** | -0.046*** | 0.006*** |
| | (0.002) | (0.001) | (0.001) |
| Log bids | | 0.046*** | -0.015*** |
| | | (0.000) | (0.000) |
| Freelancer fixed effects | Yes | Yes | Yes |
| Mean DV | 1.367 | 0.871 | 0.366 |
| No. of freelancers | 312,143 | 298,926 | 310,189 |
| R-squared | 0.654 | 0.46 | 0.583 |
| Observations | 6,633,007 | 3,790,149 | 4,430,937 |

**Appendix Table A6. Relative time estimates of the heterogeneous effect of ChatGPT on freelancer strategic positioning by freelancer ex-ante demand contraction and supply expansion exposure.**

The following table presents relative time regression estimates of the heterogeneous effect of ChatGPT on freelancer bidding activity by freelancer ex-ante demand contraction and supply expansion exposure. Robust standard errors clustered at the freelancer level in parentheses. ***, **, and * denote significance at the 1%, 5%, and 10% level, respectively.

| | (1) | (2) | (3) |
|---|---|---|---|
| Dependent variable: | Log bids | Similarity to prior behavior | Percent high-value bids |
| Demand contraction exposure x … | | | |
| June 2022 | 0.122*** | 0.024*** | -0.068*** |
| | (0.039) | (0.008) | (0.009) |
| July 2022 | 0.152*** | 0.023*** | -0.050*** |
| | (0.037) | (0.008) | (0.008) |
| August 2022 | 0.099*** | 0.019*** | -0.017** |
| | (0.035) | (0.007) | (0.008) |
| September 2022 | 0.074*** | 0.008 | -0.013* |
| | (0.028) | (0.007) | (0.007) |
| October 2022 | 0.054*** | 0.006 | -0.007 |
| | (0.019) | (0.006) | (0.005) |
| December 2022 | 0.025 | -0.013** | 0.016*** |
| | (0.018) | (0.006) | (0.005) |
| January 2023 | -0.090*** | -0.015** | 0.110*** |
| | (0.025) | (0.007) | (0.007) |
| February 2023 | -0.201*** | -0.033*** | 0.190*** |
| | (0.029) | (0.008) | (0.009) |
| March 2023 | -0.371*** | -0.054*** | 0.245*** |
| | (0.030) | (0.008) | (0.010) |
| April 2023 | -0.385*** | -0.102*** | 0.251*** |
| | (0.030) | (0.009) | (0.010) |
| May 2023 | -0.427*** | -0.127*** | 0.278*** |
| | (0.030) | (0.010) | (0.011) |
| June 2023 | -0.398*** | -0.156*** | 0.252*** |
| | (0.030) | (0.010) | (0.011) |
| July 2023 | -0.459*** | -0.164*** | 0.228*** |
| | (0.030) | (0.011) | (0.011) |
| August 2023 | -0.464*** | -0.185*** | 0.193*** |
| | (0.030) | (0.011) | (0.012) |
| September 2023 | -0.508*** | -0.199*** | 0.181*** |
| | (0.030) | (0.012) | (0.012) |
| October 2023 | -0.480*** | -0.183*** | 0.182*** |
| | (0.029) | (0.012) | (0.012) |
| November 2023 | -0.429*** | -0.245*** | 0.207*** |
| | (0.029) | (0.013) | (0.013) |
| December 2023 | -0.425*** | -0.264*** | 0.221*** |
| | (0.029) | (0.013) | (0.014) |
| January 2024 | -0.470*** | -0.281*** | 0.218*** |
| | (0.029) | (0.014) | (0.014) |
| February 2024 | -0.495*** | -0.287*** | 0.182*** |
| | (0.029) | (0.014) | (0.014) |
| March 2024 | -0.498*** | -0.297*** | 0.236*** |
| | (0.028) | (0.015) | (0.015) |
| Supply expansion exposure x … | | | |
| June 2022 | -0.278*** | -0.028*** | 0.000 |
| | (0.023) | (0.004) | (0.005) |
| July 2022 | -0.212*** | -0.027*** | -0.013** |
| | (0.022) | (0.004) | (0.005) |
| August 2022 | -0.122*** | -0.027*** | -0.016*** |
| | (0.021) | (0.004) | (0.005) |
| September 2022 | -0.009 | -0.023*** | 0.002 |
| | (0.017) | (0.004) | (0.004) |
| October 2022 | 0.002 | -0.021*** | -0.007** |
| | (0.011) | (0.003) | (0.003) |
| December 2022 | 0.033*** | 0.013*** | -0.024*** |
| | (0.011) | (0.003) | (0.003) |
| January 2023 | 0.139*** | 0.018*** | -0.037*** |
| | (0.015) | (0.004) | (0.005) |
| February 2023 | 0.271*** | 0.026*** | -0.079*** |
| | (0.018) | (0.004) | (0.006) |
| March 2023 | 0.311*** | 0.035*** | -0.107*** |
| | (0.019) | (0.004) | (0.006) |
| April 2023 | 0.315*** | 0.052*** | -0.147*** |
| | (0.019) | (0.005) | (0.006) |
| May 2023 | 0.341*** | 0.068*** | -0.170*** |
| | (0.020) | (0.005) | (0.006) |
| June 2023 | 0.259*** | 0.072*** | -0.188*** |
| | (0.020) | (0.005) | (0.006) |
| July 2023 | 0.199*** | 0.078*** | -0.174*** |
| | (0.020) | (0.005) | (0.007) |
| August 2023 | 0.066*** | 0.082*** | -0.188*** |
| | (0.020) | (0.005) | (0.007) |
| September 2023 | 0.054*** | 0.088*** | -0.198*** |
| | (0.020) | (0.006) | (0.007) |
| October 2023 | 0.061*** | 0.077*** | -0.200*** |
| | (0.020) | (0.006) | (0.007) |
| November 2023 | 0.068*** | 0.083*** | -0.210*** |
| | (0.020) | (0.006) | (0.007) |
| December 2023 | 0.052*** | 0.105*** | -0.215*** |
| | (0.019) | (0.006) | (0.007) |
| January 2024 | 0.021 | 0.110*** | -0.220*** |
| | (0.020) | (0.006) | (0.008) |
| February 2024 | 0.006 | 0.122*** | -0.221*** |
| | (0.019) | (0.007) | (0.008) |
| March 2024 | -0.031 | 0.145*** | -0.227*** |
| | (0.019) | (0.007) | (0.008) |
| Log bids | | 0.045*** | -0.014*** |
| | | (0.000) | (0.000) |
| Freelancer fixed effects | Yes | Yes | Yes |
| Month-by-year fixed effects | Yes | Yes | Yes |
| | | | |
| Mean DV | 1.367 | 0.871 | 0.366 |
| No. of freelancers | 312,143 | 298,926 | 310,189 |
| R-squared | 0.655 | 0.462 | 0.585 |
| Observations | 6,633,007 | 3,790,149 | 4,430,937 |

**Appendix Table A7. Relative time estimates of the heterogeneous effect of ChatGPT on freelancer strategic positioning by freelancer ex-ante demand contraction exposure with coarsened exact matching.**

The following table presents relative time regression estimates of the heterogeneous effect of ChatGPT on freelancer bidding activity by freelancer ex-ante demand contraction exposure with coarsened exact matching. Robust standard errors clustered at the freelancer level in parentheses. ***, **, and * denote significance at the 1%, 5%, and 10% level, respectively.

| | (1) | (2) | (3) |
|---|---|---|---|
| Dependent variable: | Log bids | Similarity to prior behavior | Percent high-value bids |
| High demand contraction exposure x … | | | |
| June 2022 | -0.002 | 0.004* | -0.006*** |
| | (0.008) | (0.002) | (0.002) |
| July 2022 | 0.006 | 0.004** | -0.003 |
| | (0.008) | (0.002) | (0.002) |
| August 2022 | -0.000 | 0.003 | 0.002 |
| | (0.007) | (0.002) | (0.002) |
| September 2022 | -0.005 | -0.000 | 0.001 |
| | (0.006) | (0.002) | (0.002) |
| October 2022 | -0.003 | -0.001 | 0.001 |
| | (0.004) | (0.001) | (0.001) |
| December 2022 | -0.003 | 0.001 | 0.003*** |
| | (0.004) | (0.002) | (0.001) |
| January 2023 | -0.039*** | 0.006*** | 0.011*** |
| | (0.005) | (0.002) | (0.002) |
| February 2023 | -0.084*** | 0.007*** | 0.016*** |
| | (0.006) | (0.002) | (0.002) |
| March 2023 | -0.122*** | 0.008*** | 0.015*** |
| | (0.007) | (0.002) | (0.002) |
| April 2023 | -0.121*** | 0.003 | 0.013*** |
| | (0.007) | (0.002) | (0.002) |
| May 2023 | -0.129*** | -0.001 | 0.013*** |
| | (0.007) | (0.002) | (0.002) |
| June 2023 | -0.130*** | -0.007** | 0.007*** |
| | (0.007) | (0.003) | (0.003) |
| July 2023 | -0.156*** | -0.006** | 0.001 |
| | (0.007) | (0.003) | (0.003) |
| August 2023 | -0.166*** | -0.008*** | -0.001 |
| | (0.007) | (0.003) | (0.003) |
| September 2023 | -0.169*** | -0.008*** | -0.001 |
| | (0.006) | (0.003) | (0.003) |
| October 2023 | -0.155*** | -0.004 | 0.002 |
| | (0.006) | (0.003) | (0.003) |
| November 2023 | -0.136*** | -0.013*** | 0.005* |
| | (0.006) | (0.003) | (0.003) |
| December 2023 | -0.132*** | -0.016*** | 0.006* |
| | (0.006) | (0.003) | (0.003) |
| January 2024 | -0.140*** | -0.016*** | 0.002 |
| | (0.006) | (0.003) | (0.003) |
| February 2024 | -0.144*** | -0.015*** | -0.006* |
| | (0.006) | (0.003) | (0.003) |
| March 2024 | -0.144*** | -0.016*** | 0.002 |
| | (0.006) | (0.004) | (0.003) |
| Supply expansion exposure x … | | | |
| June 2022 | -0.296*** | -0.014 | -0.040*** |
| | (0.052) | (0.013) | (0.013) |
| July 2022 | -0.279*** | -0.012 | -0.049*** |
| | (0.049) | (0.012) | (0.013) |
| August 2022 | -0.206*** | -0.018 | -0.021* |
| | (0.046) | (0.011) | (0.013) |
| September 2022 | -0.082** | -0.028** | 0.011 |
| | (0.038) | (0.011) | (0.010) |
| October 2022 | -0.024 | -0.037*** | -0.004 |
| | (0.026) | (0.009) | (0.007) |
| December 2022 | 0.075*** | 0.014 | 0.007 |
| | (0.024) | (0.009) | (0.007) |
| January 2023 | 0.088*** | 0.032*** | 0.026** |
| | (0.034) | (0.011) | (0.011) |
| February 2023 | 0.016 | 0.060*** | 0.022 |
| | (0.039) | (0.011) | (0.013) |
| March 2023 | -0.125*** | 0.068*** | -0.015 |
| | (0.041) | (0.012) | (0.014) |
| April 2023 | -0.145*** | 0.089*** | -0.040*** |
| | (0.041) | (0.014) | (0.015) |
| May 2023 | -0.163*** | 0.114*** | -0.077*** |
| | (0.041) | (0.014) | (0.015) |
| June 2023 | -0.219*** | 0.104*** | -0.085*** |
| | (0.041) | (0.015) | (0.015) |
| July 2023 | -0.279*** | 0.099*** | -0.101*** |
| | (0.041) | (0.015) | (0.016) |
| August 2023 | -0.341*** | 0.086*** | -0.079*** |
| | (0.041) | (0.016) | (0.017) |
| September 2023 | -0.323*** | 0.088*** | -0.102*** |
| | (0.041) | (0.016) | (0.017) |
| October 2023 | -0.337*** | 0.067*** | -0.076*** |
| | (0.040) | (0.017) | (0.017) |
| November 2023 | -0.331*** | 0.049*** | -0.095*** |
| | (0.040) | (0.018) | (0.018) |
| December 2023 | -0.325*** | 0.059*** | -0.079*** |
| | (0.040) | (0.019) | (0.018) |
| January 2024 | -0.331*** | 0.066*** | -0.095*** |
| | (0.040) | (0.019) | (0.019) |
| February 2024 | -0.351*** | 0.060*** | -0.143*** |
| | (0.039) | (0.020) | (0.019) |
| March 2024 | -0.366*** | 0.076*** | -0.156*** |
| | (0.039) | (0.021) | (0.020) |
| Log bids | | 0.068*** | -0.017*** |
| | | (0.000) | (0.000) |
| | | | |
| Freelancer fixed effects | Yes | Yes | Yes |
| Month-by-year fixed effects | Yes | Yes | Yes |
| | | | |
| Mean DV | 1.244 | 0.817 | 0.347 |
| No. of freelancers | 150,662 | 143,611 | 149,637 |
| R-squared | 0.614 | 0.431 | 0.531 |
| Observations | 3,189,325 | 1,753,343 | 2,070,483 |

**Appendix Table A8. Relative time estimates of the heterogeneous effect of ChatGPT on freelancer strategic positioning by freelancer ex-ante supply expansion exposure with coarsened exact matching.**

The following table presents relative time regression estimates of the heterogeneous effect of ChatGPT on freelancer bidding activity by freelancer ex-ante demand contraction exposure with coarsened exact matching. Robust standard errors clustered at the freelancer level in parentheses. ***, **, and * denote significance at the 1%, 5%, and 10% level, respectively.

| | (1) | (2) | (3) |
|---|---|---|---|
| Dependent variable: | Log bids | Similarity to prior behavior | Percent high-value bids |
| High supply expansion exposure x … | | | |
| June 2022 | 0.010 | -0.004** | 0.004** |
| | (0.009) | (0.002) | (0.002) |
| July 2022 | -0.003 | -0.004** | -0.001 |
| | (0.009) | (0.002) | (0.002) |
| August 2022 | 0.006 | -0.004*** | -0.004** |
| | (0.008) | (0.001) | (0.002) |
| September 2022 | 0.014** | -0.004*** | 0.001 |
| | (0.006) | (0.001) | (0.002) |
| October 2022 | 0.009** | -0.004*** | -0.002** |
| | (0.004) | (0.001) | (0.001) |
| December 2022 | -0.000 | 0.002 | -0.007*** |
| | (0.004) | (0.001) | (0.001) |
| January 2023 | 0.040*** | 0.002 | -0.012*** |
| | (0.006) | (0.001) | (0.002) |
| February 2023 | 0.091*** | 0.005*** | -0.029*** |
| | (0.007) | (0.001) | (0.002) |
| March 2023 | 0.109*** | 0.007*** | -0.044*** |
| | (0.007) | (0.002) | (0.002) |
| April 2023 | 0.121*** | 0.013*** | -0.061*** |
| | (0.008) | (0.002) | (0.002) |
| May 2023 | 0.127*** | 0.019*** | -0.064*** |
| | (0.008) | (0.002) | (0.002) |
| June 2023 | 0.103*** | 0.020*** | -0.070*** |
| | (0.008) | (0.002) | (0.002) |
| July 2023 | 0.084*** | 0.021*** | -0.070*** |
| | (0.008) | (0.002) | (0.003) |
| August 2023 | 0.051*** | 0.022*** | -0.076*** |
| | (0.008) | (0.002) | (0.003) |
| September 2023 | 0.063*** | 0.023*** | -0.081*** |
| | (0.008) | (0.002) | (0.003) |
| October 2023 | 0.077*** | 0.018*** | -0.079*** |
| | (0.008) | (0.002) | (0.003) |
| November 2023 | 0.095*** | 0.021*** | -0.084*** |
| | (0.008) | (0.002) | (0.003) |
| December 2023 | 0.107*** | 0.030*** | -0.085*** |
| | (0.008) | (0.003) | (0.003) |
| January 2024 | 0.106*** | 0.031*** | -0.083*** |
| | (0.008) | (0.003) | (0.003) |
| February 2024 | 0.110*** | 0.035*** | -0.078*** |
| | (0.008) | (0.003) | (0.003) |
| March 2024 | 0.108*** | 0.042*** | -0.080*** |
| | (0.008) | (0.003) | (0.003) |
| Demand contraction exposure x … | | | |
| June 2022 | -0.377*** | -0.036** | -0.015 |
| | (0.094) | (0.018) | (0.017) |
| July 2022 | -0.346*** | -0.034** | -0.023 |
| | (0.089) | (0.017) | (0.017) |
| August 2022 | -0.220*** | -0.023 | -0.009 |
| | (0.081) | (0.016) | (0.016) |
| September 2022 | -0.101* | -0.013 | 0.001 |
| | (0.060) | (0.015) | (0.013) |
| October 2022 | -0.090** | -0.026** | 0.009 |
| | (0.038) | (0.012) | (0.009) |
| December 2022 | -0.062 | 0.022* | -0.005 |
| | (0.038) | (0.013) | (0.010) |
| January 2023 | 0.023 | 0.012 | -0.094*** |
| | (0.055) | (0.016) | (0.015) |
| February 2023 | 0.065 | 0.057*** | -0.136*** |
| | (0.067) | (0.016) | (0.019) |
| March 2023 | 0.220*** | 0.066*** | -0.099*** |
| | (0.071) | (0.018) | (0.019) |
| April 2023 | 0.219*** | 0.094*** | -0.070*** |
| | (0.073) | (0.020) | (0.020) |
| May 2023 | 0.340*** | 0.117*** | -0.173*** |
| | (0.073) | (0.021) | (0.022) |
| June 2023 | 0.407*** | 0.130*** | -0.182*** |
| | (0.073) | (0.022) | (0.022) |
| July 2023 | 0.551*** | 0.152*** | -0.159*** |
| | (0.073) | (0.022) | (0.023) |
| August 2023 | 0.690*** | 0.167*** | -0.018 |
| | (0.073) | (0.023) | (0.023) |
| September 2023 | 0.706*** | 0.189*** | 0.017 |
| | (0.074) | (0.024) | (0.024) |
| October 2023 | 0.641*** | 0.182*** | 0.048* |
| | (0.073) | (0.025) | (0.025) |
| November 2023 | 0.540*** | 0.224*** | 0.022 |
| | (0.074) | (0.028) | (0.027) |
| December 2023 | 0.540*** | 0.243*** | 0.007 |
| | (0.073) | (0.032) | (0.029) |
| January 2024 | 0.610*** | 0.272*** | -0.046 |
| | (0.074) | (0.030) | (0.028) |
| February 2024 | 0.629*** | 0.253*** | -0.109*** |
| | (0.075) | (0.031) | (0.030) |
| March 2024 | 0.639*** | 0.289*** | -0.180*** |
| | (0.075) | (0.034) | (0.031) |
| Log bids | | 0.044*** | -0.014*** |
| | | (0.000) | (0.000) |
| Freelancer fixed effects | Yes | Yes | Yes |
| Month-by-year fixed effects | Yes | Yes | Yes |
| | | | |
| Mean DV | 1.313 | 0.873 | 0.442 |
| No. of freelancers | 147,707 | 142,126 | 146,927 |
| R-squared | 0.642 | 0.455 | 0.597 |
| Observations | 3,148,930 | 1,801,072 | 2,109,938 |

**Appendix Table A9. Relative time estimates of the heterogeneous effect of ChatGPT on freelancer strategic positioning by freelancer skill.**

The following table presents relative time regression estimates of the heterogeneous effect of ChatGPT on freelancer bidding activity by freelancer skill. Robust standard errors clustered at the freelancer level in parentheses. ***, **, and * denote significance at the 1%, 5%, and 10% level, respectively.

| | (1) | (2) | (3) |
|---|---|---|---|
| Dependent variable: | Log bids | Similarity to prior behavior | Percent high-value bids |
| Average skill pre-treatment x … | | | |
| June 2022 | 0.043*** | -0.005*** | 0.001 |
| | (0.007) | (0.002) | (0.002) |
| July 2022 | 0.049*** | -0.005*** | -0.000 |
| | (0.006) | (0.001) | (0.001) |
| August 2022 | 0.045*** | -0.004*** | 0.000 |
| | (0.006) | (0.001) | (0.001) |
| September 2022 | 0.028*** | -0.003** | 0.000 |
| | (0.005) | (0.001) | (0.001) |
| October 2022 | 0.008** | 0.001 | 0.001 |
| | (0.003) | (0.001) | (0.001) |
| December 2022 | 0.032*** | 0.001 | -0.003*** |
| | (0.003) | (0.001) | (0.001) |
| January 2023 | 0.080*** | 0.003** | -0.008*** |
| | (0.004) | (0.001) | (0.001) |
| February 2023 | 0.137*** | 0.003** | -0.016*** |
| | (0.005) | (0.001) | (0.002) |
| March 2023 | 0.166*** | 0.004*** | -0.027*** |
| | (0.005) | (0.001) | (0.002) |
| April 2023 | 0.164*** | 0.008*** | -0.031*** |
| | (0.005) | (0.002) | (0.002) |
| May 2023 | 0.177*** | 0.009*** | -0.033*** |
| | (0.005) | (0.002) | (0.002) |
| June 2023 | 0.181*** | 0.012*** | -0.030*** |
| | (0.005) | (0.002) | (0.002) |
| July 2023 | 0.186*** | 0.015*** | -0.031*** |
| | (0.005) | (0.002) | (0.002) |
| August 2023 | 0.181*** | 0.019*** | -0.034*** |
| | (0.005) | (0.002) | (0.002) |
| September 2023 | 0.179*** | 0.016*** | -0.036*** |
| | (0.005) | (0.002) | (0.002) |
| October 2023 | 0.178*** | 0.022*** | -0.037*** |
| | (0.005) | (0.002) | (0.002) |
| November 2023 | 0.166*** | 0.024*** | -0.041*** |
| | (0.005) | (0.002) | (0.002) |
| December 2023 | 0.151*** | 0.026*** | -0.041*** |
| | (0.005) | (0.002) | (0.002) |
| January 2024 | 0.141*** | 0.028*** | -0.042*** |
| | (0.005) | (0.002) | (0.002) |
| February 2024 | 0.126*** | 0.030*** | -0.040*** |
| | (0.005) | (0.002) | (0.002) |
| March 2024 | 0.108*** | 0.029*** | -0.043*** |
| | (0.005) | (0.002) | (0.002) |
| Log bids | | 0.045*** | -0.015*** |
| | | (0.000) | (0.000) |
| | | | |
| Freelancer fixed effects | Yes | Yes | Yes |
| Modal pre-period work category x date fixed effects | Yes | Yes | Yes |
| | | | |
| Mean DV | 1.570 | 0.880 | 0.389 |
| No. of freelancers | 223,384 | 219,112 | 222,913 |
| R-squared | 0.658 | 0.456 | 0.584 |
| Observations | 4,800,894 | 3,107,253 | 3,515,942 |

**Appendix Table A10. Relative time estimates of the heterogeneous effect of ChatGPT on freelancer strategic positioning by freelancer skill with coarsened exact matching.**

The following table presents relative time regression estimates of the heterogeneous effect of ChatGPT on freelancer bidding activity by freelancer skill with coarsened exact matching. Robust standard errors clustered at the freelancer level in parentheses. ***, **, and * denote significance at the 1%, 5%, and 10% level, respectively.

| | (1) | (2) | (3) |
|---|---|---|---|
| Dependent variable: | Log bids | Similarity to prior behavior | Percent high-value bids |
| High skill pre-treatment x … | | | |
| June 2022 | 0.029** | -0.002 | 0.003 |
| | (0.011) | (0.003) | (0.003) |
| July 2022 | 0.018* | -0.002 | 0.004 |
| | (0.011) | (0.002) | (0.002) |
| August 2022 | 0.012 | -0.002 | 0.002 |
| | (0.010) | (0.002) | (0.002) |
| September 2022 | 0.011 | -0.001 | 0.004* |
| | (0.008) | (0.002) | (0.002) |
| October 2022 | 0.005 | 0.002 | 0.002 |
| | (0.005) | (0.002) | (0.001) |
| December 2022 | 0.027*** | 0.000 | -0.003** |
| | (0.005) | (0.002) | (0.002) |
| January 2023 | 0.071*** | -0.000 | -0.014*** |
| | (0.008) | (0.002) | (0.002) |
| February 2023 | 0.113*** | -0.003 | -0.027*** |
| | (0.009) | (0.002) | (0.003) |
| March 2023 | 0.138*** | -0.003 | -0.042*** |
| | (0.010) | (0.002) | (0.003) |
| April 2023 | 0.142*** | -0.001 | -0.045*** |
| | (0.010) | (0.003) | (0.003) |
| May 2023 | 0.167*** | -0.001 | -0.046*** |
| | (0.010) | (0.003) | (0.003) |
| June 2023 | 0.183*** | 0.002 | -0.042*** |
| | (0.010) | (0.003) | (0.003) |
| July 2023 | 0.204*** | 0.010*** | -0.044*** |
| | (0.011) | (0.003) | (0.003) |
| August 2023 | 0.210*** | 0.014*** | -0.052*** |
| | (0.010) | (0.003) | (0.003) |
| September 2023 | 0.226*** | 0.010*** | -0.050*** |
| | (0.010) | (0.003) | (0.003) |
| October 2023 | 0.234*** | 0.014*** | -0.051*** |
| | (0.010) | (0.003) | (0.004) |
| November 2023 | 0.238*** | 0.021*** | -0.053*** |
| | (0.011) | (0.004) | (0.004) |
| December 2023 | 0.229*** | 0.018*** | -0.054*** |
| | (0.011) | (0.004) | (0.004) |
| January 2024 | 0.222*** | 0.020*** | -0.056*** |
| | (0.011) | (0.004) | (0.004) |
| February 2024 | 0.213*** | 0.026*** | -0.052*** |
| | (0.011) | (0.004) | (0.004) |
| March 2024 | 0.200*** | 0.026*** | -0.059*** |
| | (0.011) | (0.004) | (0.004) |
| Log bids | | 0.051*** | -0.015*** |
| | | (0.001) | (0.001) |
| | | | |
| Freelancer fixed effects | Yes | Yes | Yes |
| Modal pre-period work category x date fixed effects | Yes | Yes | Yes |
| | | | |
| Mean DV | 1.401 | 0.859 | 0.392 |
| No. of freelancers | 101,194 | 98,504 | 100,842 |
| R-squared | 0.640 | 0.457 | 0.592 |
| Observations | 2,159,230 | 1,308,865 | 1,511,164 |